\documentclass{aa}
\usepackage{graphicx}
\usepackage{txfonts}
\usepackage{url}
\usepackage{color}
\usepackage{multirow,bigdelim}
\newcommand\kms{{\rm\,km\,s^{-1}}}

\newcommand\teff{T_{\rm eff}}

\usepackage{hyperref}
\hypersetup{
     breaklinks,
     colorlinks   = true,
     citecolor    = blue,
     urlcolor     = blue,
     linkcolor    = blue
}

\begin{document}

\title{
Exploring the Milky Way stellar disk
}
\subtitle{
Carbon, nitrogen, oxygen, sulphur, potassium, and copper abundances \\for 714 F- and G-type dwarf stars in the solar neighbourhood
}

\titlerunning{Exploring C, N, O, S, K, and Cu in the Milky Way stellar disk}

\author{Thomas Bensby}

\institute{Lund Observatory, Division of Astrophysics, Department of Physics, Lund University, Box 118, SE-221\,00 Lund, Sweden \\
\email{thomas.bensby@fysik.lu.se}
}


\date{Received 26 June 2026 / Accepted 20 August 2026}

\abstract{Elemental abundances in long-lived dwarf stars provide important constraints on the chemical evolution of the Galactic disk. While several large spectroscopic surveys provide abundances for vast stellar samples, small homogeneous  samples remain essential for defining precise abundance trends, calibrating survey abundance scales, and linking stellar chemistry to Galactic evolution.}
{We aim to determine abundances of carbon, nitrogen, sulphur, potassium, and copper for 714 nearby F- and G-type dwarf and subgiant stars, and to re-derive oxygen abundances using updated corrections for departures from the assumption of local thermodynamic equilibrium. We also revised the stellar ages using modern stellar evolutionary models together with {\sl Gaia} DR3 photometry and astrometry. These elements extend the chemical inventory of our previous studies and provide new constraints on the relative enrichment histories of the Galactic thin and thick disks.}
{Abundances were derived through spectral line synthesis using high-resolution and high signal-to-noise spectra, as well as the stellar parameters from our previous homogeneous analysis. Departures from local thermodynamic equilibrium were considered for all elements. We examined abundance trends with metallicity, oxygen, and stellar age, and compared the results with previous observational studies of Galactic disk stars.}
{The $\alpha$-element behaviour of oxygen is confirmed, with old stars defining an enhanced sequence relative to young stars. Sulphur closely follows oxygen, while potassium shows a broadly $\alpha$-like behaviour in [K/Fe] but residual trends relative to oxygen. Carbon shows a modest separation in [$X$/Fe], and there is no separation for nitrogen. Both carbon and nitrogen show much clearer population differences in [$X$/O]. Copper displays a strong metallicity dependence and clear separation between old and young populations when compared to oxygen. We also find that [O/Mg] is not constant, demonstrating that oxygen and magnesium provide complementary rather than interchangeable reference scales. Quantitative comparisons of all elements analysed in our studies show that carbon, oxygen, sulphur, and potassium rank among the most age-sensitive elements in the sample and provide strong discrimination between old and young disk populations.}
{The new abundance measurements substantially expand the diagnostic power of this local stellar sample. The results show that abundance ratios relative to oxygen, together with precise stellar ages, reveal population differences that are partly hidden in traditional [$X$/Fe] trends. The expanded abundance inventory provides a homogeneous reference dataset for studies of Galactic chemical evolution, Galactic archaeology, and large spectroscopic surveys.}
   \maketitle
\nolinenumbers

\section{Introduction}

Understanding how galaxies form and evolve remains a central challenge in astrophysics. A powerful way to address this question is through Galactic archaeology, which reconstructs the assembly history of the Milky Way from the dynamical and chemical properties of its long-lived stars \citep{freeman2002}. Chemical tagging, in this context, relies on the premise that stars born in the same molecular gas cloud share a common chemical fingerprint of the gas from which they formed billions of years ago. The atmospheres of late F- and G-type dwarf stars, with expected lifetimes of the same order as, or longer than, the current age of the Universe are particularly useful.  Their detailed abundance patterns provide a fossil record of the Galaxy’s past, encoding the relative contributions of nucleosynthetic processes operating on different timescales. By measuring these abundances with high precision, one can trace the enrichment history of the interstellar medium, identify stars that originated in common birth environments, and distinguish between populations that formed in situ and those that accreted from disrupted satellite galaxies.

Theory and observation have shown that no single element, or even small set of elements, is sufficient to capture the complexity of the Galaxy’s enrichment. Different nucleosynthetic channels such as core-collapse supernovae of type II (SNe\,II), supernovae of type Ia (SNe\,Ia), asymptotic giant branch (AGB) stars, rapidly rotating massive stars, and the weak and main s process, operate at different physical conditions and produce distinct elemental abundance patterns \citep[e.g.][]{nomoto2013,kobayashi2020}. A broad suite of elemental abundances is therefore required to disentangle these contributions. 

The distribution of stars in high-dimensional chemical space provides a powerful means of identifying stellar birth sites and tracing disrupted structures \citep{ting2015,ness2018,horta2023}. The {\sl Gaia} mission \citep{gaia_prusti2016}  has revealed that the Milky Way contains multiple overlapping stellar components, including both native disk and halo stars and accreted structures \citep[e.g.][]{belokurov2018,helmi2018}. Astrometry and kinematic information alone, however, is not sufficient to distinguish these populations, since debris from disrupted satellites can overlap significantly with the orbits of stars that formed within the Galaxy \citep{jeanbaptiste2017,das2020}. Chemistry provides the missing dimension where multi-dimensional abundance patterns can separate stars with similar kinematics but distinct origins, as demonstrated for several halo substructures \citep{horta2023}. Furthermore, combining detailed abundances with precise stellar ages offers a chronological sequence of enrichment events, which has recently been used to reconstruct the order of major accretion episodes in the Milky Way \citep{giribaldi2023}.

While large spectroscopic surveys such as GALAH \citep[Galactic Archaeology with HERMES,][]{desilva2015,martell2017}, APOGEE \citep[Apache Point Observatory Galactic Evolution Experiment,][]{majewski2017}, {\sl Gaia}-ESO \citep{gilmore2022,randich2022}, and the upcoming The Four-metre Multi-Object Spectroscopic Telescope (4MOST) Milky Way Disk and Bulge high- and low-resolution (4MIDABLE-HR/LR) surveys \citep{bensby2019,chiappini2019} have, and will, deliver chemical abundances for very large stellar samples, their interpretation relies on comparison with smaller benchmark datasets analysed homogeneously and at high precision. Such reference samples are important for placing survey abundances on a consistent scale, and for testing analysis pipelines.  One such dataset is the sample of 714 F- and G-type dwarf and subgiant stars by \citet[][Paper I]{bensby2014}, which was designed to explore the age and abundance structure of the Galactic disk in the solar neighbourhood. It provided detailed abundances for 13 elements spanning $\alpha$ elements, odd-Z elements, and iron-peak species (O, Na, Mg, Al, Si, Ca, Ti, Cr, Fe, Ni, Zn, Y, and Ba). Later papers expanded the chemical inventory considerably: odd iron-peak elements (Sc, V, Mn, and Co) were added in \citet[][Paper II]{battistini2015}, neutron-capture elements (Sr, Zr, La, Ce, Nd, Sm, and Eu) were added in \citet[][Paper III]{battistini2016}, and Li was added in \citet[][Paper IV]{bensby2018}. 

Despite this extensive coverage, several nucleosynthetically important elements have not yet been included in the series. These elements -- carbon, nitrogen, sulphur, potassium, and copper -- probe enrichment channels that are either poorly constrained or incompletely represented by the standard set of $\alpha$ elements, iron-peak elements, and neutron-capture species commonly analysed in dwarf stars. Carbon and nitrogen trace contributions from AGB stars and the primary–secondary behaviour of nitrogen, an area where Galactic chemical evolution models continue to face challenges \citep{henry2000b,vincenzo2018}. Sulphur is an $\alpha$ element that allows direct comparison between stellar abundances and measurements in high-redshift damped Ly-$\alpha$ systems, providing a bridge between local and early-Universe chemical enrichment \citep{nissen2007,matrozis2013}. Potassium is one of the most poorly reproduced elements in chemical evolution models, with a strong sensitivity to metallicity-dependent yields and possible contributions from hypernovae \citep{zhang2006b,romano2010}. Copper is influenced by both the weak s-process in massive stars and potentially SNe\,Ia, and it may offer complementary constraints to previously studied iron-peak and neutron-capture elements \citep{mishenina2002,reddy2003}. These elements have spectral features in our spectra, but they were not included in earlier analyses due to observational difficulties such as weak or blended lines, hyperfine structure, or strong non-local thermodynamic equilibrium (NLTE) effects. The inclusion of these elements enhances the diagnostic power of the sample for Galactic archaeology and provides a more complete reference dataset. For this fifth paper, we derived abundances of carbon, nitrogen, sulphur, potassium, and copper for all 714 stars in the sample.  We also re-evaluated the oxygen abundances from Paper~I taking NLTE effects into account.

\section{Abundance analysis and new ages}

The sample of 714 stars, mainly F- and G-type dwarf and subgiant stars, in the solar neighbourhood is the effort of several observing runs with high-resolution spectrographs at the main observing facilities around the World. Full details of the observations and abundance analysis are given in \cite{bensby2014}.

\subsection{Abundance determination}

The abundance analysis was done through spectral line synthesis based on a $\chi^2$-minimisation routine between the observed spectrum and a synthetic spectrum. The synthetic spectra were calculated with PySME \citep{wehrhahn2023,jian2026} which is the python implementation of the spectroscopy made easy (SME) software \citep{valenti1996,piskunov2017}.

PySME needs atomic line data, stellar parameters, broadening parameters, and model stellar atmospheres. For the latter we use the one-dimensional plane-parallel MARCS model atmospheres from \cite{gustafsson2008}. The stellar parameters ($\teff$, $\log g$, [Fe/H], and microturbulence) are taken from  \citep{bensby2014}, where the sample and the details of the determination of stellar parameters are fully described. Briefly, the stellar parameters were  based on equivalent width measurements and one-dimensional (1D), plane-parallel, LTE model stellar atmospheres calculated with the Uppsala MARCS code \citep{gustafsson1975,edvardsson1993,asplund1997}.  The effective temperature $(\teff$) was determined by requiring excitation balance of abundances from Fe\,{\sc i} lines, and the surface gravity ($\log g$) was determined by requiring ionisation balance between abundances from Fe\,{\sc i} and Fe\,{\sc ii} lines. The microturbulence parameter ($\xi_{\rm t}$) was obtained by requiring that abundances from Fe\,{\sc i} lines are independent of line strength. In every step of the analysis, NLTE corrections from \cite{lind2012} were applied to the abundances from individual Fe\,{\sc i} lines.

In addition to atomic line broadening, the observed line profile is broadened by the instrument, the line-of-sight component of the stellar rotation ($v_{\rm rot} \cdot \sin i$), and small- and large-scale motions in the stellar atmosphere (microturbulence, $\xi_{\rm t}$, and macroturbulence, $v_{\rm macro}$, respectively). The instrument broadening is set by the resolving power of the spectrograph ($R$) and is treated with a Gaussian profile, while $v_{\rm rot} \cdot \sin i$ and $v_{\rm macro}$ are jointly accounted for with a radial-tangential (RAD-TAN) profile. The RAD-TAN broadening values was taken from \cite{bensby2018}, where we analysed the Li line at 6707\,{\AA}. 

Atomic line data and atomic line broadening parameters were gathered from the VALD database \citep{vald_1,vald_2,vald_3,vald_4,vald_5,vald_6}.  When querying VALD we used the `extract stellar' option, using the solar stellar parameters ($\teff=5750$\,K, $\log g = 4,4$, $\rm [Fe/H]=0$, and $\xi_{\rm t}=1.00\,\kms$) and including all lines with estimated depths greater than 0.005, in a wavelength range of about $\pm 10$\,{\AA} around the line centre. These lists then includes both the spectral lines for the species of interest  as well as nearby and blending lines. In Table~\ref{tab:atomdata} we list the spectral lines that were analysed.   

It should be noted that VALD gives too many components for the three sulphur lines analysed in this work. These lines correspond to transitions between the (3p$^3$($^4$S$^{\circ}$)4p,{}$^5$P$_J$) and (3p$^3$($^4$S$^\circ$)5d,{}$^5$D$^{\circ}_{J'}$) terms. The allowed components are ($J=1\rightarrow J'=0,1,2$), ($J=2\rightarrow J'=1,2,3$), and ($J=3\rightarrow J'=2,3,4$) giving a nine fine-structure component system grouped around 6743, 6748, and 6757\,{\AA}. For the 6743\,{\AA} line the $J=1 \rightarrow J=0$ and $J=1 \rightarrow J=2$ transitions are duplicated in VALD; for the 6748\,{\AA} line the $J=2 \rightarrow J=2$ transition is duplicated; and for the $J=3 \rightarrow J=2$ transition is duplicated. These duplications were removed. Also, the wavelengths and oscillator strengths of these sulphur lines are highly uncertain, and VALD does not contain the most recent updates.

Therefore, atomic data for the \ion{S}{i} multiplet near 6750\,\AA\ were compiled using the recent calculations of \cite{li2026}.  Wavelengths were determined from experimental \ion{S}{i} energy levels in the NIST Atomic Spectra Database; these levels incorporate the revised level optimisation of \cite{civis2024}. These experimental energies were used by \cite{li2026} to calculate the corresponding Ritz vacuum wavelengths. For our purposes we quote the corresponding wavelengths in standard air. Oscillator strengths were adopted from the fine-tuned multi-configuration Dirac–Hartree–Fock/relativistic-configuration-interaction (MCDHF/RCI) calculations of \cite{li2026}. We adopt the fine-tuned Babushkin-gauge values, ($\log gf_{\rm B,ft}$), from their machine-readable transition table. The new oscillator strengths differ systematically from values commonly adopted in older stellar line lists. In particular, the three components near 6757\,\AA\ have ($\log gf=-1.857$), ($-1.012$), and ($-0.427$), compared with the commonly used Kurucz-based values of approximately ($-1.76$), ($-0.90$), and ($-0.31$).

Examples of the spectral line synthesis fits for the individual diagnostics used in the abundance analysis are shown in Figs.~\ref{fig:clines}-\ref{fig:culines} in the appendix. These figures illustrate the quality of the fits for representative stars spanning the range of stellar parameters covered by the sample and demonstrate the consistency of the abundances derived from the individual spectral lines.

\begin{table}[t]
\caption{Columns in the online table of abundance results table\tablefootmark{$\dagger$}.}
\label{tab:cds}
\centering
\small
\setlength{\tabcolsep}{4.5mm}
\begin{tabular}{lll}
\hline\hline
\noalign{\smallskip}
Column & Unit & Description \\
\hline
\noalign{\smallskip}
HIP & – & HIP identifier \\
$T_{\rm eff}$ & K & Effective temperature \\
e$T_{\rm eff}$ & K & Effective temperature  uncertainty\\
$\log g$ & cgs & Surface gravity \\
e$\log g$ & cgs & Surface gravity uncertainty\\
$[\mathrm{Fe/H}]$ & Sun & Iron abundance \\
e$[\mathrm{Fe/H}]$ & dex & Iron abundance uncertainty\\
Age & Gyr & Adopted stellar age \\
eAge & Gyr & Age uncertainty \\
Age source & - & This study or \cite{bensby2014} \\ 
$[\mathrm{C/H}]$ & Sun & Carbon abundance ratio \\
$[\mathrm{N/H}]$ & " & Nitrogen abundance ratio \\
$[\mathrm{O/H}]$ & " & Oxygen abundance ratio \\
$[\mathrm{S/H}]$ & " & Sulphur abundance ratio \\
$[\mathrm{K/H}]$ & " & Potassium abundance ratio \\
$[\mathrm{Cu/H}]$ & " & Copper abundance ratio \\
e[C/H] & dex & Carbon abundance uncertainty \\
e[N/H] & " & Nitrogen abundance uncertainty \\
e[O/H] & " & Oxygen abundance uncertainty \\
e[S/H] & " & Sulphur abundance uncertainty \\
e[K/H] & " & Potassium abundance uncertainty \\
e[Cu/H] & " & Copper abundance uncertainty \\
\hline
\end{tabular}
\tablefoot{
\tablefoottext{$\dagger$}{
The full table is only available at the CDS.
}}
\end{table}

\subsection{NLTE effects}

Departures from LTE arise because the statistical equilibrium of atoms in stellar atmospheres is governed by the balance between radiative and collisional processes, making realistic descriptions of collisions with neutral hydrogen particularly important \citep[e.g.][]{barklem2016}. For carbon, NLTE effects in late-type stars have been investigated by \cite{fabbian2006} and \cite{alexeeva2015}, who showed that the high-excitation \ion{C}{i} lines can experience significant departures from LTE, particularly at low metallicities. For oxygen, similar effects are present for the \ion{O}{i} triplet lines around 777\,nm, where the need for substantial NLTE corrections have been demonstrated \citep{fabbian2009a,amarsi2016,amarsi2020b}.

Corrections for NLTE effects are generally smaller for sulphur than for oxygen, but can become significant for some of the stronger near-infrared lines. \cite{takeda2005_zn} investigated the formation of several sulphur lines and showed that the magnitude of the NLTE corrections depends strongly on the specific transitions employed.

The \ion{K}{i} resonance line at 7699\,\AA\ has long been recognised as strongly affected by NLTE effects \citep[e.g.][]{takeda1996,zhang2006a}, and more recent calculations by \cite{reggiani2019} confirmed the importance of NLTE corrections across a wide range of stellar parameters.

Copper presents extensive hyperfine structure \citep{ankush2014} and NLTE effects \citep{shi2014,yan2015,korotin2018,andrievsky2018}. Both effects must be taken into account in order to obtain reliable copper abundances.

Departures from NLTE were considered for all species analysed. The model atoms edopted in this work use modern descriptions of inelastic collisions with neutral hydrogen based on asymptotic two-electron potentials combined with the Landau–Zener formalism for the collision dynamics, rather than the classical Drawin approximation \citep[e.g.][]{barklem2016,lind2024}. NLTE departure coefficients for carbon, nitrogen, oxygen, and potassium were calculated by \citet{amarsi2020b}, for copper by \citet{caliskan2025}, and for sulphur by \citet{amarsi2025}. The importance of these improved hydrogen collision data for deriving reliable Galactic abundance trends has been demonstrated for several elements, including copper \citep{caliskan2025}. The departure coefficients are available from Zenodo for direct ingestion into the PySME spectrum synthesis code.\footnote{\url{https://doi.org/10.5281/zenodo.3888393})}.
 
\begin{figure*}
\centering
\resizebox{0.95\hsize}{!}{
\includegraphics[trim={0 18mm 0 0}]{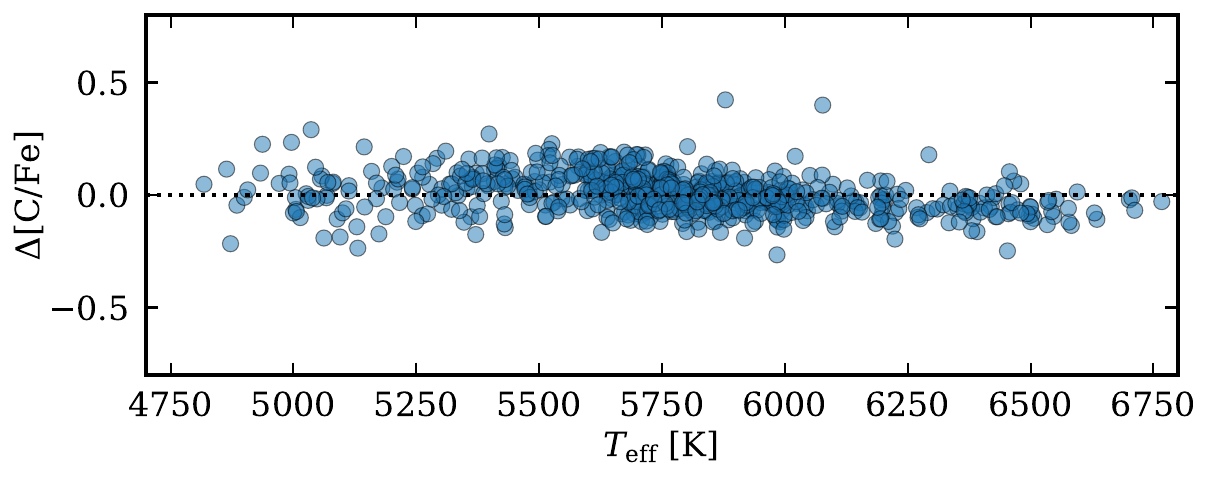}
\includegraphics[trim={0 18mm 0 0}]{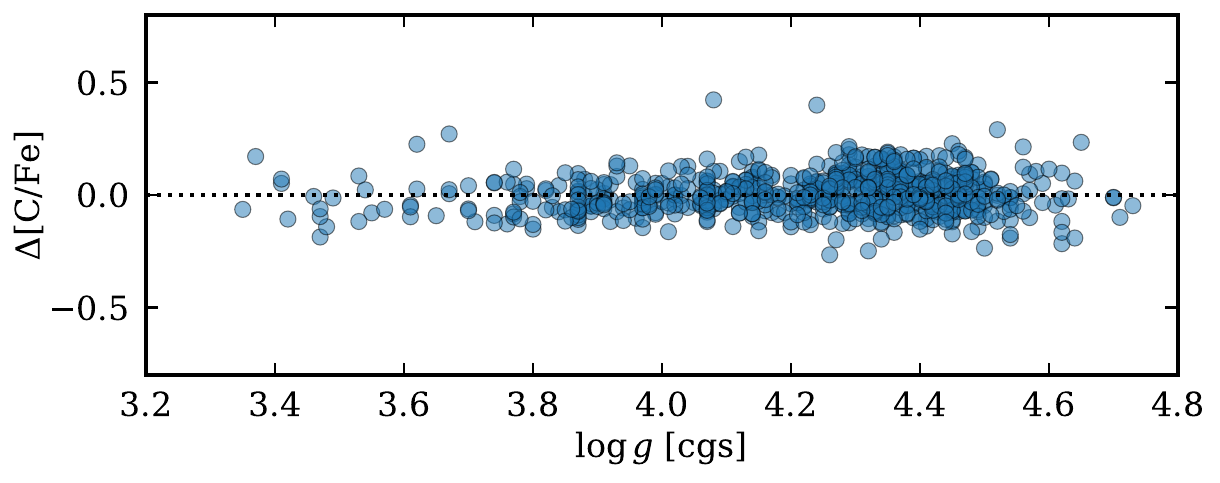}}
\resizebox{0.95\hsize}{!}{
\includegraphics[trim={0 18mm 0 0}]{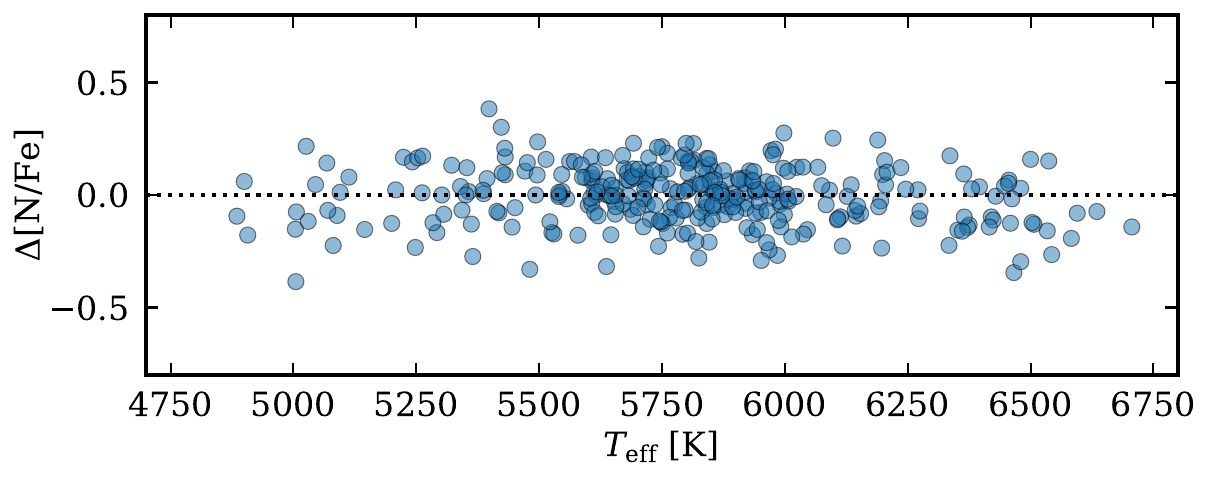}
\includegraphics[trim={0 18mm 0 0}]{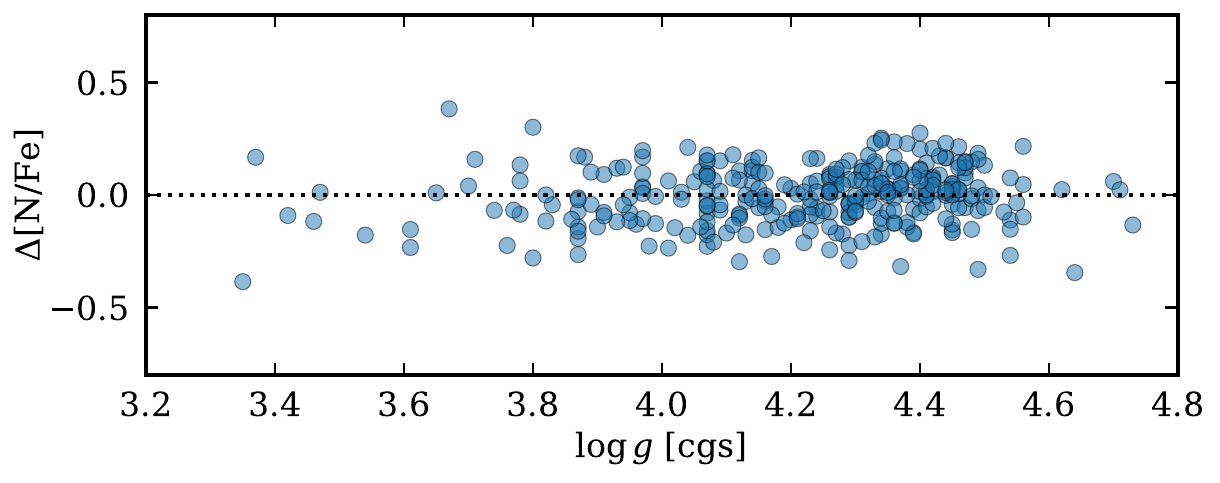}}
\resizebox{0.95\hsize}{!}{
\includegraphics[trim={0 18mm 0 0}]{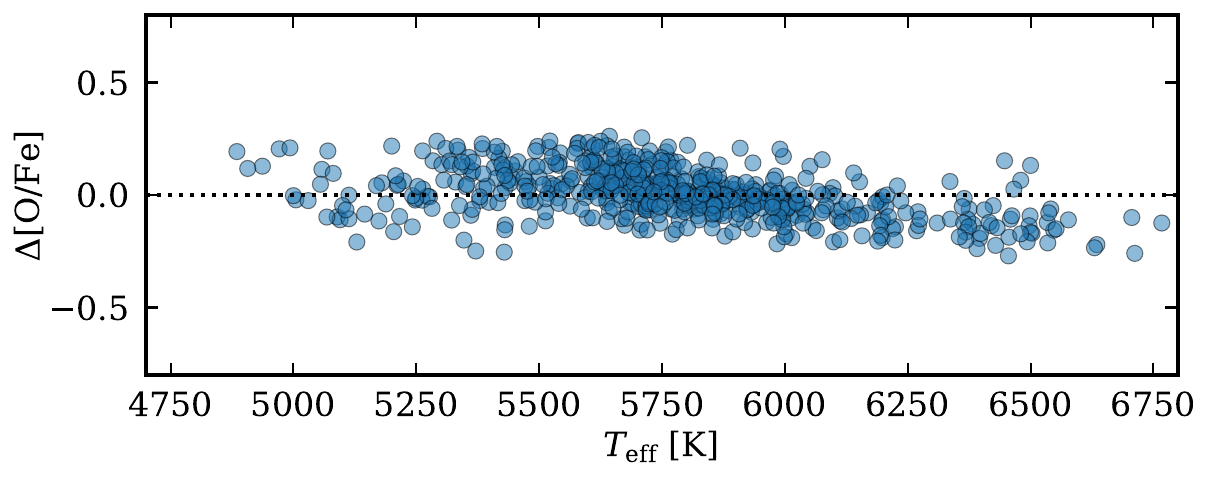}
\includegraphics[trim={0 18mm 0 0}]{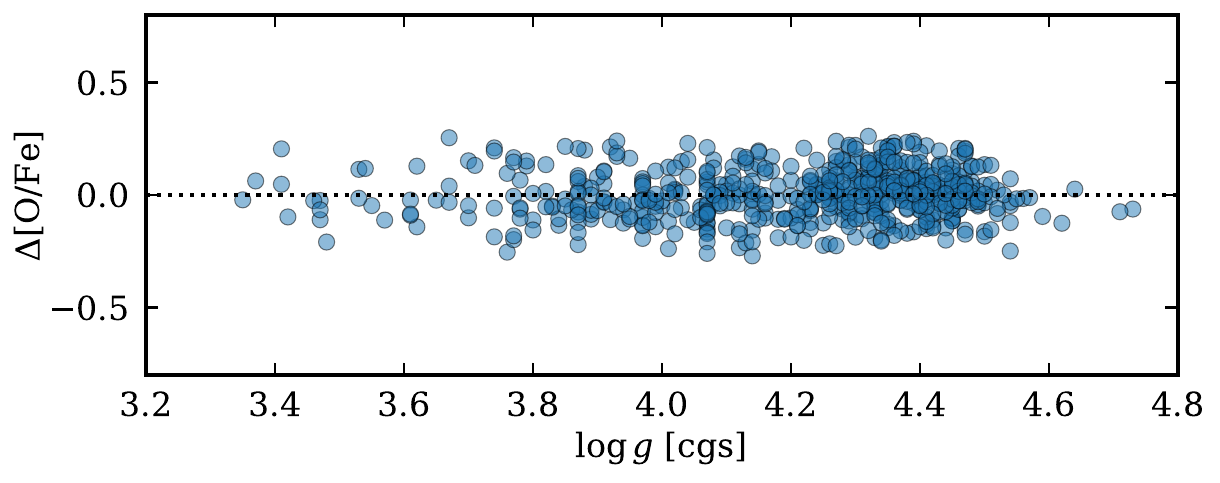}}
\resizebox{0.95\hsize}{!}{
\includegraphics[trim={0 18mm 0 0}]{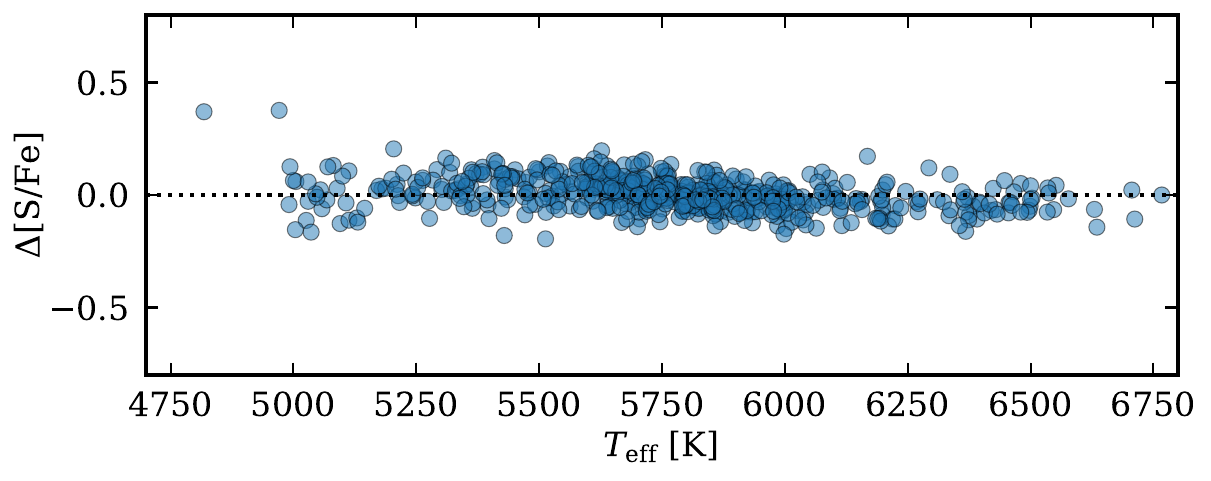}
\includegraphics[trim={0 18mm 0 0}]{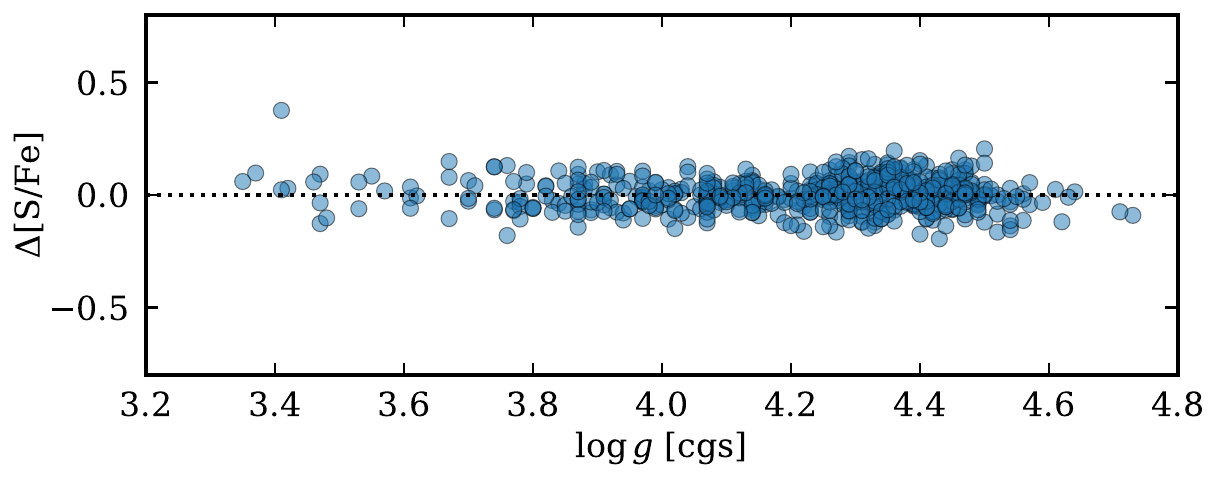}}
\resizebox{0.95\hsize}{!}{
\includegraphics[trim={0 18mm 0 0}]{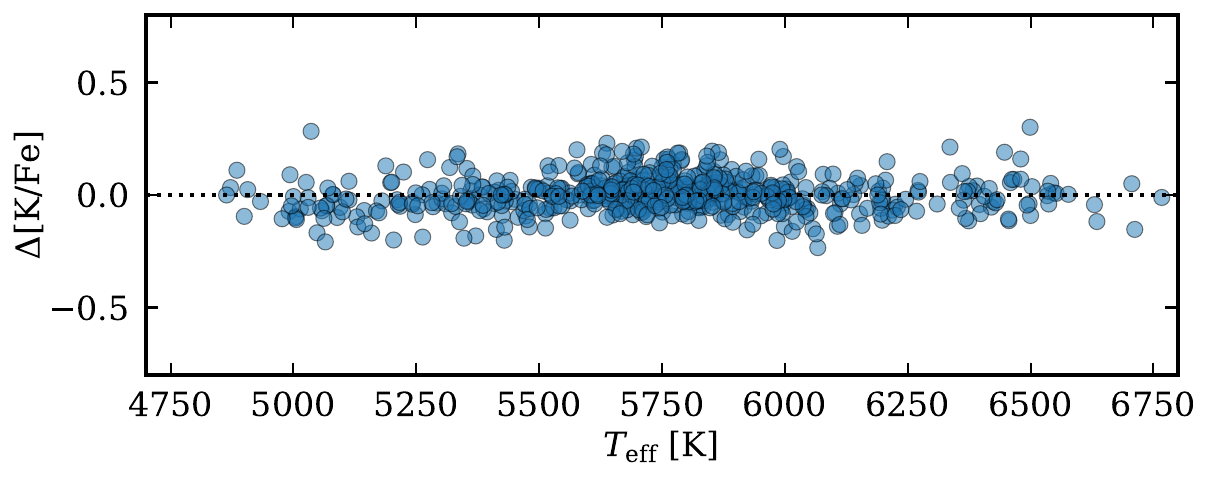}
\includegraphics[trim={0 18mm 0 0}]{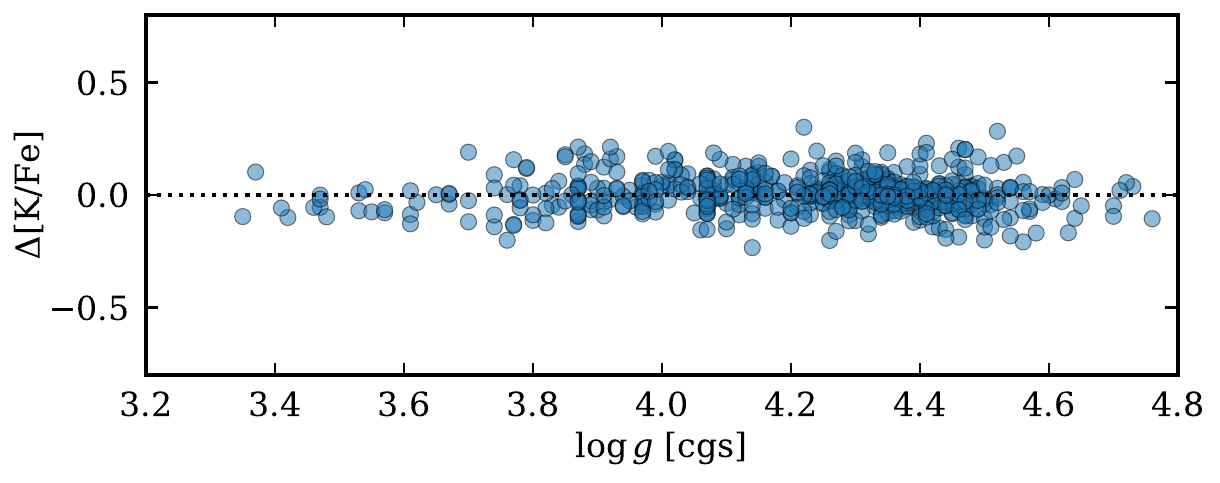}}
\resizebox{0.95\hsize}{!}{
\includegraphics{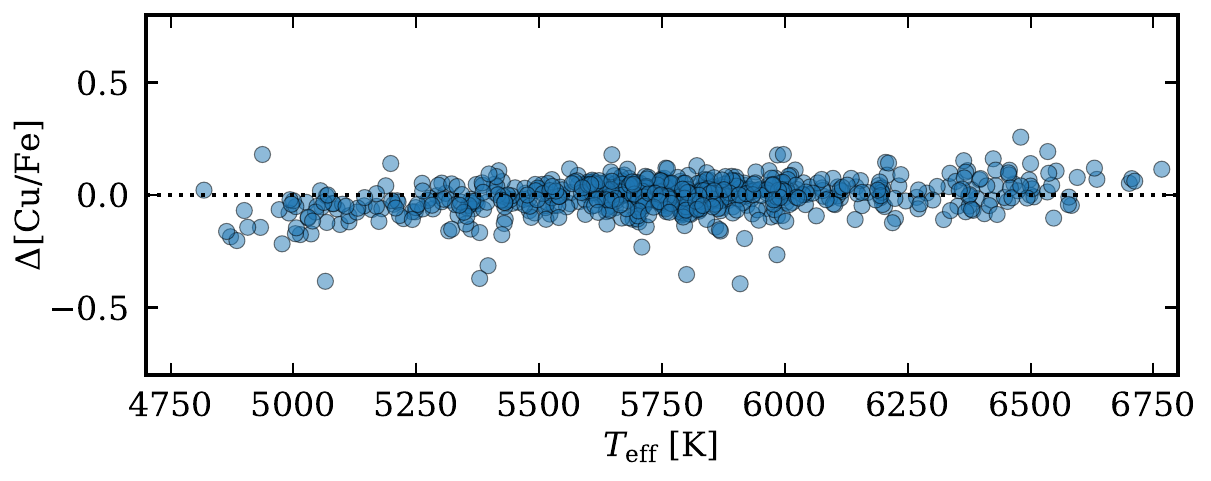}
\includegraphics{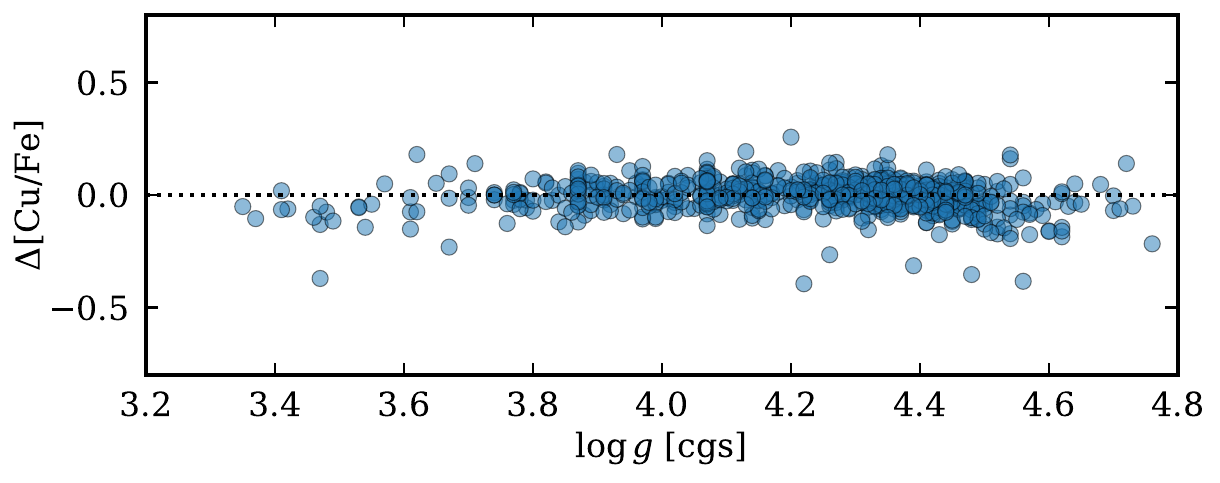}}
\caption{Residual abundance ratios, $\Delta[\mathrm{X/Fe}]$, as functions of effective temperature (left column) and surface gravity (right column). The residual abundances were calculated relative to the running median abundance trend as a function of metallicity.
\label{fig:xfe_teff}
}
\end{figure*}

\subsection{Solar analysis and abundance normalisation}
\label{sec:solaranalysis}

As in Papers~I--IV, stellar abundances were determined differentially to the Sun. For that four independent solar spectra were analysed using exactly the same line lists, model atmospheres, atomic data, line synthesis procedure, and NLTE treatment as for the stellar sample. For each spectral line, a solar abundance was determined from each solar spectrum and the final adopted solar abundance was taken as the median value. The solar abundances, and the 1$\sigma$ standard deviations, are listed in Table~\ref{tab:atomdata}.

The solar abundances derived from the four spectra show good internal consistency. For most lines the spectrum-to-spectrum scatter is below 0.05\,dex, with several lines exhibiting standard deviations of only 0.01--0.03\,dex. The \ion{O}{i} triplet lines are particularly consistent, yielding nearly identical abundances in all four spectra. Similarly small scatter is found for most of the carbon and copper lines. The largest line-to-line differences occur for sulphur, where the three multiplets yield somewhat different absolute abundances, although each multiplet individually remains highly consistent between the four solar spectra.

The abundances were then normalised on a line-by-line basis relative to the corresponding solar abundance. This strictly differential approach minimises systematic uncertainties arising from oscillator strengths, continuum placement, model atmospheres, and line formation calculations, and ensures a homogeneous abundance scale. Although some of the solar abundances derived here differ from the recommended absolute abundance scale of \citet{grevesse2007}, these offsets have negligible impact on the stellar abundance trends discussed in Sect.~\ref{sec:results} because all stellar abundances are determined differentially with respect to the same solar reference spectra.

The solar analysis described above could not be carried out for nitrogen as the lines were too weak in our solar spectra. To normalise the abundances for the two \ion{N}{i} features we therefore matched their respective [N/Fe]-[FeH] trends so that the average [N/Fe] values were close to solar ($\approx$0) at $\rm [Fe/H]=0$. The N abundances from the 7468\,{\AA} line were offset by $-0.18$\,dex and the N abundances from the CN+N feature at 8629\,{\AA} were offset by $-0.05$\,dex. These corrected abundances were then used to calculate the mean [N/H] abundances, assuming a standard solar N abundance of 7.78 as given by \cite{grevesse2007}.

\subsection{Error analysis}

Abundance uncertainties were estimated by repeating the analysis after perturbing the stellar parameters ($T_{\rm eff}$, $\log g$, [Fe/H], and $\xi_{\rm t}$) one at a time by their adopted uncertainties. The abundance differences relative to the nominal solution were then added in quadrature, assuming the individual contributions to be independent. This assumption is not exact, but it provides a standard estimate of the internal abundance uncertainties.

For elements with multiple measured lines ($N$), the standard error of the mean line abundance (line-to-line scatter divided by $\sqrt{N}$) was added in quadrature to the parameter-induced uncertainty. For elements based on one or very few lines, the uncertainty is dominated by the parameter sensitivity and the quality of the spectral fit. The uncertainties are reported in Table~\ref{tab:cds}.

The uncertainties above represent random errors arising from uncertainties in the adopted stellar parameters and the abundance measurements themselves. Additional systematic uncertainties may arise from the use of one-dimensional model atmospheres, the adopted NLTE calculations, atomic data, continuum placement, and details of the spectral synthesis procedure. These effects are difficult to quantify reliably and are not included in the error estimates presented here. However, as discussed in Sect.~\ref{sec:solaranalysis}, systematic uncertainties should be partially mitigated due to the differential line-by-line analysis to the Sun.

\subsection{Abundances versus stellar parameters}

The derived abundances should not exhibit systematic trends with the stellar parameters. To assess the robustness of the analysis, we examined abundance residuals relative to the running median abundance trends as functions of effective temperature and surface gravity. The residual abundances are defined as
\begin{equation}
\Delta[\mathrm{X/Fe}] =
[\mathrm{X/Fe}] -
[\mathrm{X/Fe}]_{\rm median,\,\,([\mathrm{Fe/H}])},
\end{equation}
where the second term represents the running median abundance trend as a function of metallicity. For all elements, the residuals remain centred around zero over the parameter range covered by the sample (see Fig.~\ref{fig:xfe_teff}). No strong systematic trends with either effective temperature or surface gravity are apparent.

The largest scatter is seen for nitrogen, reflecting both the weakness of the available nitrogen features and the smaller number of stars with measurable N abundances. Potassium and copper, despite their known sensitivity to NLTE effects and hyperfine structure, respectively, show no evidence for strong residual parameter trends. Likewise, the carbon, oxygen, and sulphur remain stable across the full range of effective temperatures and surface gravities. It is therefore  unlikely that the abundance trends presented in Sect.~\ref{sec:results} are  driven by residual systematic effects related to the stellar parameters.

\begin{figure}
\centering
\resizebox{0.95\hsize}{!}{
\includegraphics{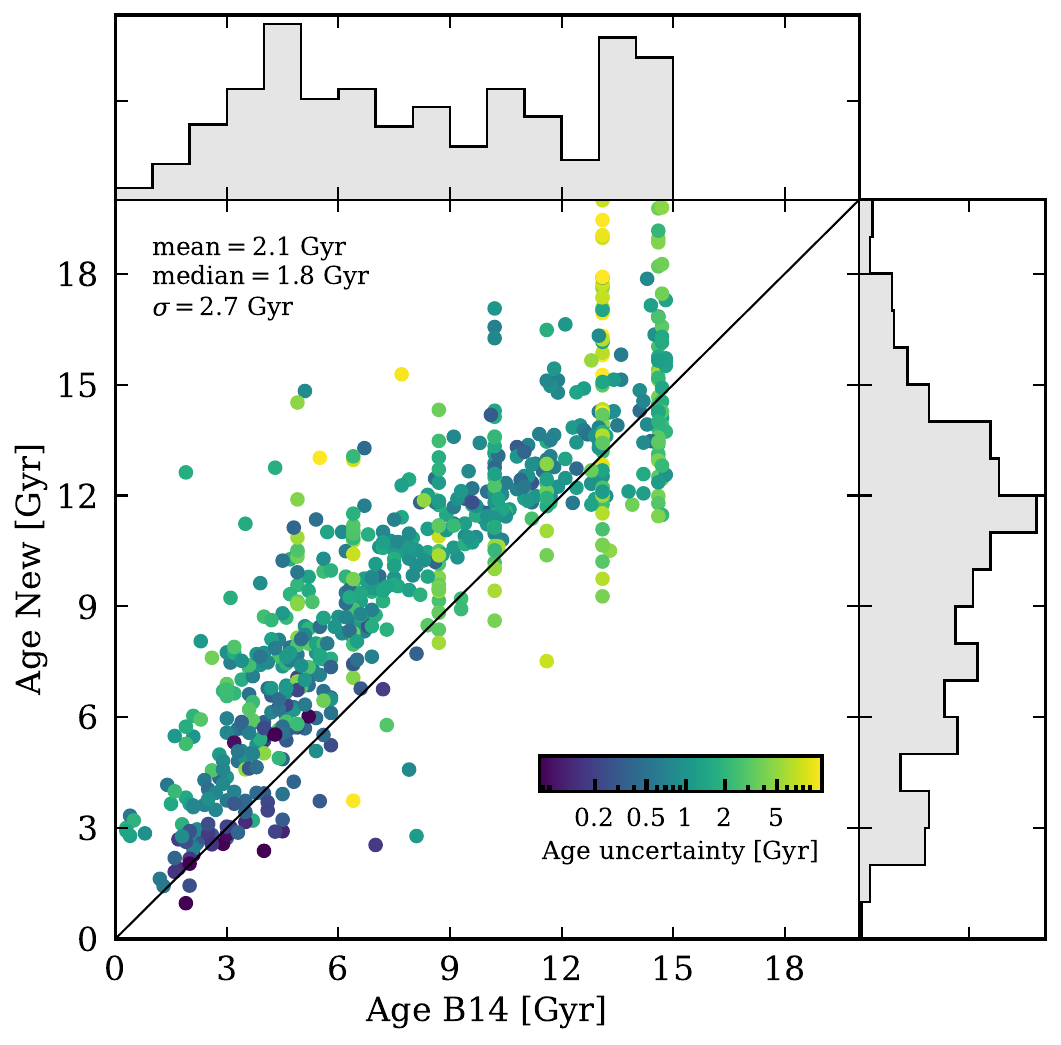}}
\caption{Comparison of the new age estimates using the MIST isochrones to the previous age estimates from \cite{bensby2014} for the 697 stars present and with non-negative parallaxes in {\sl Gaia} DR3. The mean and median age differences (new minus old) and the rms scatter is indicated. The points are colour-coded by the age uncertainties of the new ages.
\label{fig:newage}
}
\end{figure}

\subsection{New stellar ages}
\label{sec:ages}

Stellar ages were determined in \cite{bensby2014} using a grid of $\alpha$-enhanced Yonsei–Yale (Y$^2$) isochrones \citep{demarque2004}, adopting $\rm [\alpha/Fe] = 0$ for $\rm [Fe/H] > 0$, $\rm [\alpha/Fe] = -0.3\times[Fe/H]$ for $\rm -1 \le [Fe/H] \le 0$, and $\rm [\alpha/Fe] = +0.3$ for $\rm [Fe/H] < -1$. Taking the uncertainties in effective temperature, surface gravity, and metallicity into account, an age probability distribution (APD) was constructed for each star, and the most probable age was estimated as described by \cite{melendez2012}.

In this work we re-determine the stellar ages using the Neural Network Estimator of Stellar Times (NEST) package \citep{boin2026}. that makes use of modern stellar evolutionary models together with the precise astrometric and photometric information available from {\sl Gaia}. We adopt the MIST isochrones \citep{mist1,mist2} and use metallicities from the present abundance analysis together with absolute $G$ magnitudes and colours derived from {\sl Gaia} DR3 \citep{GDR3_vallenari2023}. Uncertainties were propagated from the {\sl Gaia} photometric and parallax uncertainties. To account for the effects of $\alpha$-enhancement, we convert the metallicities to global metallicities using the prescription by \cite{salaris1993};
\begin{equation}
\rm [M/H] = [Fe/H] + \log(0.638 \cdot 10^{[\alpha/Fe]} + 0.362), 
\end{equation}
where $\alpha$ is the mean abundance of O, Mg, Si, Ca, and Ti.

Of the 714 stars in the sample, 706 are present in {\sl Gaia} DR3 and 697 have reliable parallax measurements. As all stars are located within  $\sim$100\,pc of the Sun, interstellar extinction is negligible and has been ignored. For each star, NEST was evaluated 100 times and the final age was taken as the mean of the resulting age distribution. The associated age uncertainty was estimated from the dispersion of the individual realisations.

Figure~\ref{fig:newage} compares the new ages with those presented in \cite{bensby2014}. The two age scales are strongly correlated and preserve the overall ranking of stars from young to old ages. The new ages are, however, systematically older, with a mean age difference of $+2.1$\,Gyr and a median difference of $+1.8$\,Gyr (new minus old ages). The rms scatter around the mean difference is 2.7\,Gyr. The median age uncertainty is approximately 1–2\,Gyr, although a subset of stars exhibit substantially larger uncertainties. These larger uncertainties primarily reflect regions of the HR diagram where isochrones of different ages overlap more strongly, making the age determination less precise.

The comparison further reveals that the new age scale produces a smoother age distribution at old ages and largely removes the pronounced clustering visible in the original age estimates. In particular, the accumulation of stars near the upper age boundary of the Y$^2$ grid is significantly reduced, suggesting that the new determinations are less affected by discretisation and edge effects in the underlying stellar models.

We note that the standard MIST isochrones extend to ages of approximately 12.7 Gyr. While NEST is able to extrapolate beyond this limit, ages close to and above the oldest isochrones should be interpreted with caution, as they are no longer directly constrained by the underlying stellar evolution calculations.

For the 697 stars with the necessary {\sl Gaia} information available, we adopt the new NEST ages throughout this work. For the remaining 17 stars we retain the ages from \cite{bensby2014}.

\begin{figure}
\centering
\resizebox{0.95\hsize}{!}{
\includegraphics{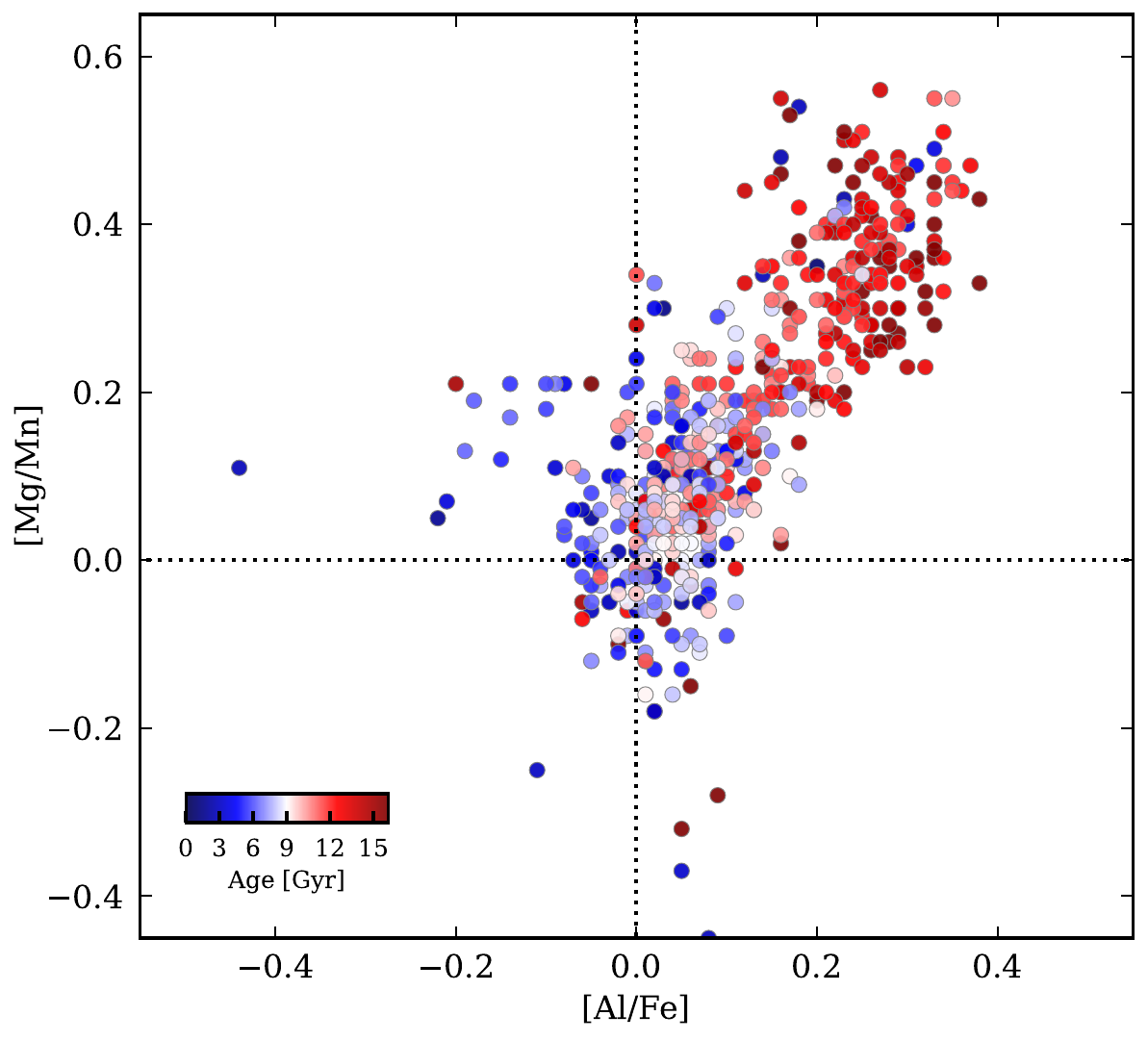}}
\caption{[Mg/Mn] versus [Al/Fe].  The data points have been colour-coded by the estimated ages of the stars. Abundances are from our previous studies \citep{bensby2014, battistini2015}.
\label{fig:mgmn}
}
\end{figure}
\begin{figure*}
\centering
\resizebox{0.95\hsize}{!}{
\includegraphics[trim={0 18mm 0 0}]{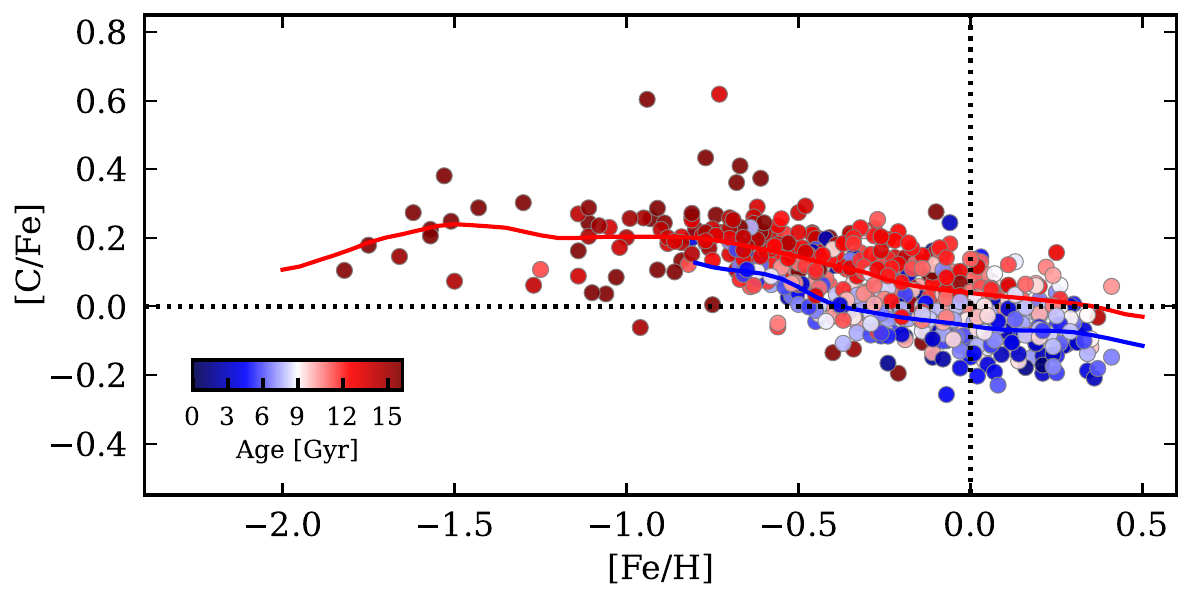}
\includegraphics[trim={0 18mm 0 0}]{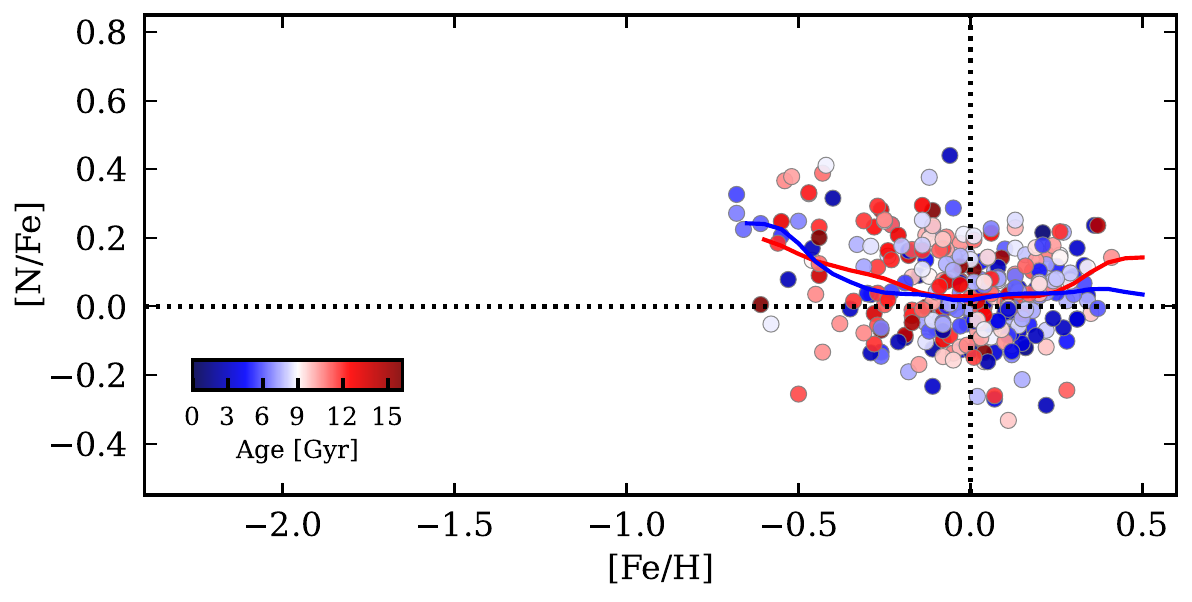}}
\resizebox{0.95\hsize}{!}{
\includegraphics[trim={0 18mm 0 0}]{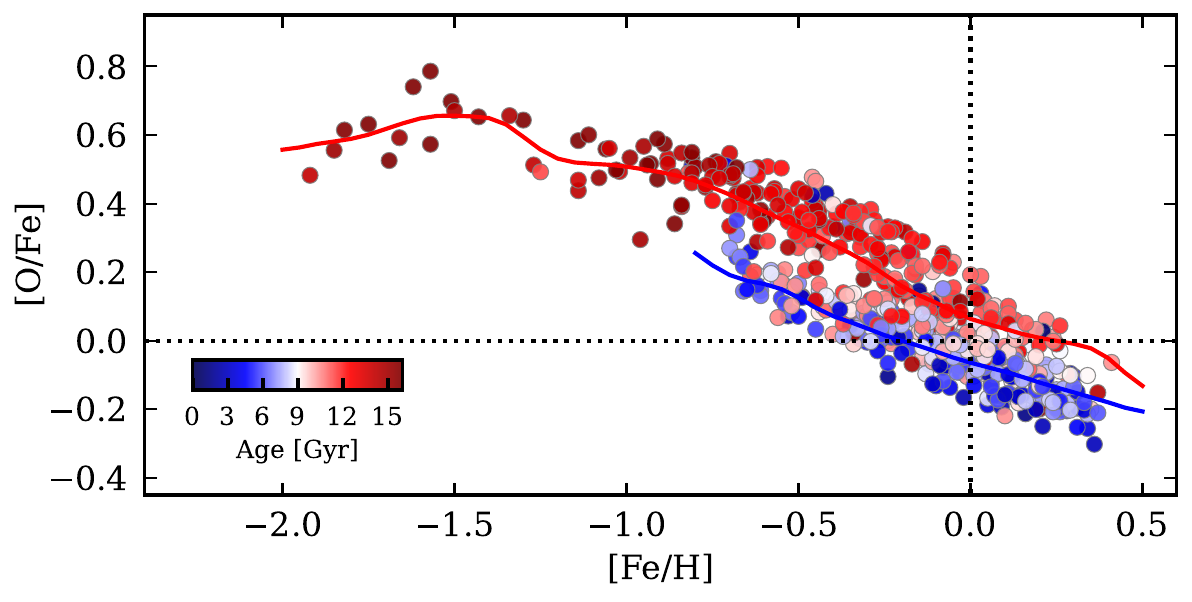}
\includegraphics[trim={0 18mm 0 0}]{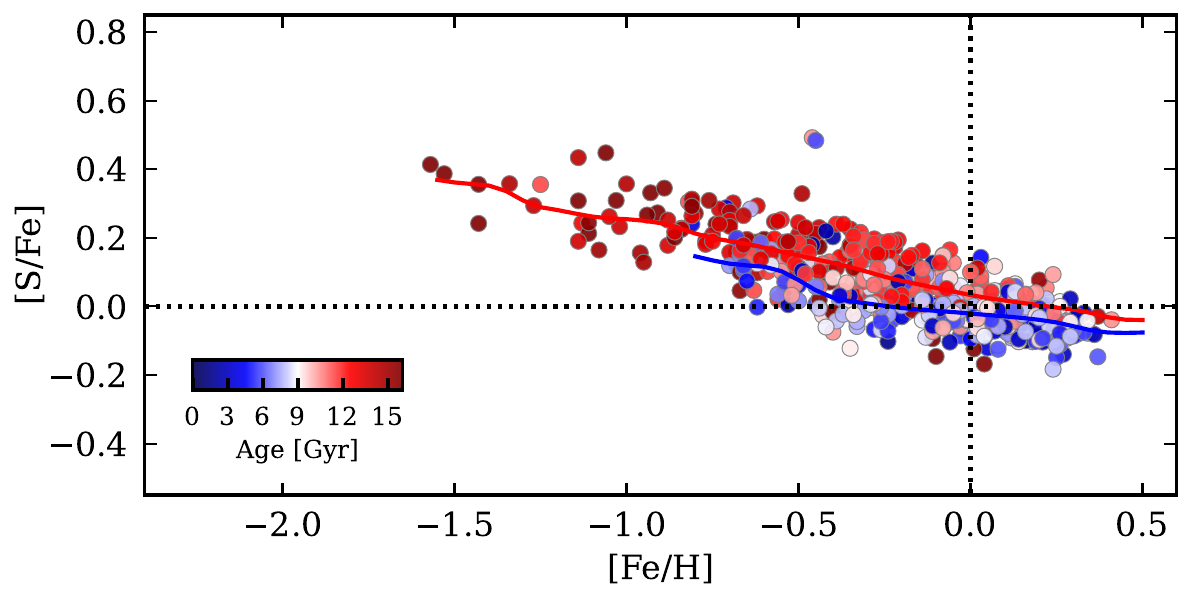}}
\resizebox{0.95\hsize}{!}{
\includegraphics{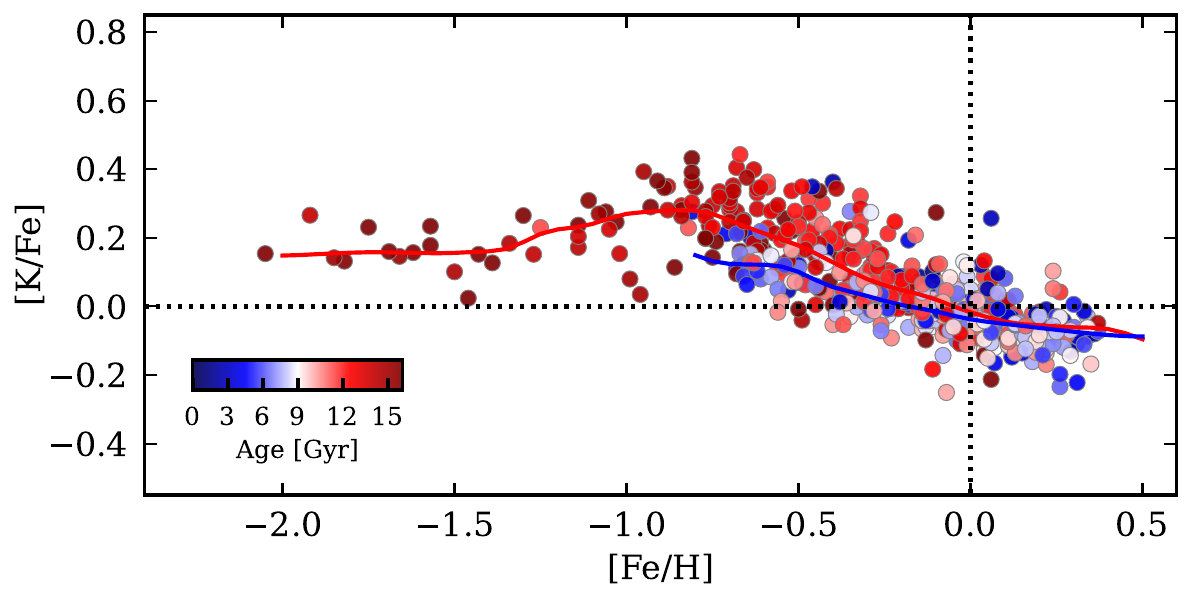}
\includegraphics{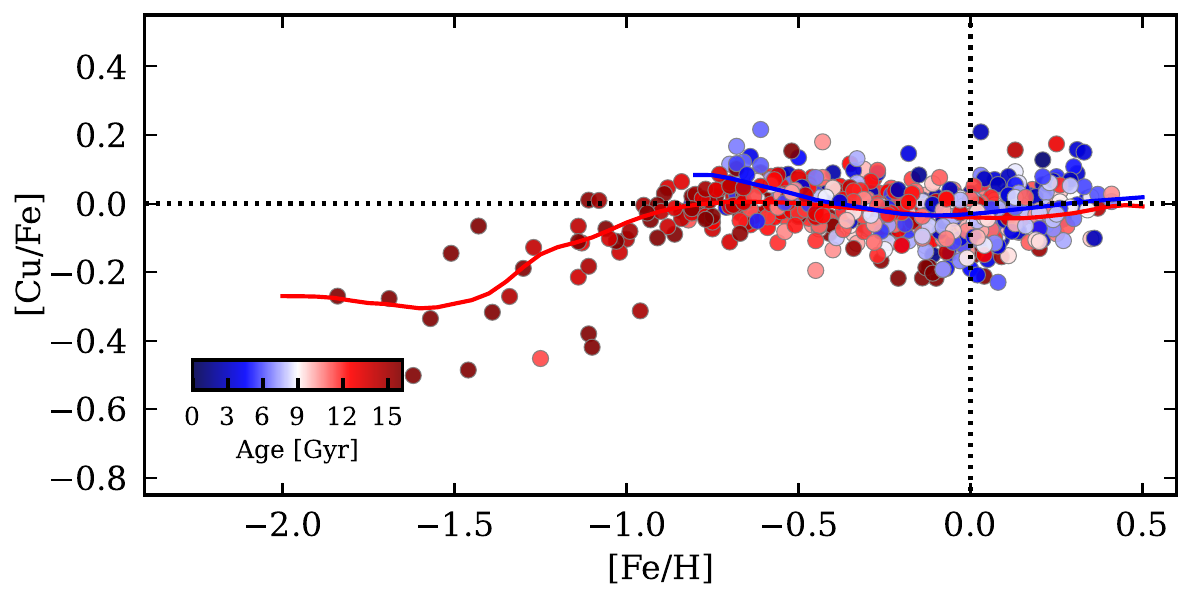}}
\caption{Abundance trends of [$X$/Fe] versus [Fe/H].  The data points have been colour-coded by the estimated ages of the stars.  The red and blue lines mark the smoothed running mean for stars that have ages greater than 9.5\,Gyr, and less than 8.5\,Gyr, respectively.
\label{fig:xfe}
}
\end{figure*}

\section{A clean disk sample}

Figure~\ref{fig:mgmn} shows the [Mg/Mn] versus [Al/Fe] abundance plane, which in recent years has been widely used as a chemical diagnostic to identify stars that may have been accreted during past merger events between the Milky Way and dwarf satellite galaxies. Differences in star formation timescales and nucleosynthetic enrichment between in situ and accreted systems lead to characteristic abundance signatures in this plane, with accreted stars typically exhibiting low aluminium abundances, $\mathrm{[Al/Fe] \lesssim -0.2}$, combined with enhanced $\mathrm{[Mg/Mn]}$ ratios, $\mathrm{[Mg/Mn] > 0}$ \citep{das2020, hawkins2015}. Furthermore, \cite{feuillet2022} identified a population of stars with disk-like kinematics but chemical signatures suggestive of an accreted origin. These stars also display low [Al/Fe] abundances but occupy a region with $\mathrm{[Mg/Mn] \lesssim 0}$, distinct from the classical halo accreted locus, illustrating that accreted populations can overlap with the disk in kinematic space while remaining chemically distinct \citep{jeanbaptiste2017}.

The current sample is essentially devoid of stars in both of these chemically defined accreted regions. Likewise, the region commonly occupied by halo populations, characterised by $\mathrm{[Mg/Mn] \gtrsim 0.3}$ and $\mathrm{[Al/Fe] \lesssim 0}$, is also empty. The absence of stars in these regions strongly suggests that the fraction of accreted stars in the sample is very small and therefore appears consistent with a sample tracing a predominantly in situ population of the Milky Way thin and thick disks.

\begin{figure*}
\centering
\resizebox{0.95\hsize}{!}{
\includegraphics[trim={0 18mm 0 0}]{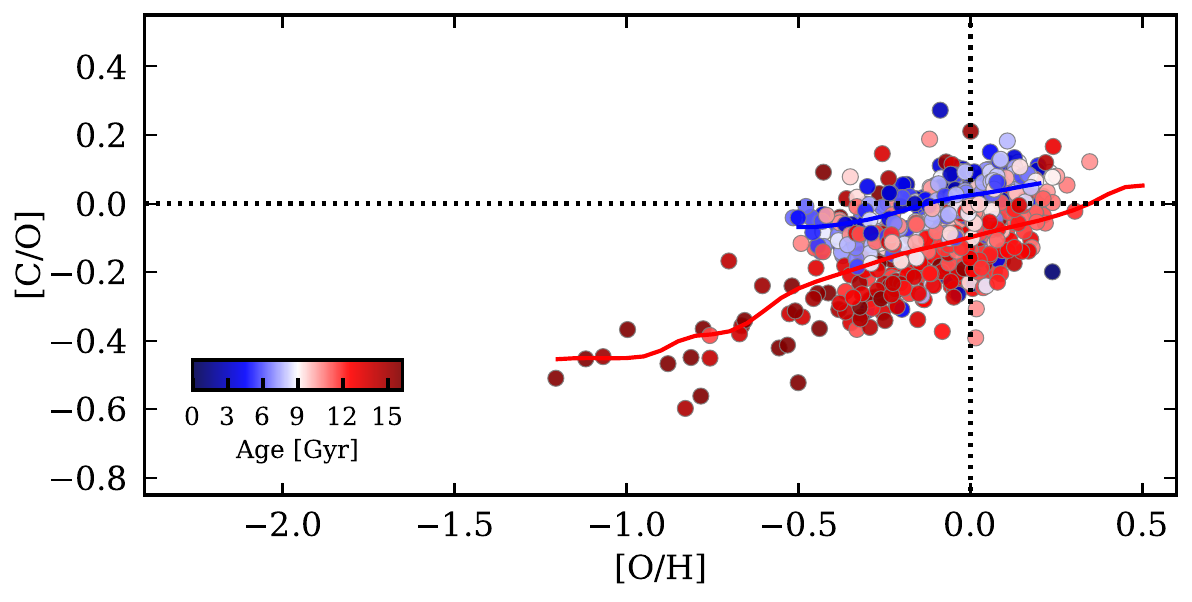}
\includegraphics[trim={0 18mm 0 0}]{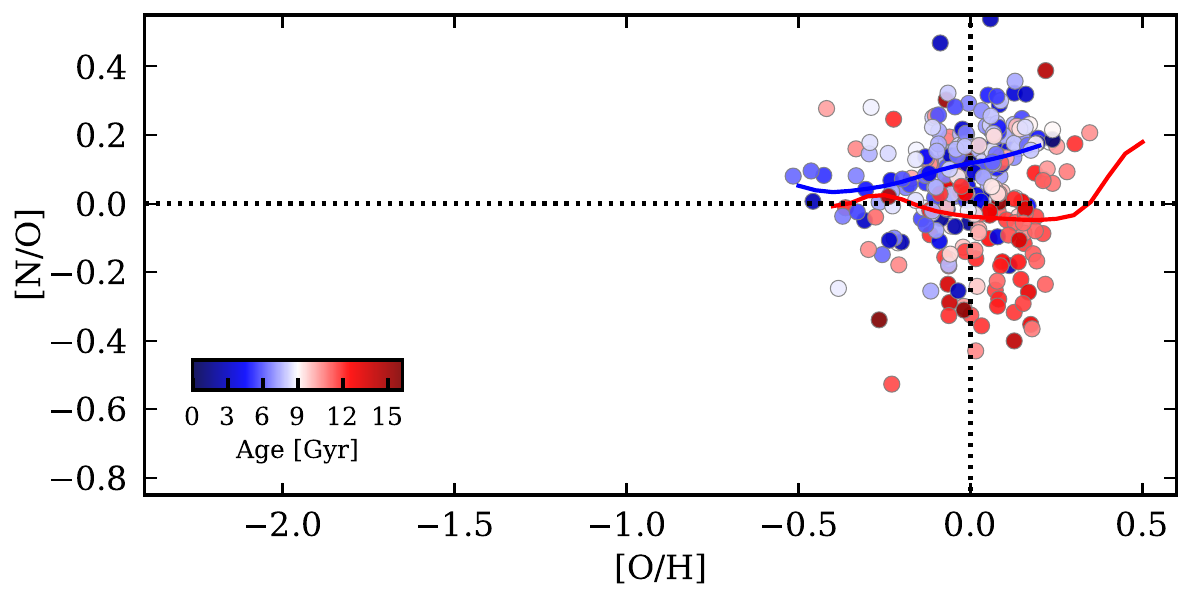}}
\resizebox{0.95\hsize}{!}{
\includegraphics[trim={0 18mm 0 0}]{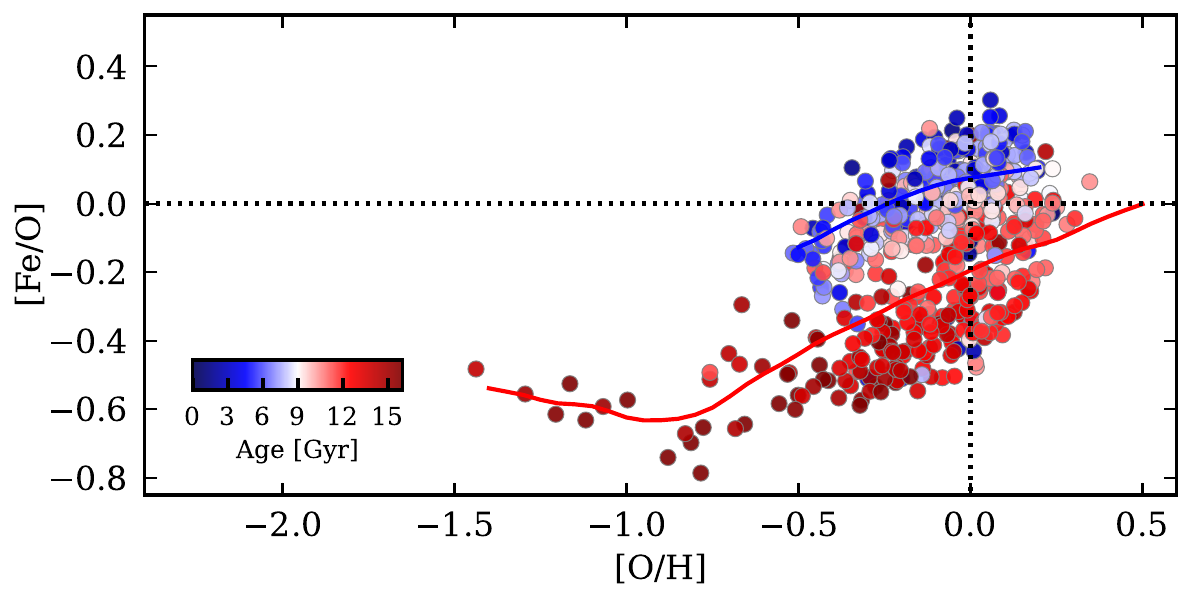}
\includegraphics[trim={0 18mm 0 0}]{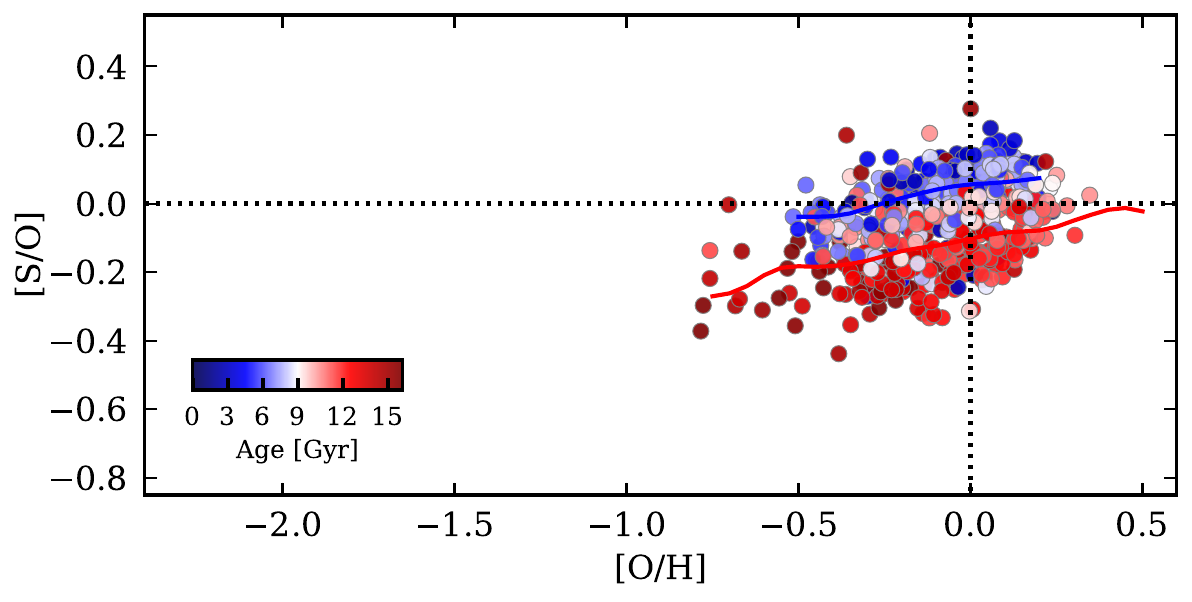}}
\resizebox{0.95\hsize}{!}{
\includegraphics{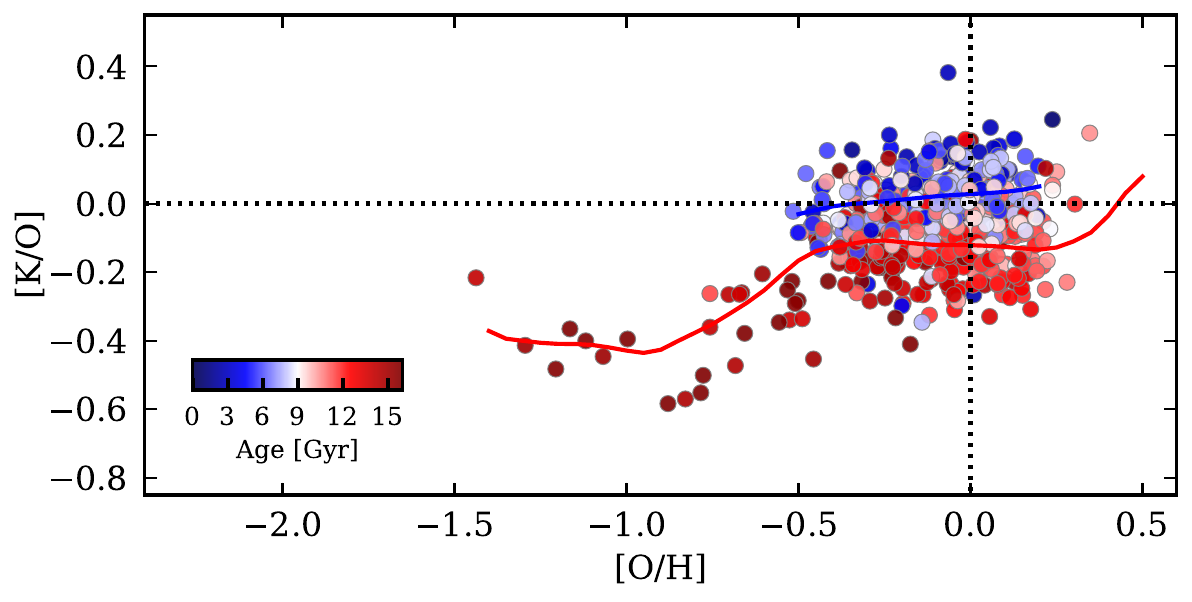}
\includegraphics{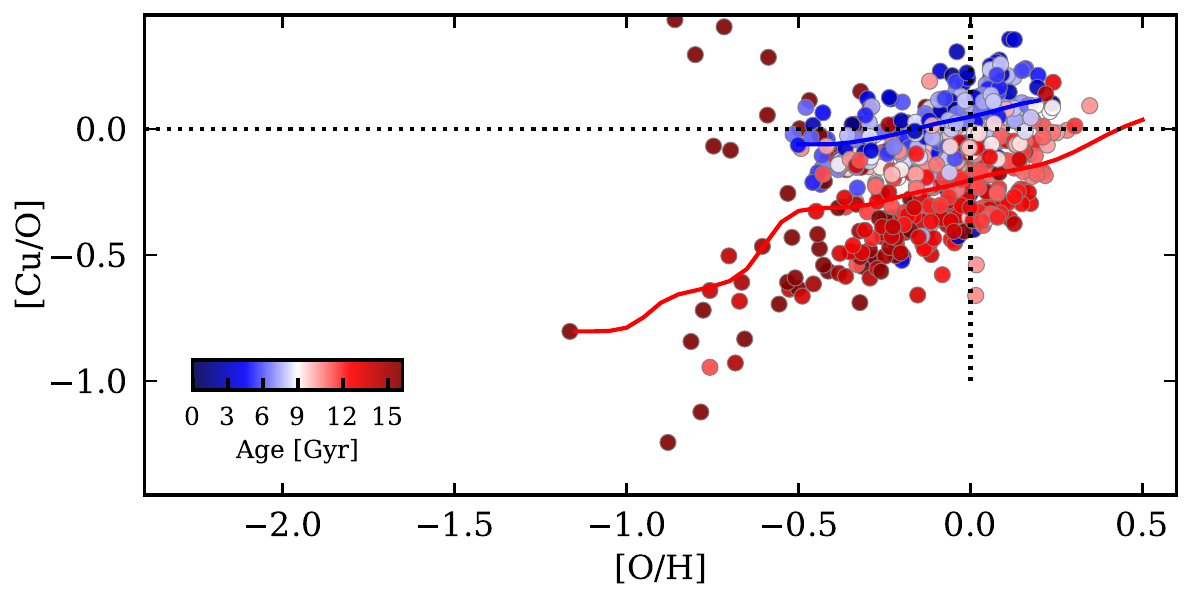}}
\caption{Abundance trend of [$X$/O] versus [O/H]. The data points have been colour-coded  by the estimated ages of the stars. The red and blue lines mark the smoothed running mean for stars that have ages greater than 9.5\,Gyr, and less than 8.5\,Gyr, respectively.
\label{fig:xo}
}
\end{figure*}

\section{Results}
\label{sec:results}

\subsection{Abundance trends versus iron and oxygen}

The abundance trends are shown in Fig.~\ref{fig:xfe} with iron as reference element and in Fig.~\ref{fig:xo} with oxygen as reference element. Furthermore, Figs.~\ref{fig:clines}-\ref{fig:culines} in the appendix show abundance trends based on abundances from the individual lines. They demonstrate that the individual carbon lines yield highly consistent abundance trends; that the atomic \ion{N}{i} feature at 7468\,\AA\ and the blended CN+N feature at 8629\,\AA\ produce broadly compatible abundance trends; that all three \ion{O}{i} triplet lines produce nearly identical abundance trends with very low internal scatter; that the different sulphur lines around 6743–6757\,\AA\ produce consistent abundance trends; that the \ion{K}{i} line at 7699,\AA\ yields a coherent abundance sequence over the full metallicity range; and that the three \ion{Cu}{i} lines at 5105, 5218, and 5782\,\AA\ produce consistent abundance trends.

\begin{figure*}
\centering
\resizebox{0.95\hsize}{!}{
\includegraphics[trim={0 18mm 0 0}]{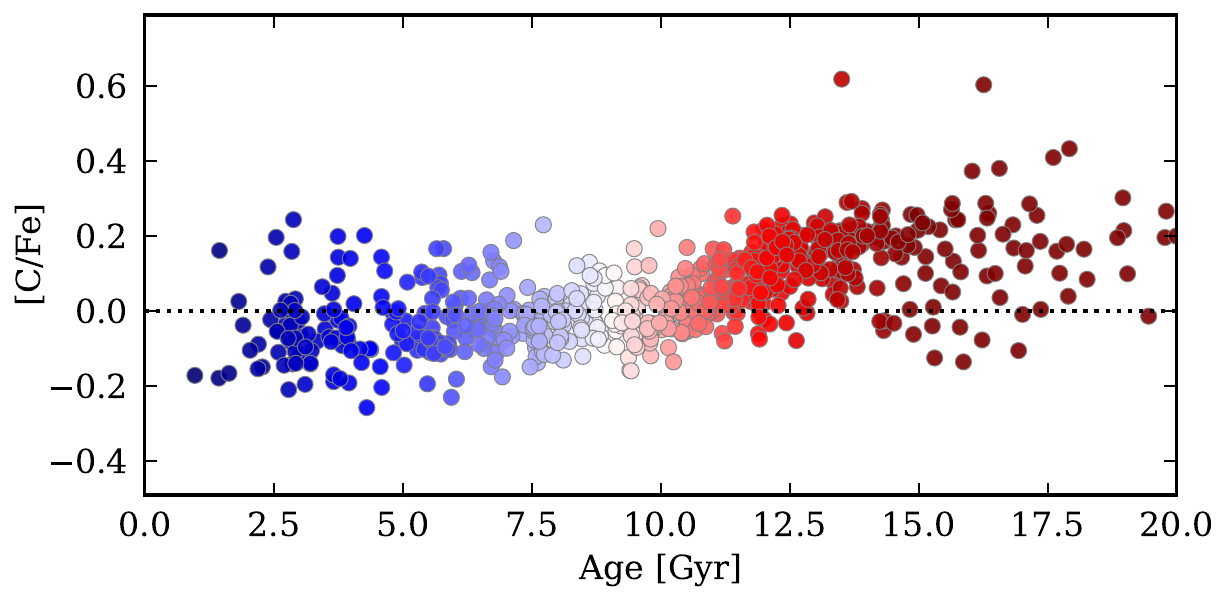}
\includegraphics[trim={0 18mm 0 0}]{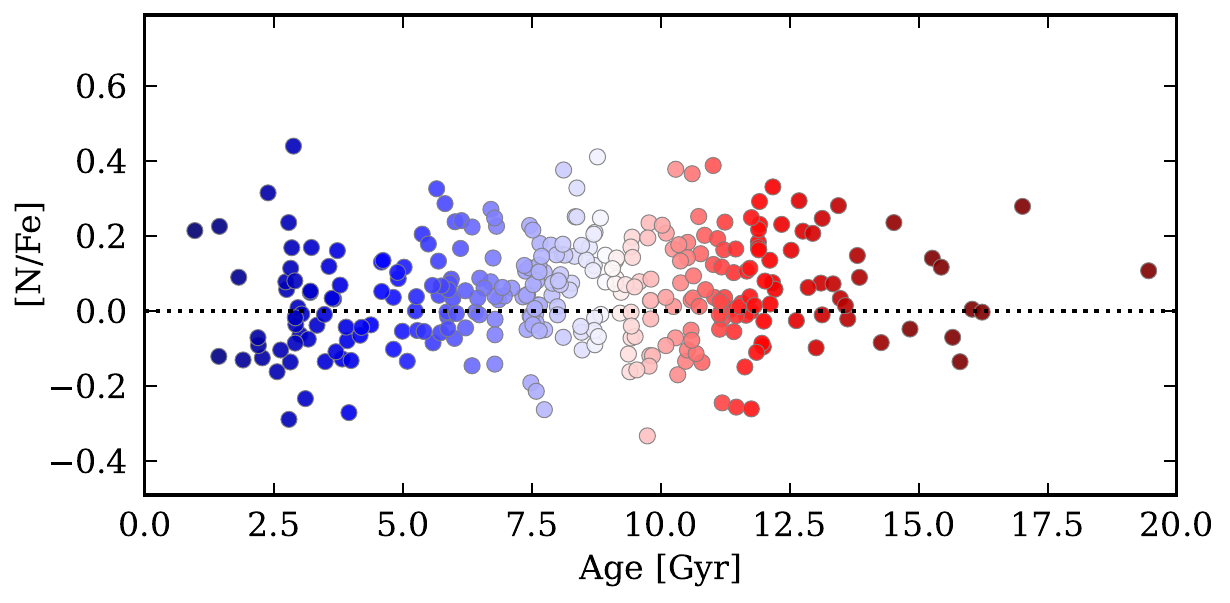}}
\resizebox{0.95\hsize}{!}{
\includegraphics[trim={0 18mm 0 0}]{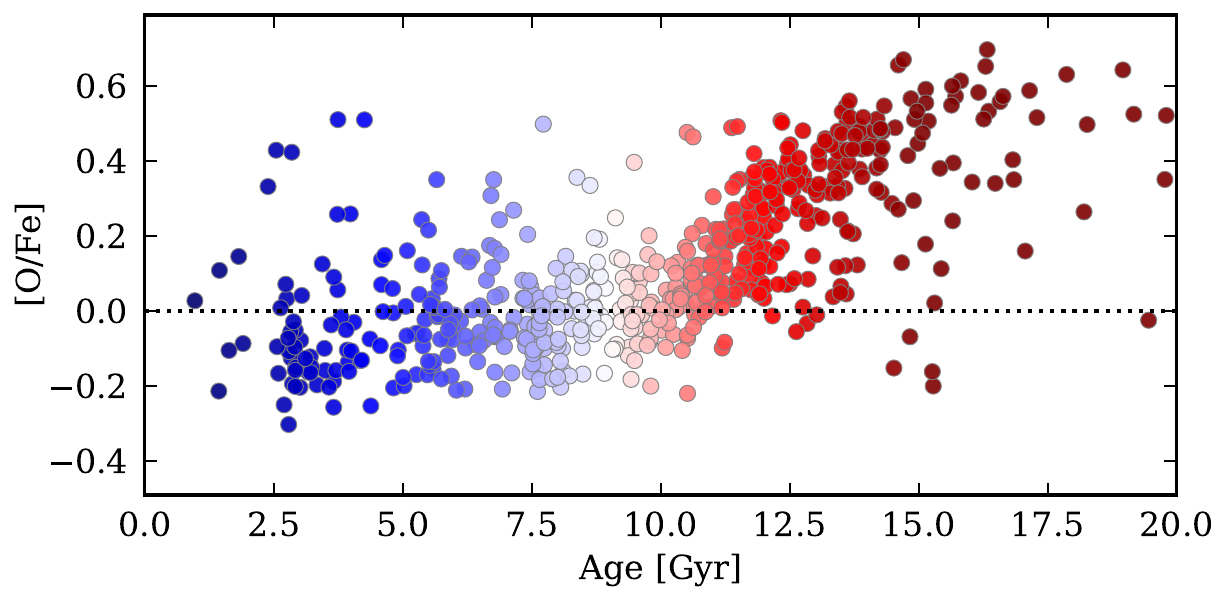}
\includegraphics[trim={0 18mm 0 0}]{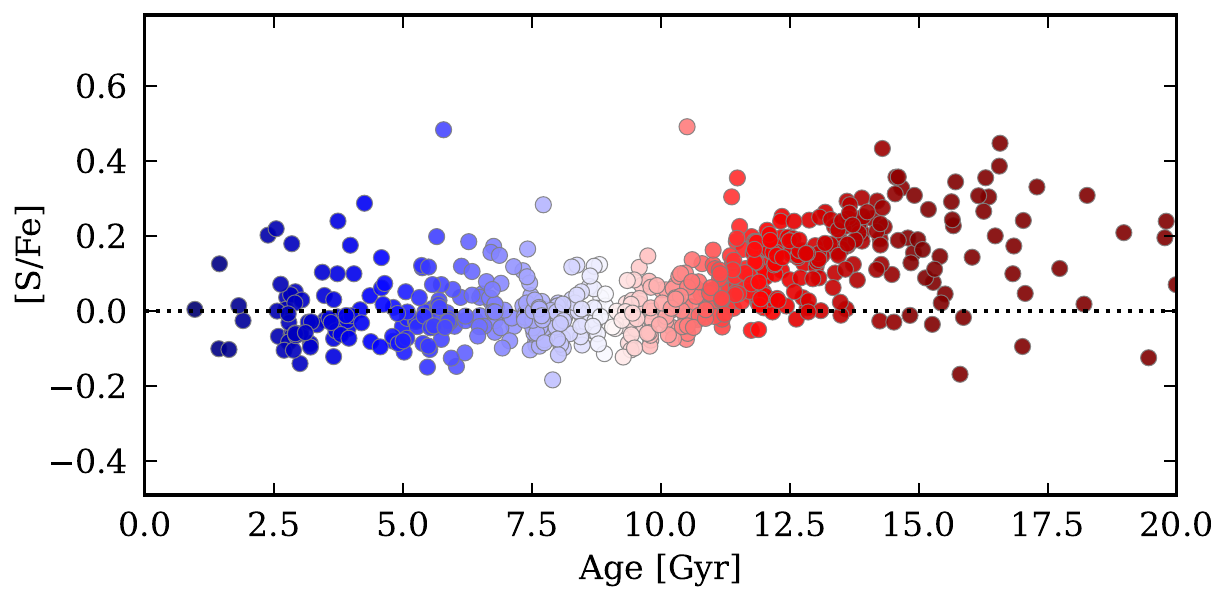}}
\resizebox{0.95\hsize}{!}{
\includegraphics{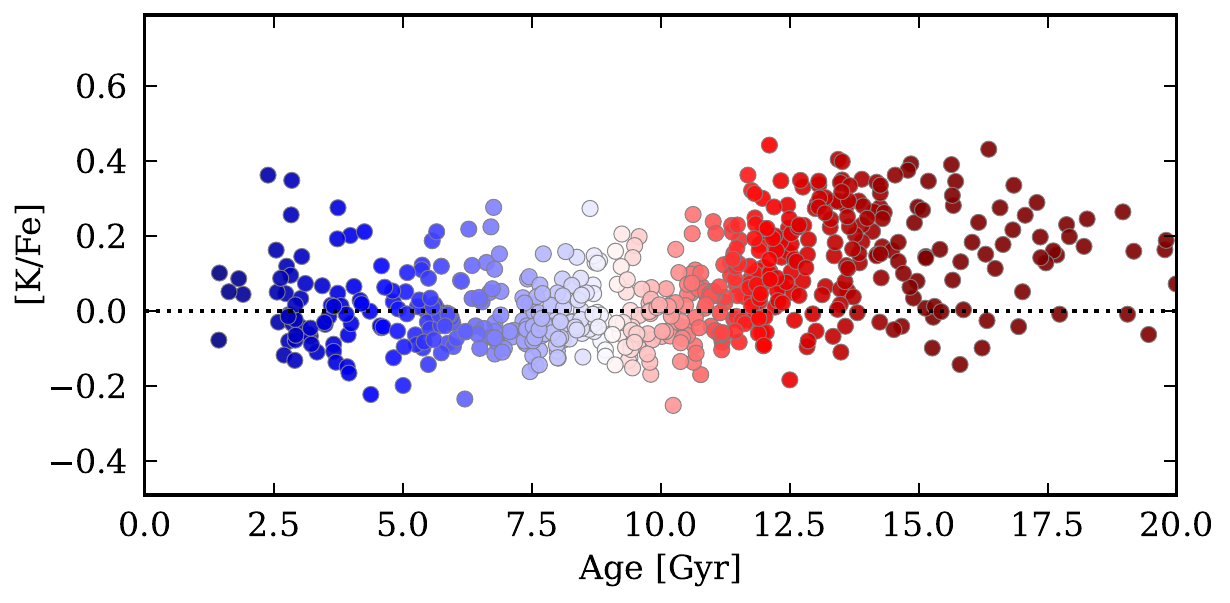}
\includegraphics{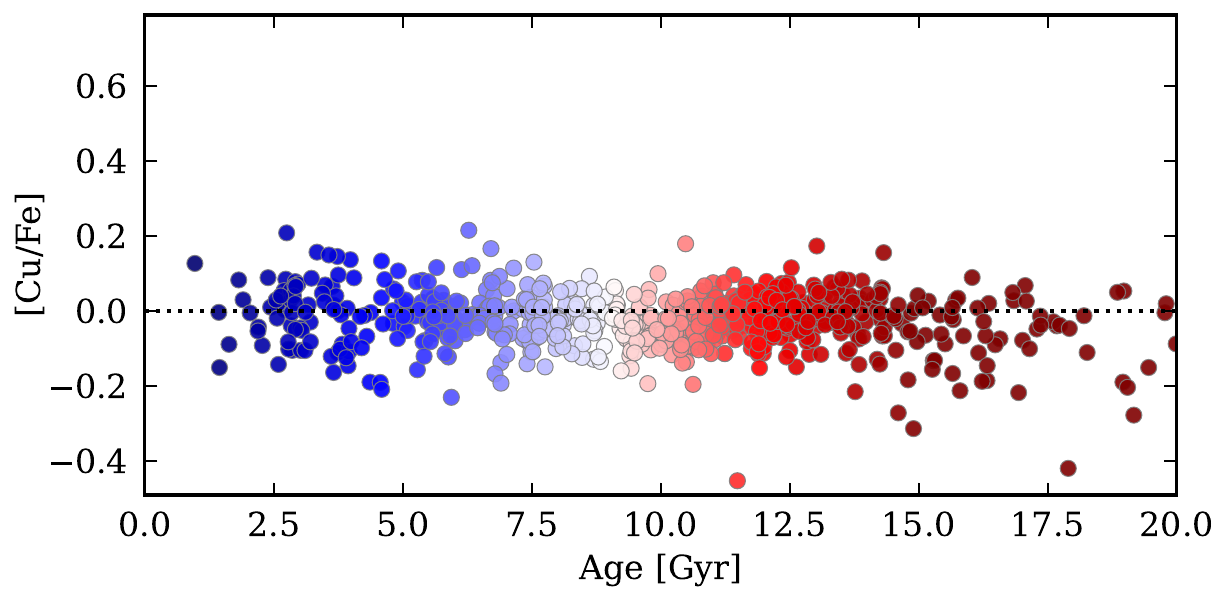}}
\caption{Abundance trends of [$X$/Fe] versus stellar age.  The data points have been colour-coded by the estimated ages of the stars. 
\label{fig:xfeage}
}
\end{figure*}

\paragraph{\textsl{Carbon:}}

The [C/Fe] ratios display a mild enhancement at sub-solar metallicities and gradually decline towards solar and super-solar metallicities. The separation between the old and young stellar populations is relatively modest, with substantial overlap over most of the metallicity range. This behaviour is in good agreement with previous studies of nearby thin and thick disk stars based on both forbidden and permitted carbon lines. In particular, the overall trends closely resemble those found by \cite{gustafsson1999,reddy2003,reddy2006} and \cite{bensby2006}, who showed that the thin and thick disks exhibit largely overlapping [C/Fe] trends. Similar abundance behaviour has also been reported in large samples of nearby FGK stars by \cite{petigura2011,delgadomena2021} and \cite{stonkute2020}.

The [C/O] trends shown in Fig.~\ref{fig:xo} reveal a significantly clearer separation between the old and young stellar populations than seen in [C/Fe]. Older stars define systematically lower [C/O] ratios than younger stars at the same oxygen abundance, consistent with the trends identified by \cite{akerman2004,fabbian2009b,nissen2013} and \cite{nissen2014}. In particular, the increase in [C/O] with increasing [O/H] and the distinct behaviour of different stellar populations closely resemble the results by \cite{nissen2014}.

Our trends also show good agreement with the {\sl Gaia}-ESO study by \cite{franchini2020}, as well as with carbon abundances derived for microlensed bulge dwarfs by \cite{bensby2021}. The consistency between these studies, despite the use of different abundance diagnostics and stellar populations, supports the robustness of the carbon abundance scale adopted here.

\paragraph{\textsl{Nitrogen:}}

Nitrogen abundances could only be determined for a subset of the sample due to the intrinsic weakness of the nitrogen features. Nevertheless, the [N/Fe] ratios remain close to solar over most of the metallicity range, with indications of a mild increase toward super-solar metallicities. The [N/O] ratios shown in Fig.~\ref{fig:xo} exhibit a clearer increase with increasing [O/H].

The observed trends are broadly consistent with the previous analysis of nitrogen in nearby disk stars by \cite{reddy2003}. The increasing [N/O] trend with metallicity also agrees with the {\sl Gaia}-ESO results by \cite{magrini2018}, who analysed nitrogen in different Galactic populations. At lower metallicities our trends are qualitatively consistent with the NH-band study of \cite{israelian2004b}, although the present sample contains relatively few stars in this metallicity regime. The agreement between these studies is notable given the very different abundance diagnostics employed.

\paragraph{\textsl{Oxygen:}}

The revised oxygen abundances show the well-established behaviour of a canonical $\alpha$ element. Older stars define an enhanced [O/Fe] sequence extending toward low metallicities, while younger stars show a lower sequence approaching solar ratios at [Fe/H]$\approx0$. The separation between the two populations is particularly clear in the interval $-0.8 \lesssim$ [Fe/H] $\lesssim -0.2$.

The trends are in agreement with previous studies of nearby disk stars, including our earlier analysis of the forbidden \ion{O}{i} line at 6300\,\AA\ in \cite{bensby2004}, as well as more recent large-scale analyses such as \cite{franchini2021}.

\paragraph{\textsl{Sulphur:}}

Sulphur closely follows the behaviour expected for an $\alpha$ element. The thick-disk stars exhibit enhanced [S/Fe] ratios at low metallicities, while the thin-disk sequence declines smoothly toward solar and super-solar metallicities. The old stellar population remains systematically enhanced relative to younger stars over most of the metallicity range.

The trends are consistent with previous analyses of sulphur in Galactic disk stars by \cite{chen2002,reddy2003,reddy2006,caffau2005,duffau2017,costasilva2020}, and \cite{perdigon2021}. In particular, we confirm the tendency reported by \cite{perdigon2021} that [S/Fe] continues to decrease at super-solar metallicities. At lower metallicities, our trends agree well with halo-star studies \citep{israelian2001b,ryde2004,nissen2004,nissen2007}. The trends are furthermore compatible with the giant-star study by \cite{takeda2016}, as well as with the microlensed bulge dwarf abundances by \cite{lucertini2022} and the open-cluster abundances by \cite{lucertini2023}, indicating a broadly consistent sulphur evolution across different Galactic stellar populations.

\paragraph{\textsl{Potassium:}}

The potassium abundances display a behaviour broadly similar to that of the $\alpha$ elements, with enhanced [K/Fe] ratios in older, metal-poor stars and declining abundance ratios toward solar metallicity. The thick-disk stars define a mildly elevated sequence relative to the younger thin-disk population.

These results agree well with previous analyses of potassium in nearby disk and halo stars \citep{takeda2002,reddy2003,reddy2006,zhang2006b,andrievsky2010}. In particular, we confirm the enhancement of [K/Fe] in metal-poor stars and the gradual decline toward solar metallicity.

\paragraph{\textsl{Copper:}}

The [Cu/Fe] abundance ratios increase from sub-solar values at low metallicity toward approximately solar values around [Fe/H]$\approx-0.7$. Older stars tend to occupy slightly lower [Cu/Fe] ratios than younger stars at a given metallicity, producing a modest but visible age dependence. The [Cu/O] ratios shown in Fig.~\ref{fig:xo} reveal an even clearer separation between the old and young stellar populations.

The trends are broadly consistent with previous studies of copper in nearby disk stars \citep{sneden1991,mishenina2002,reddy2003,reddy2006,nissen2016,delgadomena2017}. Similar behaviour was also reported by \cite{shi2018} for disk and halo stars. The copper trends further agree with the bulge giant-star analyses by \cite{xu2019} and \cite{ernandes2020}, both of which found that the bulge and disk populations follow broadly similar copper abundance sequences over a wide metallicity interval.

\subsection{Abundances versus age}

The abundance trends of [$X$/Fe] versus age are shown in Fig.~\ref{fig:xfeage}. Here we see that oxygen exhibits the strongest age dependence of the elements analysed in this work. The [O/Fe] ratios increase steadily with stellar age, reaching values of about $+0.5$\,dex among the oldest stars in the sample. Carbon and sulphur show clearly detectable correlations with age. Stars younger than about 8-10,Gyr tend to have lower [C/Fe] and [S/Fe] ratios than older stars, and the oldest stars define a distinct sequence at enhanced [C/Fe] and [S/Fe].  The age trends are somewhat weaker than for oxygen, consistent with the more modest separation between the old and young stellar populations in the [C/Fe]–[Fe/H] and [S/Fe]-[Fe/H] planes. Potassium also displays an age dependence. Older stars show enhanced [K/Fe] ratios while younger stars cluster around solar values. The trend is qualitatively similar to those for oxygen, carbon, and sulphur, although the amplitude is smaller and the scatter somewhat larger.  

Nitrogen displays the largest scatter of all elements considered here. A weak increase in [N/Fe] with age may be present. This behaviour likely reflects both the smaller number of stars with measurable nitrogen abundances and the uncertainties associated with the weak nitrogen features. Copper shows no dependence on age, suggesting that age is not the dominant parameter governing the [Cu/Fe] abundance ratio.

The abundance-age relations demonstrate that the elements analysed in this work exhibit markedly different evolutionary behaviour. Oxygen, sulphur, and carbon show the strongest correlations with age, potassium displays an intermediate behaviour, while nitrogen and copper exhibit substantially weaker age dependence. The transition around ages of 8--10\,Gyr, corresponding to the approximate age boundary adopted in this work, is visible particularly for oxygen, sulphur, carbon, and potassium.

The age trends therefore reinforce the picture emerging from Figs~\ref{fig:xfe} and \ref{fig:xo}: elements that exhibit only modest separation in [$X$/Fe] can reveal much stronger evolutionary signatures when examined relative to oxygen or when stellar ages are taken into account. Similar conclusions have been reached in studies combining detailed abundances and ages \citep[][]{masseron2015,ho2017}, and more recently by the {\sl Gaia}-ESO survey \citep{viscasillasvazquez2022}. In addition, \cite{nissen2020} identified two distinct sequences in several abundance-age relations for nearby solar-type stars, separated at an age of approximately 7\,Gyr on their age scale. The transition observed in the present sample occurs at somewhat older ages, consistent with the systematic offset between our revised NEST ages and the ASTEC-based ages adopted by Nissen et al., but may reflect the same transition between the dominant phases of thin- and thick-disk evolution. The present results extend this picture by showing that sulphur, potassium, and carbon exhibit behaviour similar to oxygen, whereas nitrogen and copper display much weaker age dependence.

\begin{figure}
\centering
\resizebox{0.95\hsize}{!}{
\includegraphics{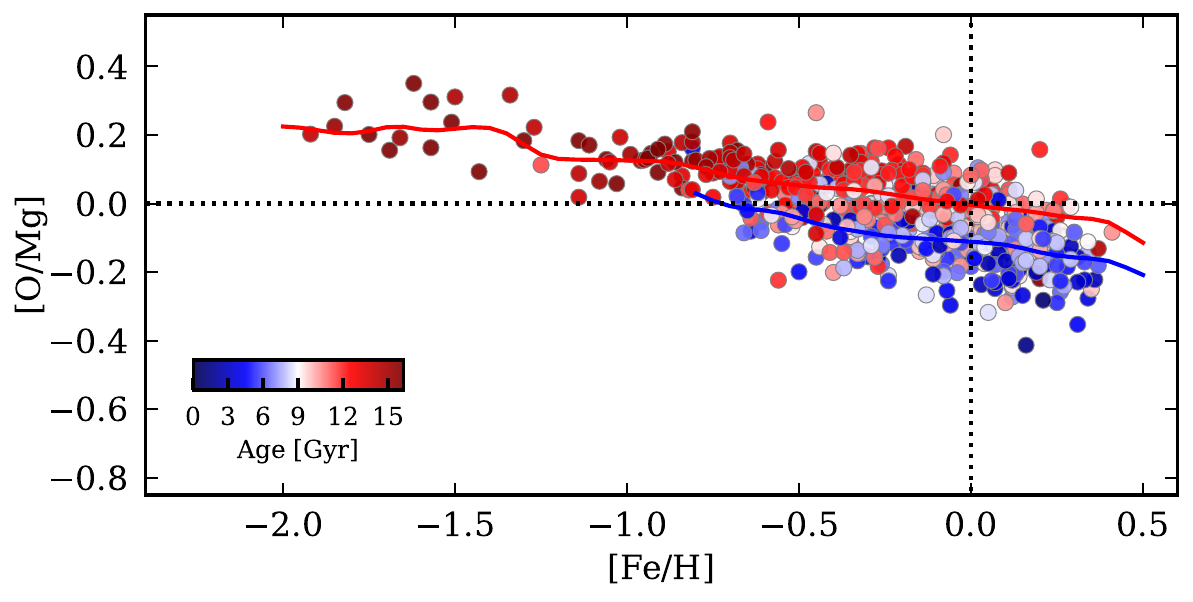}}
\caption{Abundance trend of [O/Mg] versus [Fe/H]. The data points have been colour-coded  by the estimated ages of the stars. The red and blue lines mark the smoothed running mean for stars that have ages greater than 9.5\,Gyr, and less than 8.5\,Gyr, respectively.
\label{fig:omgfe}
}
\end{figure}

\section{Discussion}

\subsection{Galactic chemical evolution}

The carbon and nitrogen abundance trends illustrate the importance of considering reference elements other than iron when investigating Galactic chemical evolution. While the [C/Fe] and [N/Fe] trends show only modest separation between the old and young disk populations (Figs.~\ref{fig:xfe} and \ref{fig:xfeage}), much clearer differences emerge when oxygen is used as reference elements (Fig.~\ref{fig:xo}).

For carbon, the weak separation seen in [C/Fe] is consistent with earlier findings by \cite{bensby2006} and \cite{nissen2014}, who showed that the thin and thick disks largely overlap in this abundance plane despite the strong separation observed in [O/Fe]. However, [C/O] reveals a clear offset between the old and young stellar populations, with younger stars exhibiting systematically higher abundance ratios. The age dependence seen in Fig.~\ref{fig:xfeage} supports the same picture, indicating a gradual increase in carbon relative to the $\alpha$ elements over the lifetime of the Galactic disk. Such behaviour is qualitatively consistent with chemical evolution models in which carbon receives important contributions on longer timescales than oxygen, although the importance of the different carbon production sites remains uncertain \citep[e.g.][]{mattsson2010}. Additional contributions from rotating massive stars have also been suggested to influence the evolution of carbon and other CNO elements in Galactic chemical evolution models \citep{romano2019}. The origin of Galactic carbon remains an active topic of investigation, with contributions from both massive stars and longer-lived stellar populations likely required to reproduce the observed abundance patterns \citep[see][for a recent review]{gustafsson2022}.

Nitrogen displays a similar, but even more pronounced, behaviour. The [N/Fe] ratios remain close to solar over the metallicity range, while [N/O] increases with metallicity. The stronger separation between stellar populations in this abundance plane suggests that the information carried by nitrogen is largely masked when iron is used as the reference element. Similar behaviour has been discussed in Galactic chemical evolution studies of the disk \citep{chiappini2003,grisoni2021,johnson2023, romano2022}, as well as in the bulge and disk by \cite{cescutti2009}, who showed that abundance ratios involving carbon and oxygen are particularly sensitive to the relative contributions from different stellar populations.

\begin{figure*}
\centering
\resizebox{0.95\hsize}{!}{
\includegraphics{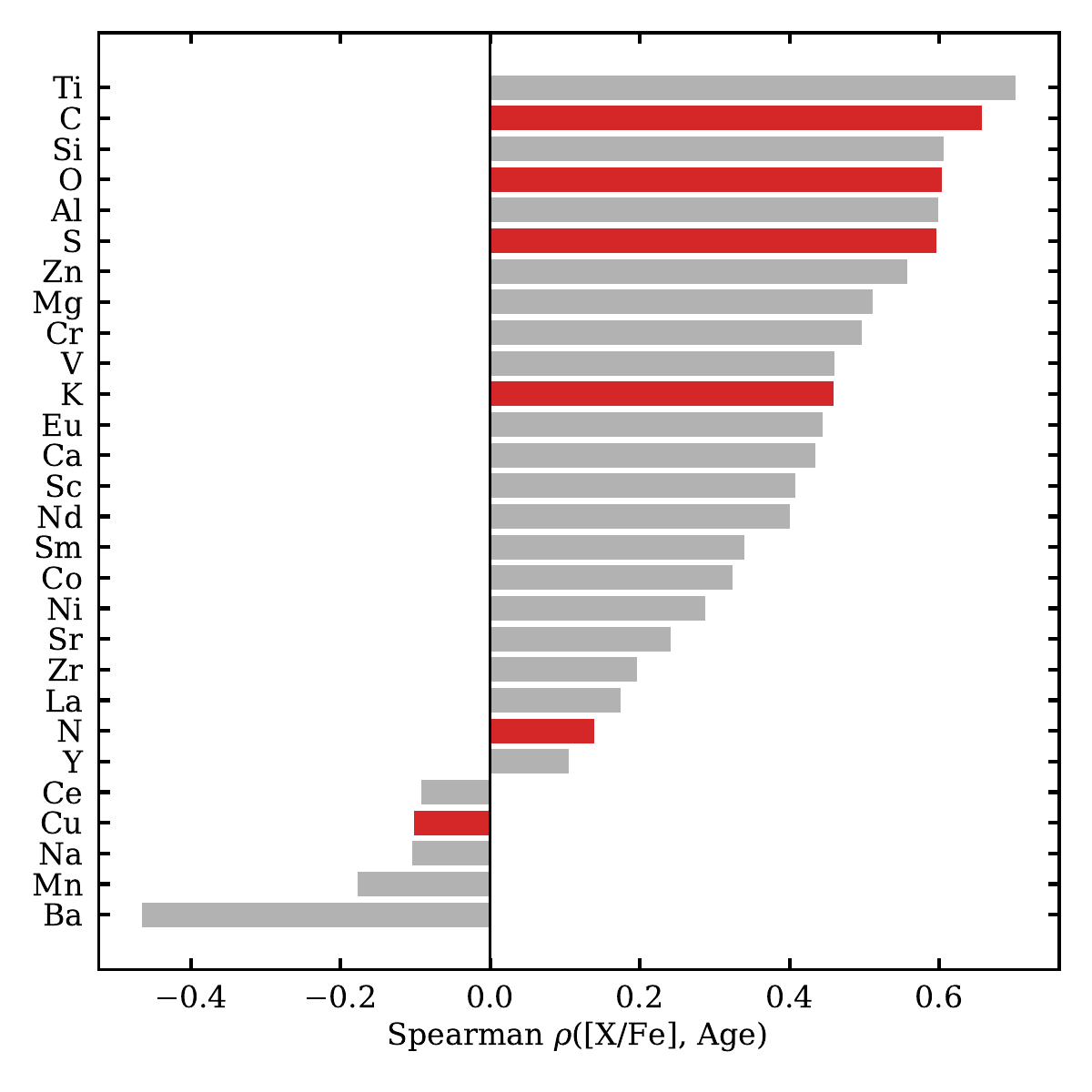}
\includegraphics{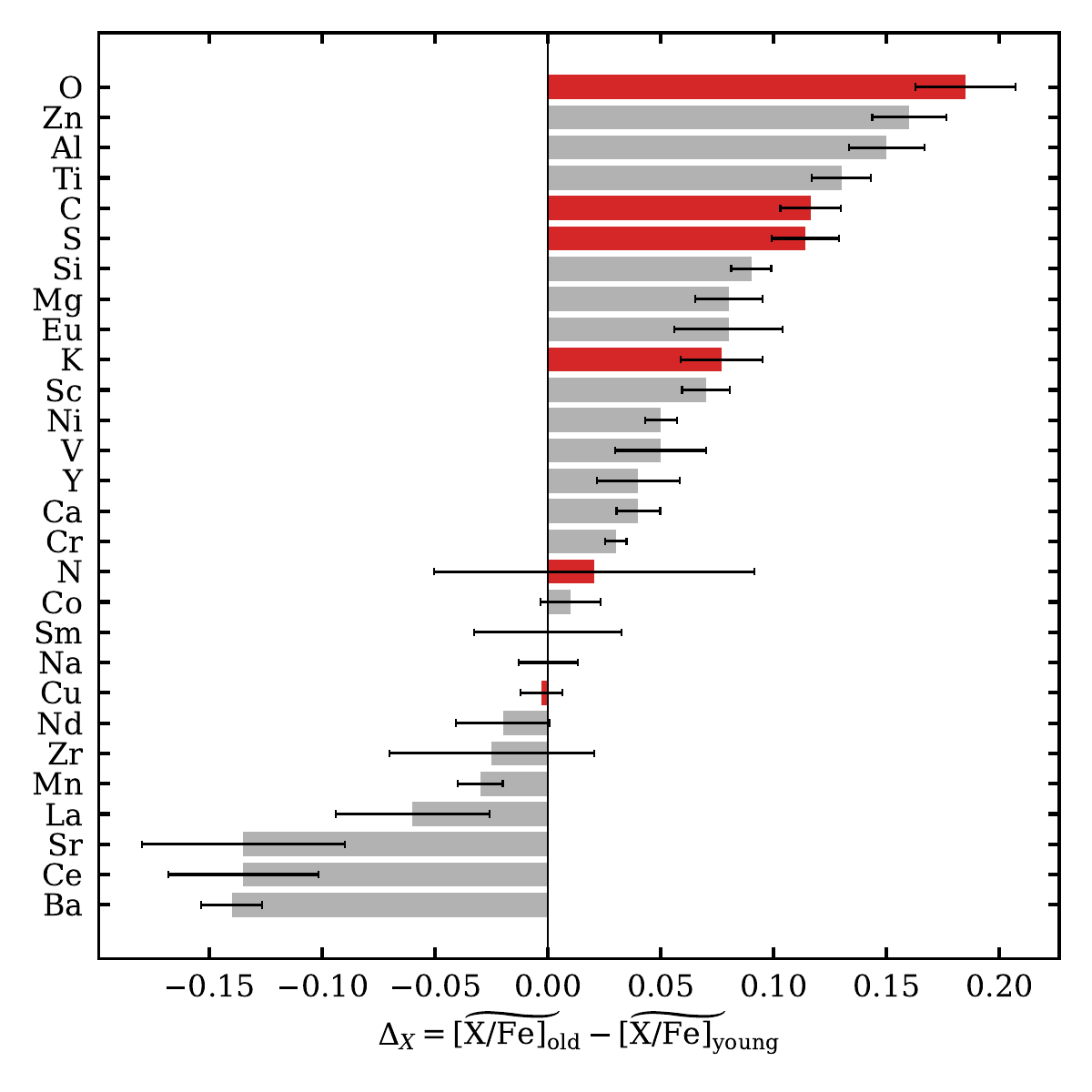}}
\caption{{\sl Left:} Spearman rank correlation coefficient between stellar age and [$X$/Fe] for all elements analysed in Papers~I–V. Positive values indicate abundance ratios that increase with age, while negative values indicate abundance ratios that decrease with age.  {\sl Right:} Difference in median abundance ratio between stars older and younger than 10\,Gyr, $\Delta_X = \widetilde{[\mathrm{X/Fe}]}_{\rm old}-\widetilde{[\mathrm{X/Fe}]}_{\rm young}$, computed for stars in the metallicity interval $-0.6 < \mathrm{[Fe/H]} < -0.2$. Error bars show the 1-$\sigma$ uncertainties estimated from 1000 bootstrap realisations. Positive values indicate elements that are enhanced in the old stellar population, while negative values indicate elements that are enhanced in the younger population. The elements introduced in the present work (C, N, O, S, K, and Cu) are highlighted in red in both plots.
\label{fig:spearman}
}
\end{figure*}

The abundance trends presented in Fig.~\ref{fig:omgfe} demonstrate that oxygen and magnesium should not be regarded as interchangeable reference elements. Although both elements are classified as products of massive stars and used interchangeably as tracers of core-collapse supernova enrichment, the present sample reveals systematic differences between their evolutionary behaviour. The [O/Mg] ratio is not constant but varies systematically with both metallicity (and stellar age). A similar conclusion was reached by \cite{magrini2017}, who found that oxygen and magnesium do not trace identical enrichment histories. The oldest stars define an elevated [O/Mg] sequence that gradually declines toward younger populations and higher metallicities. This trend demonstrates that oxygen and magnesium do not evolve in lockstep throughout the history of the Galactic disk. Consequently, abundance ratios relative to oxygen and magnesium provide complementary rather than redundant information.

The [S/Fe] trend closely resembles the behaviour of [O/Fe], and the small scatter and weak age dependence indicate that sulphur traces the enrichment history of the classical $\alpha$ elements remarkably well. This conclusion is consistent with previous studies of sulphur in disk and halo stars \citep[e.g.][]{nissen2004,nissen2007,perdigon2021,lucertini2022}.

Potassium behaves differently. In the conventional [K/Fe] plane it displays the characteristic appearance of an $\alpha$-like element, with enhanced abundance ratios in old stars and declining ratios toward solar metallicity. However,  [K/O] is not completely flat, indicating that potassium does not evolve strictly in parallel with oxygen. The fact that these trends remain visible when oxygen is adopted as reference element suggests that potassium carries additional information about the chemical evolution of the disk beyond that encoded in the classical $\alpha$ elements. Similar difficulties in reproducing Galactic potassium abundances have previously been noted in chemical evolution studies \citep[e.g.][]{zhang2006b,romano2010}.

Copper exhibits one of the strongest relative abundance evolutions among the elements in this work. While [Cu/Fe] increases steadily with metallicity, the separation between stellar populations becomes more pronounced in [Cu/O]. This suggests that copper evolution is more effectively revealed when compared to elements predominantly produced by massive stars than when compared to iron. Similar behaviour has been reported in previous studies of nearby disk stars \citep[e.g.][]{nissen2016,delgadomena2017} and in bulge giant stars \citep{xu2019}. The present results therefore reinforce the view that copper occupies an intermediate position between the classical $\alpha$ elements and iron-peak species, reflecting a more complex enrichment history than either group individually. Chemical evolution calculations by \cite{romano2007} suggested that massive stars play a dominant role in the production of copper, a scenario qualitatively consistent with the close connection between copper and oxygen seen in the present data.

\subsection{The value of additional abundance dimensions}

The abundance trends discussed in the previous sections demonstrate that the newly analysed elements provide information that complements that obtained from the elements investigated in Papers~I–IV. To quantify this more directly, Fig.~\ref{fig:spearman} compares the age sensitivity and population-separation power of all elements currently available in the sample.

The left-hand panel shows the Spearman rank correlation coefficient between stellar age and [$X$/Fe]. The Spearman coefficient measures the strength of a monotonic relation between two variables without assuming a linear dependence and is therefore well suited for the abundance-age relations shown in Fig.~\ref{fig:xfeage}. Elements with large positive coefficients exhibit abundance ratios that systematically increase with age, while negative coefficients indicate abundance ratios that decrease with age.

The right-hand panel shows the quantity
\begin{equation}
\Delta_X =
\widetilde{[\mathrm{X/Fe}]}_{\rm old}
-
\widetilde{[\mathrm{X/Fe}]}_{\rm young},
\end{equation}

where the tildes denote median abundance ratios. The old and young populations were defined as stars older and younger than 9\,Gyr, respectively, and only stars in the metallicity interval $\rm -0.6 < [Fe/H] < -0.2$ were considered. Restricting to this metallicity range minimises the influence of the overall abundance-metallicity trends and instead highlights the abundance differences between the two age populations. The uncertainties were estimated through bootstrap resampling.

Titanium and carbon exhibit the strongest positive age correlations, followed closely by silicon, oxygen, aluminium, and sulphur. Potassium also rank among the strong age tracers, while nitrogen displays only a modest positive correlation and copper a weak negative correlation. These results together with the abundance-age relations shown in Fig.~\ref{fig:xfeage} demonstrate that the newly added elements span a wide range of age sensitivities.

The population-separation metric yields a similar ranking. Oxygen exhibits the largest abundance difference between the old and young populations, followed by zinc, aluminium, titanium, carbon, and sulphur. Potassium also shows a clear distinction between the two populations, while nitrogen and copper exhibit smaller separation. At the opposite end of the distribution, several neutron-capture elements, particularly strontium, barium, and cerium, are enhanced in the younger population.

The two diagnostics provide complementary information. The Spearman coefficient measures the overall age sensitivity across the full sample, whereas $\Delta_X$ quantifies the ability of an abundance ratio to distinguish between the old and young disk populations. The broadly consistent rankings obtained from the two methods indicate that the newly added elements contribute meaningful and largely independent information to the chemical characterisation of the Galactic disk.

These results demonstrate that the inclusion of carbon, nitrogen, oxygen, sulphur, potassium, and copper substantially expands the diagnostic power of the sample. In particular, carbon, oxygen, sulphur, and potassium emerge as valuable tracers of Galactic disk evolution that complement the abundance information already available from Papers~I–IV. This is consistent with recent work showing that the usefulness of an abundance dataset is determined not only by the number of measured elements, but also by the degree to which those elements provide independent chemical information \citep[e.g.][]{spina2022}

\section{Summary and conclusion}

In this fifth paper of the series exploring the chemical evolution of the Galactic disk, we have derived abundances of carbon, nitrogen, sulphur, potassium, and copper for 714 nearby F- and G-type dwarf and subgiant stars, re-determined oxygen abundances using updated NLTE corrections, and re-derived stellar ages using the NEST framework and {\sl Gaia} DR3 photometry and astrometry. The stellar sample and stellar parameters are adopted from \citet{bensby2014}, placing the new abundances on the same homogeneous scale as the elements analysed in Papers~I–IV. Our main results can be summarised as follows:

\begin{itemize}

\item Carbon exhibits only a modest separation between the old and young disk populations in the [C/Fe]–[Fe/H] plane, consistent with previous studies of nearby disk stars. However, much clearer population differences are revealed in the [C/O] abundance plane. Carbon also emerges as one of the strongest age-sensitive abundance ratios in the sample.

\item Nitrogen abundances could only be determined for a subset of stars due to the weakness of the available spectral features. While [N/Fe] remains approximately solar over the metallicity range, [N/O] increases systematically with metallicity.

\item The revised oxygen abundances confirm the classical $\alpha$ element behaviour of oxygen. Old stars define a high-[O/Fe] sequence distinct from the younger disk population.

\item Sulphur closely follows oxygen throughout the metallicity range covered by the sample, indicating that sulphur traces the enrichment history of the $\alpha$ elements remarkably well. Sulphur also ranks among the strongest age-sensitive abundance ratios in the sample.

\item Potassium displays $\alpha$-like behaviour in the [K/Fe] plane, but residual trends remain visible in [K/O]. Potassium therefore does not evolve strictly in parallel with oxygen and provides information that is complementary to that of the $\alpha$ elements.

\item Copper shows a strong metallicity dependence and one of the clearest separations between old and young populations when oxygen is adopted as reference element, although [Cu/Fe] itself shows no correlation with age.

\item The [O/Mg] ratio is not constant but varies systematically with metallicity and age. Oxygen and magnesium therefore provide complementary rather than interchangeable reference scales for studies of Galactic chemical evolution.

\item The abundance-age relations demonstrate that oxygen, carbon, sulphur, and potassium exhibit some of the strongest age dependencies among all elements available in the sample. In contrast, nitrogen and copper show substantially weaker age sensitivity in the [$X$/Fe] plane.

\end{itemize}

The importance of combining multiple light-element abundance tracers has recently been reviewed by \cite{randich2021}, who highlighted the diagnostic power of the carbon, nitrogen, and oxygen for Galactic archaeology. The addition of copper, nitrogen, sulphur, potassium, and copper, together with the revised oxygen abundances presented here, substantially expands the chemical dimensionality of the sample established in Paper~I. It now contains abundance information spanning the major nucleosynthetic channels operating in the Milky Way and provides homogeneous measurements for more than thirty elements. The results demonstrate that abundance ratios relative to oxygen, combined with precise stellar ages, provide a powerful framework for disentangling the chemical evolution of the Galactic disk. The expanded abundance inventory is particularly relevant in the context of upcoming surveys such as 4MIDABLE-HR, where the simultaneous analysis of large numbers of chemical species is expected to provide powerful constraints on the formation and evolution of the Milky Way \citep[e.g.][]{storm2026}.

\section*{Data availability}

The full Table~\ref{tab:cds} is only available at the CDS via anonymous ftp to \url{cdsarc.u-strasbg.fr} (130.79.128.5) or via
\url{http://cdsarc.u-strasbg.fr/viz-bin/qcat?J/A+A/XXX/AXXX}

\begin{acknowledgement}

T.B. acknowledges financial support by grants No.~2024-04990 from the Swedish Research Council.  The study is based on data obtained with the European Southern Observatory telescopes (Proposal ID:s 87.B-0600, 88.B-0349, 89.B-0047, 90.B-0204, 91.B-0289, 92.B-0626, 93.B-0700), and the Magellan Clay telescope at the Las Campanas observatory. This work has made use of the VALD database, operated at Uppsala University, the Institute of Astronomy RAS in Moscow, and the University of Vienna. We thank the referee, Dr. Anish Amarsi, for valuable comments and suggestions that improved the analysis and text of the paper.

\end{acknowledgement}

\bibliographystyle{aa}
\bibliography{referenser}

\begin{appendix}

\section{Atomic line data}
\label{sec:atomdata}

In Table~\ref{tab:atomdata} we list the spectral lines that were analysed.   For each line we give the wavelength (air), the $\log gf$ value, the lower excitation energy ($\chi_{\rm l}$), and the absolute abundance ($\log \epsilon (X)$). For lines that show hyperfine structure, these components are listed as well. As the analysis is based on four different solar spectra, the median value is given and within parenthesis the 1-$\sigma$ standard deviation is given. For comparison, the last column gives the standard solar value as given by \cite{grevesse2007}.

\begin{table}[h]
\centering
\setlength{\tabcolsep}{2.5mm}
\caption{
Spectral lines investigated in this study. 
\label{tab:atomdata}
}
\tiny
\begin{tabular}{ccrccc}
\hline\hline
\noalign{\smallskip}
Elem. & $\lambda$ & $\log (gf)$ & $\chi_{\rm l}$
      & $\log \epsilon (X)$ & $\log \epsilon (X)$ \\
      & ({\AA})   &             & (eV)
      & NLTE               & Sun G07 \\
\noalign{\smallskip}
\hline
\noalign{\smallskip}

\ion{C}{i} & 5052.167 & $-1.303$ & 7.685 & 8.37 (0.03) & 8.39 \\[2pt]
\ion{C}{i} & 5380.337 & $-1.616$ & 7.685 & 8.39 (0.01) & 8.39 \\[2pt]
\ion{C}{i} & 6587.610 & $-1.003$ & 8.537 & 8.29 (0.04) & 8.39 \\[2pt]
\ion{C}{i} & 7111.470 & $-1.085$ & 8.640 & 8.32 (0.07) & 8.39 \\[2pt]
\ion{C}{i} & 7113.180 & $-0.773$ & 8.647 & 8.33 (0.02) & 8.39 \\[2pt]
\ion{C}{i} & 7115.170 & $-0.934$ & 8.643 & 8.31 (0.04) & 8.39 \\
            & 7115.190 & $-1.473$ & 8.640 &             &      \\[2pt]
\ion{C}{i} & 7116.990 & $-0.907$ & 8.647 & 8.39 (0.01) & 8.39 \\[2pt]
\ion{C}{i} & 8335.141 & $-0.437$ & 7.685 & 8.34 (0.01) & 8.39 \\[2pt]

\hline
\noalign{\smallskip}

\ion{N}{i} & 7468.31  & $-0.190$ & 10.336 & $-0.18^{\ast}$ & 7.78 \\[2pt]
CN         & 8629.071 & $-1.135$ & 1.066  & $-0.05^{\ast}$ & 7.78 \\
\ion{N}{i} & 8629.235 & $0.077$  & 10.69  &         &      \\[2pt]

\hline
\noalign{\smallskip}

\ion{O}{i} & 7771.944 & $ 0.369$ & 9.146 & 8.63 (0.01) & 8.66 \\[2pt]
\ion{O}{i} & 7774.166 & $ 0.223$ & 9.146 & 8.64 (0.01) & 8.66 \\[2pt]
\ion{O}{i} & 7775.388 & $ 0.002$ & 9.146 & 8.64 (0.05) & 8.66 \\[2pt]

\hline
\noalign{\smallskip}

\ion{S}{i} & 6743.473 & $-1.383$ & 7.8663 & 7.16 (0.03) & 7.14 \\
            & 6743.547 & $-1.030$ & 7.8663 &             &      \\
            & 6743.658 & $-1.139$ & 7.8663 &             &      \\[2pt]

\ion{S}{i} & 6748.591 & $-1.507$ & 7.8677 & 7.21 (0.01) & 7.14 \\
            & 6748.702 & $-0.917$ & 7.8677 &             &      \\
            & 6748.851 & $-0.713$ & 7.8677 &             &      \\[2pt]

\ion{S}{i} & 6756.871 & $-1.857$ & 7.8699 & 7.21 (0.01) & 7.14 \\
            & 6757.021 & $-1.012$ & 7.8699 &             &      \\
            & 6757.178 & $-0.427$ & 7.8699 &             &      \\[2pt]

\hline
\noalign{\smallskip}

\ion{K}{i} & 7698.958 & $-0.659$ & 0.0000 & 5.14 (0.06) & 5.08 \\
            & 7698.959 & $-1.358$ & 0.0000 &             &      \\
            & 7698.967 & $-0.659$ & 0.0000 &             &      \\
            & 7698.968 & $-0.659$ & 0.0000 &             &      \\

\noalign{\smallskip}
\hline
\end{tabular}
\tablefoot{
\tablefoottext{$\ast$}{
Note that we have not determined the solar abundances for N. The values given here are the
offsets that have been applied in order to normalise the [N/Fe] trends to the Sun at $\rm [Fe/H]=0$ (see further discussion in Sect.~\ref{sec:solaranalysis}).
}}
\end{table}

\setcounter{table}{0}

\begin{table}[h]
\centering
\setlength{\tabcolsep}{2.5mm}
\caption{
{\it continued }
\label{tab:atomdata}
}
\tiny
\begin{tabular}{ccrccc}
\hline\hline
\noalign{\smallskip}
Elem. & $\lambda$ & $\log (gf)$ & $\chi_{\rm l}$
      & $\log \epsilon (X)$ & $\log \epsilon (X)$ \\
      & ({\AA})   &             & (eV)
      & NLTE               & Sun G07 \\
\noalign{\smallskip}
\hline
\noalign{\smallskip}

\ion{Cu}{i} & 5105.48902 & $-4.257$ & 1.3889 & 4.18 (0.05) & 4.21 \\
             & 5105.49287 & $-3.303$ & 1.3889 &             &      \\
             & 5105.49491 & $-3.257$ & 1.3889 &             &      \\
             & 5105.49656 & $-4.433$ & 1.3889 &             &      \\
             & 5105.50178 & $-3.190$ & 1.3889 &             &      \\
             & 5105.50563 & $-2.935$ & 1.3889 &             &      \\
             & 5105.51103 & $-3.906$ & 1.3889 &             &      \\
             & 5105.51466 & $-2.952$ & 1.3889 &             &      \\
             & 5105.51661 & $-2.906$ & 1.3889 &             &      \\
             & 5105.51683 & $-3.287$ & 1.3889 &             &      \\
             & 5105.51792 & $-4.082$ & 1.3889 &             &      \\
             & 5105.52205 & $-2.685$ & 1.3889 &             &      \\
             & 5105.52275 & $-2.839$ & 1.3889 &             &      \\
             & 5105.52639 & $-2.584$ & 1.3889 &             &      \\
             & 5105.53673 & $-2.936$ & 1.3889 &             &      \\
             & 5105.54156 & $-2.334$ & 1.3889 &             &      \\
             & 5105.54599 & $-2.479$ & 1.3889 &             &      \\
             & 5105.56406 & $-2.128$ & 1.3890 &             &      \\

\noalign{\smallskip}

\ion{Cu}{i} & 5218.19042 & $-1.351$ & 3.8167 & 4.16 (0.06) & 4.21 \\
             & 5218.19058 & $-1.000$ & 3.8167 &             &      \\
             & 5218.19255 & $-1.029$ & 3.8167 &             &      \\
             & 5218.19255 & $-1.397$ & 3.8167 &             &      \\
             & 5218.19261 & $-0.678$ & 3.8167 &             &      \\
             & 5218.19261 & $-1.046$ & 3.8167 &             &      \\
             & 5218.19641 & $-0.428$ & 3.8167 &             &      \\
             & 5218.19641 & $-0.933$ & 3.8167 &             &      \\
             & 5218.19641 & $-2.000$ & 3.8167 &             &      \\
             & 5218.19657 & $-0.779$ & 3.8167 &             &      \\
             & 5218.19657 & $-1.284$ & 3.8167 &             &      \\
             & 5218.19657 & $-2.351$ & 3.8167 &             &      \\
             & 5218.20146 & $-2.176$ & 3.8167 &             &      \\
             & 5218.20146 & $-0.222$ & 3.8167 &             &      \\
             & 5218.201   & $-1.030$ & 3.8167 &             &      \\
             & 5218.202   & $-2.527$ & 3.8167 &             &      \\
             & 5218.202   & $-0.573$ & 3.8167 &             &      \\
             & 5218.202   & $-1.381$ & 3.8167 &             &      \\

\noalign{\smallskip}

\ion{Cu}{i} & 5782.03635 & $-3.620$ & 1.6422 & 4.21 (0.01) & 4.21 \\
             & 5782.04490 & $-3.921$ & 1.6422 &             &      \\
             & 5782.05702 & $-3.222$ & 1.6422 &             &      \\
             & 5782.06622 & $-3.269$ & 1.6422 &             &      \\
             & 5782.07404 & $-3.570$ & 1.6422 &             &      \\
             & 5782.08532 & $-2.871$ & 1.6422 &             &      \\
             & 5782.08768 & $-3.222$ & 1.6422 &             &      \\
             & 5782.09979 & $-3.222$ & 1.6422 &             &      \\
             & 5782.11377 & $-2.871$ & 1.6422 &             &      \\
             & 5782.12505 & $-2.871$ & 1.6422 &             &      \\
             & 5782.15545 & $-2.775$ & 1.6422 &             &      \\
             & 5782.17723 & $-2.424$ & 1.6422 &             &      \\

\noalign{\smallskip}
\hline
\end{tabular}
\end{table}

\section{Abundance trends from individual spectral lines}

The abundance trends presented in Sect.~\ref{sec:results} are based on mean abundances derived from all available spectral lines for each element. In order to assess the internal consistency of the abundance analysis, Figs.~\ref{fig:clines}–\ref{fig:culines} show the abundance trends obtained from the individual lines used in the analysis.

In general, the individual lines reproduce the same overall abundance behaviour as the mean abundances presented in Sect.~\ref{sec:results}. The line-to-line agreement is particularly good for carbon, oxygen, and sulphur, where the different diagnostics define very similar abundance sequences over the full metallicity range covered by the sample. Nitrogen exhibits larger scatter, reflecting the intrinsic weakness of the available nitrogen features and the smaller number of stars for which reliable abundances could be determined. Potassium is represented by a single line, while the three copper lines show generally consistent abundance trends despite their differing sensitivity to blending and hyperfine structure. The overall consistency between the individual diagnostics demonstrates that the abundance trends discussed in this paper are not driven by any single spectral feature but are reproduced by multiple independent abundance indicators.

\begin{figure*}[h]
\centering
\resizebox{0.85\hsize}{!}{
\includegraphics[trim={0 9.5mm 0 0}]{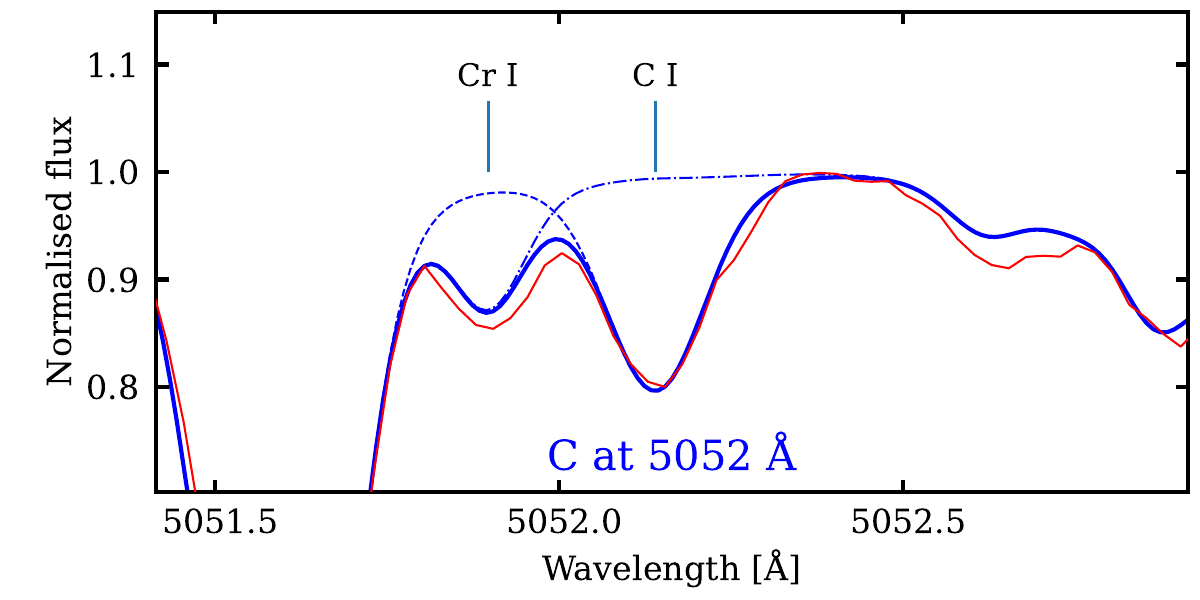}
\hspace{15mm}
\includegraphics[trim={0 9.5mm 0 0}]{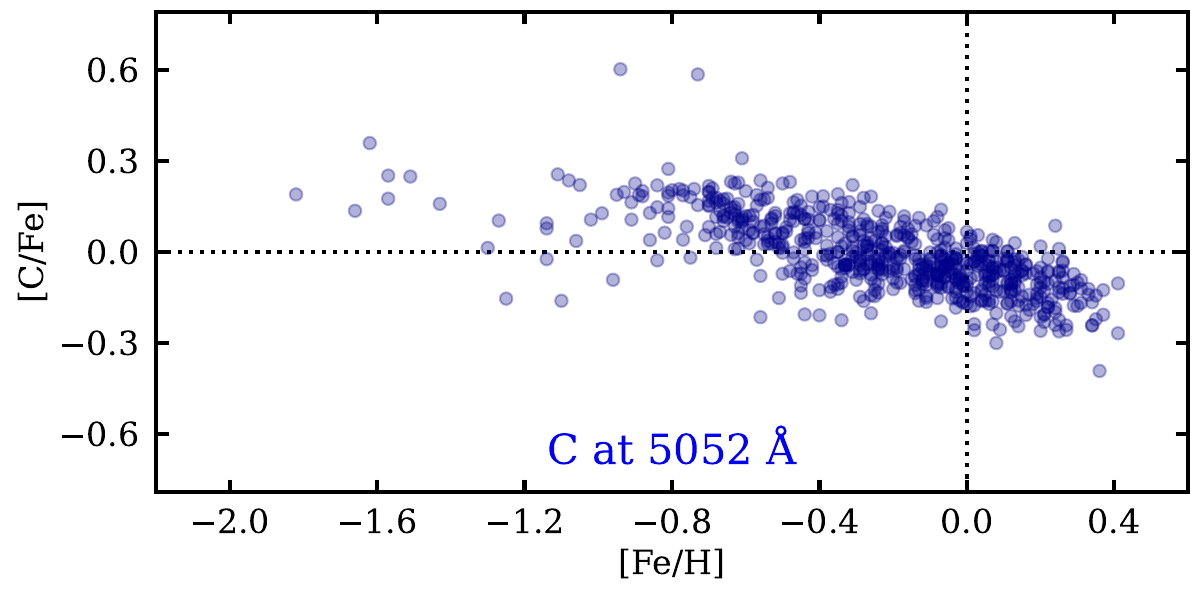}}
\resizebox{0.85\hsize}{!}{
\includegraphics[trim={0 9.5mm 0 0}]{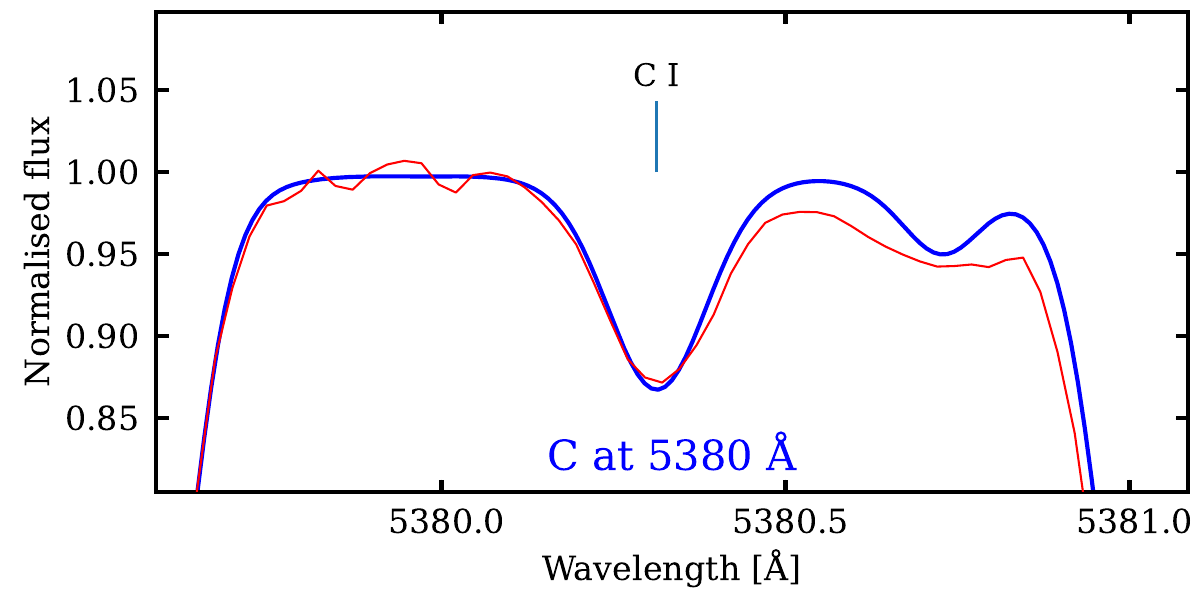}
\hspace{15mm}
\includegraphics[trim={0 9.5mm 0 0}]{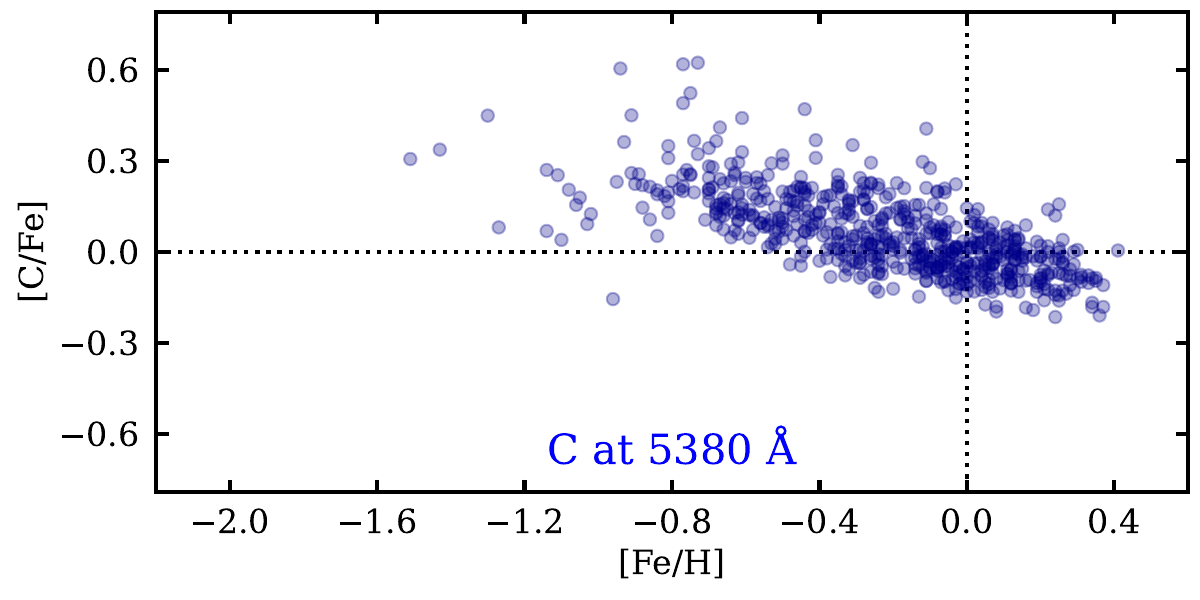}}
\resizebox{0.85\hsize}{!}{
\includegraphics[trim={0 9.5mm 0 0}]{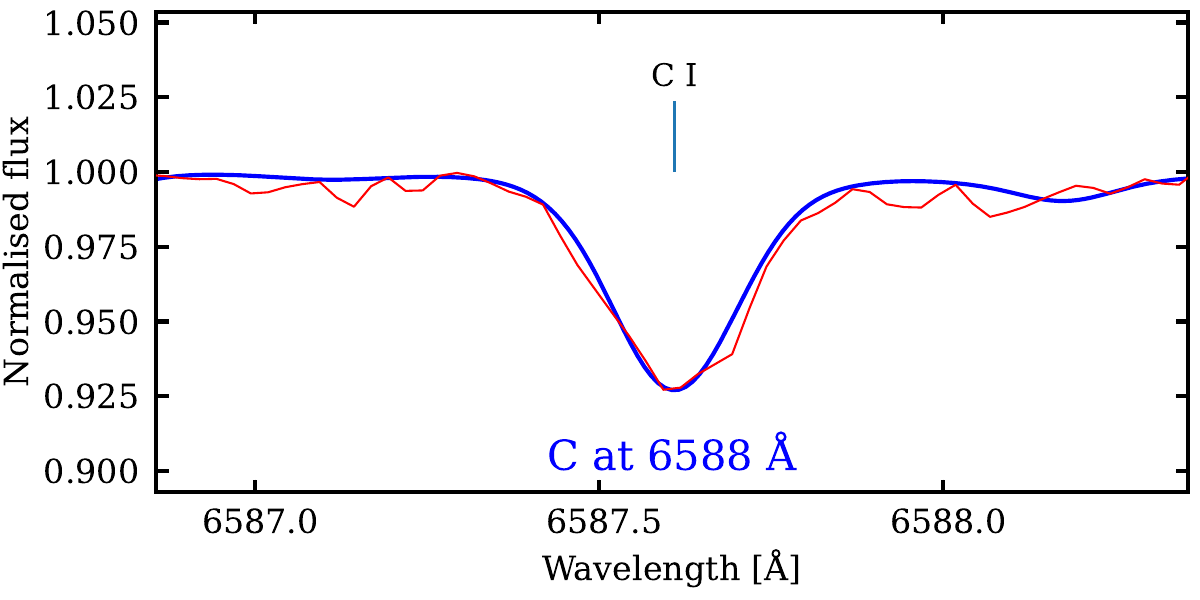}
\hspace{15mm}
\includegraphics[trim={0 9.5mm 0 0}]{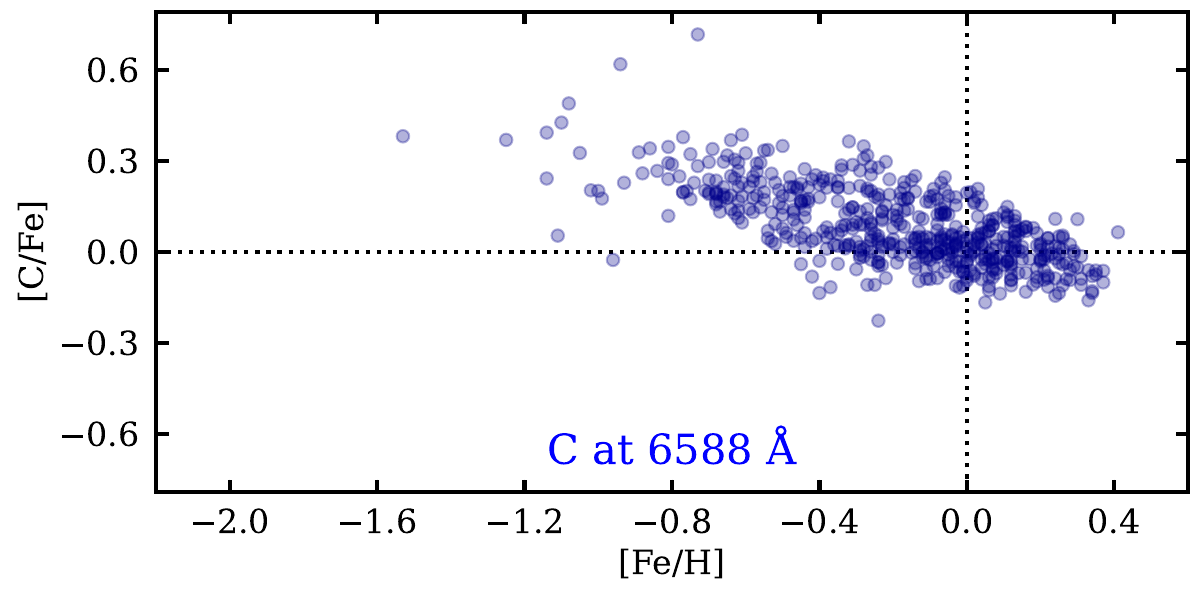}}
\resizebox{0.85\hsize}{!}{
\includegraphics[trim={0 9.5mm 0 0}]{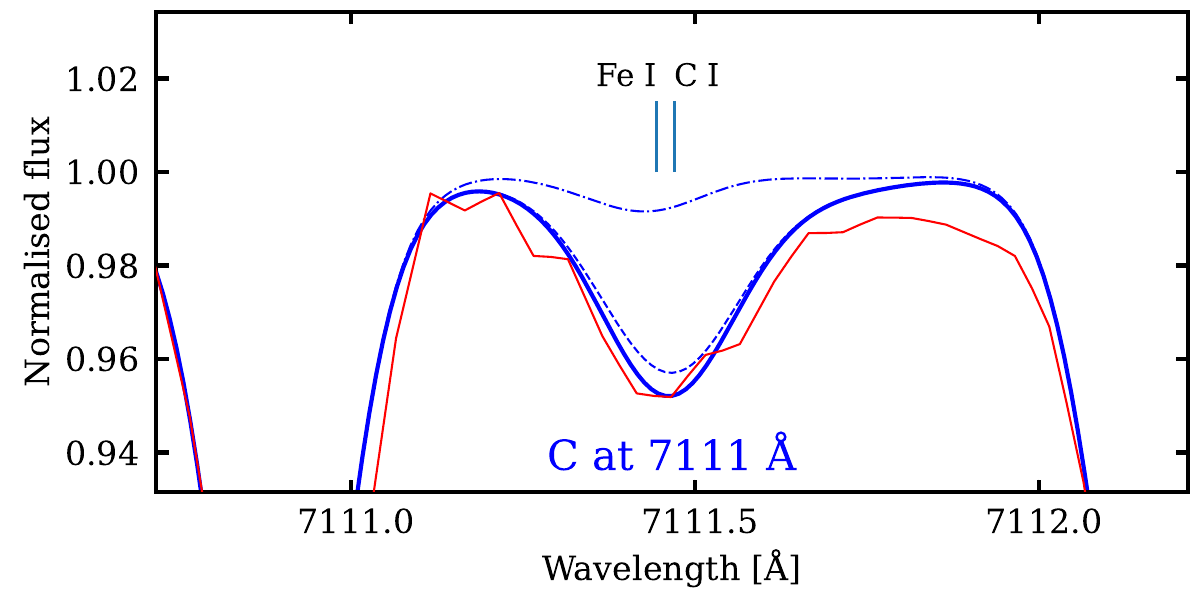}
\hspace{15mm}
\includegraphics[trim={0 9.5mm 0 0}]{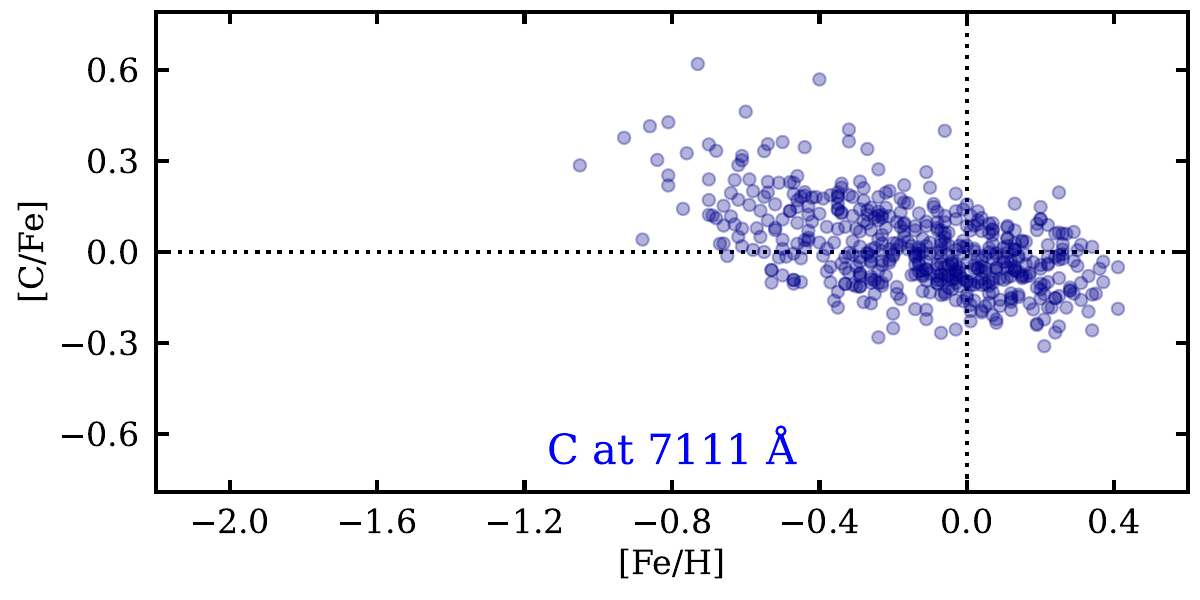}}
\resizebox{0.85\hsize}{!}{
\includegraphics[trim={0 0 0 0}]{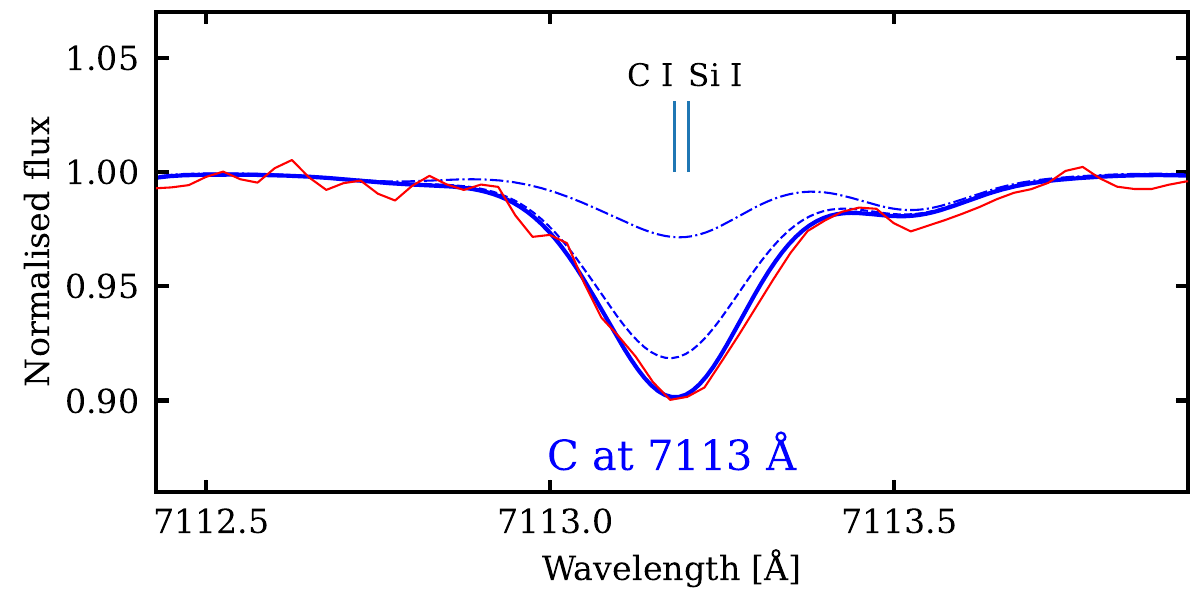}
\hspace{15mm}
\includegraphics[trim={0 0 0 0}]{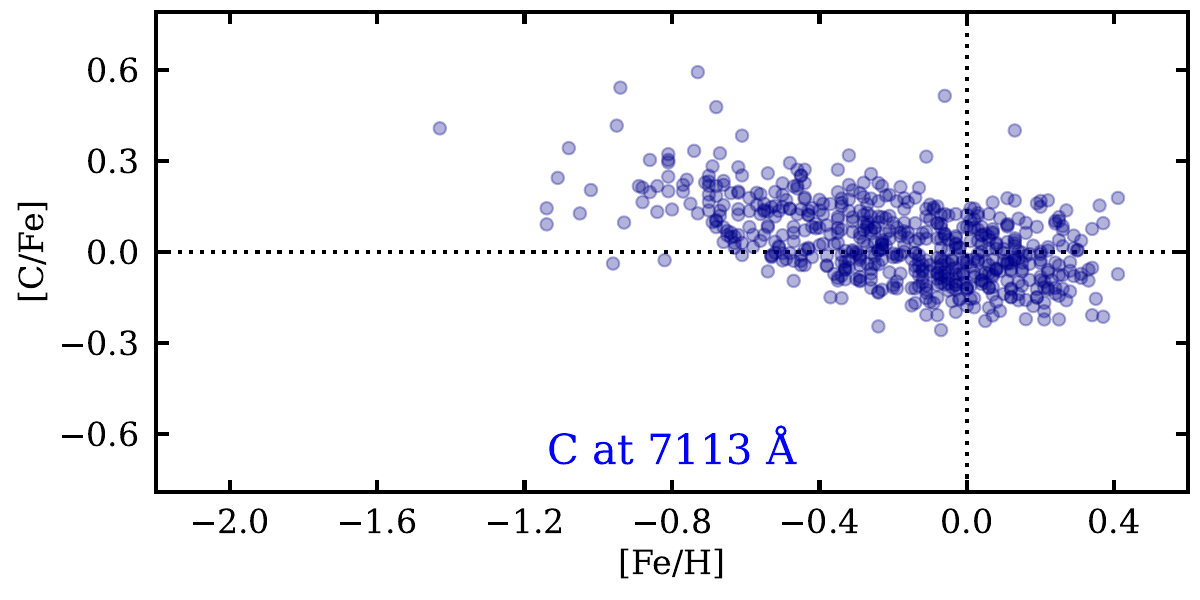}}
\caption{Analysis of carbon. {\sl Left column:} The observed spectrum for HIP~13341 (red lines) and the best fitting synthetic spectrum (solid blue line). The dotted line shows the spectrum when the blending lines are removed, hence highlighting the carbon contribution to the joint line profile, and the dash-dotted line shows the spectrum when the carbon line is removed, hence showing the strength of the blending lines.  {\sl Right column:} The [C/Fe] versus [Fe/H] trend based on the individual lines shown on the left-hand side.
\label{fig:clines}
}
\end{figure*}
\setcounter{figure}{0}
\begin{figure*}
\centering
\resizebox{0.9\hsize}{!}{
\includegraphics[trim={0 9.5mm 0 0}]{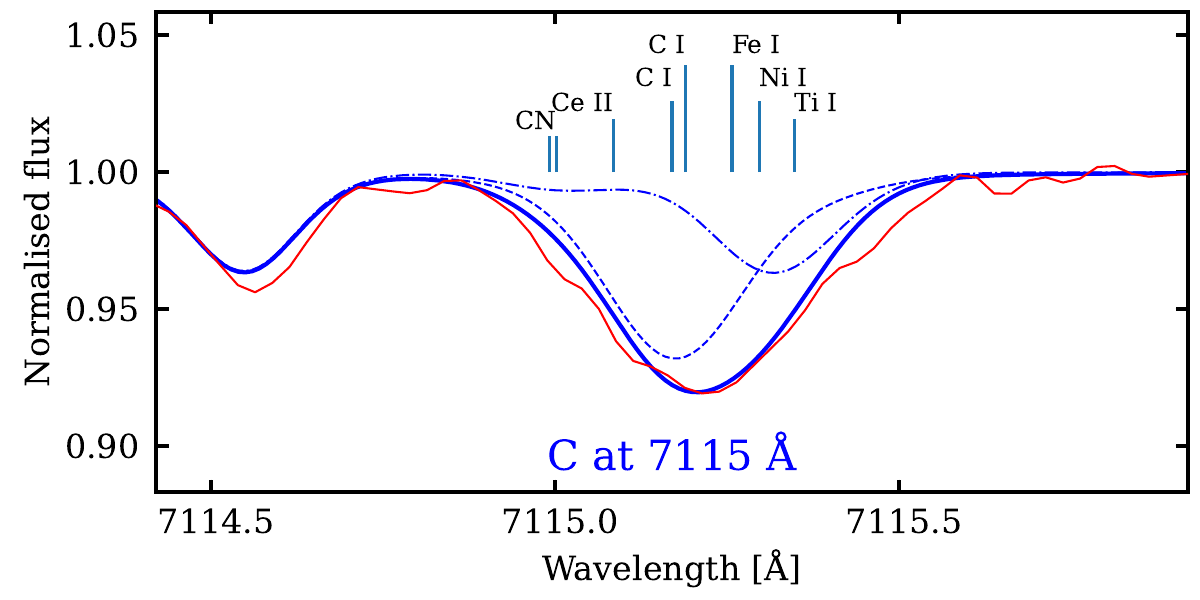}
\hspace{15mm}
\includegraphics[trim={0 9.5mm 0 0}]{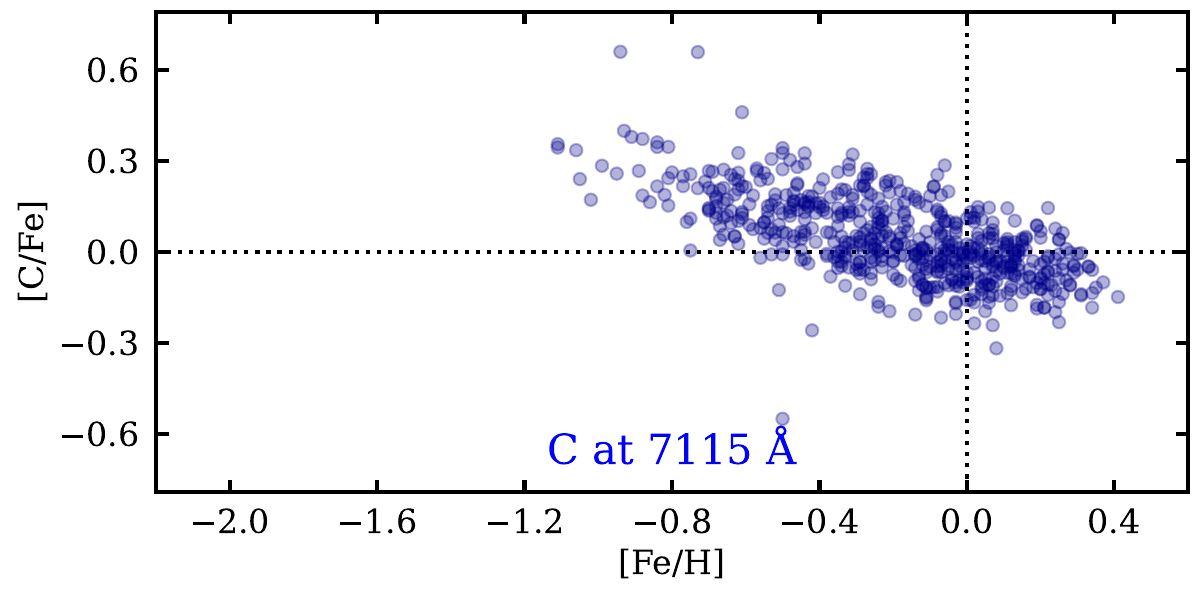}}
\resizebox{0.9\hsize}{!}{
\includegraphics[trim={0 9.5mm 0 0}]{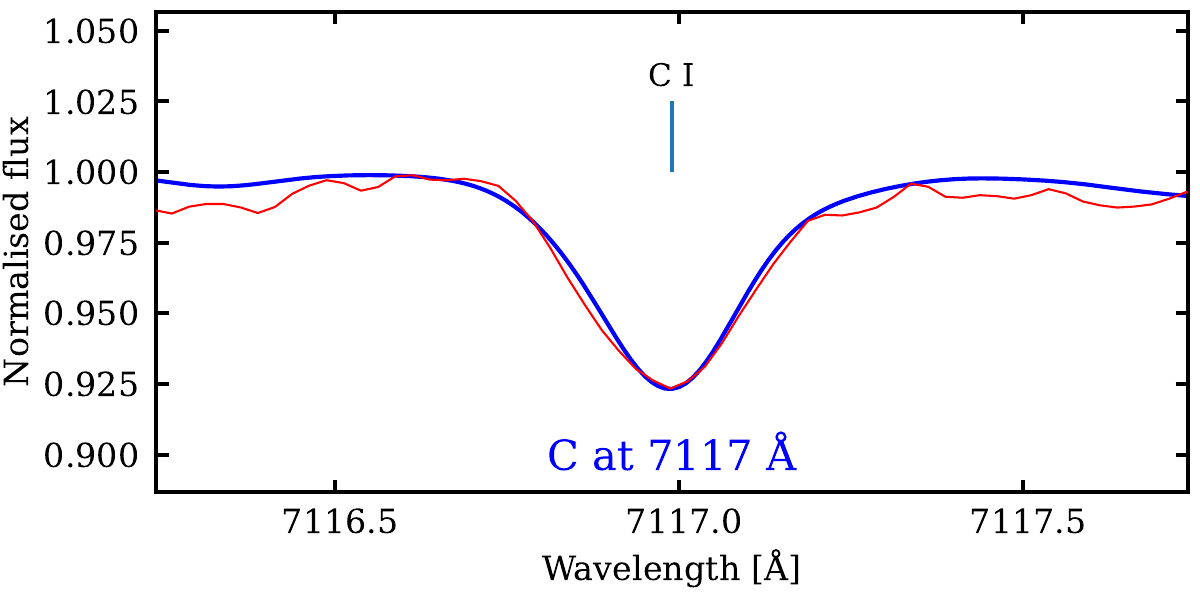}
\hspace{15mm}
\includegraphics[trim={0 9.5mm 0 0}]{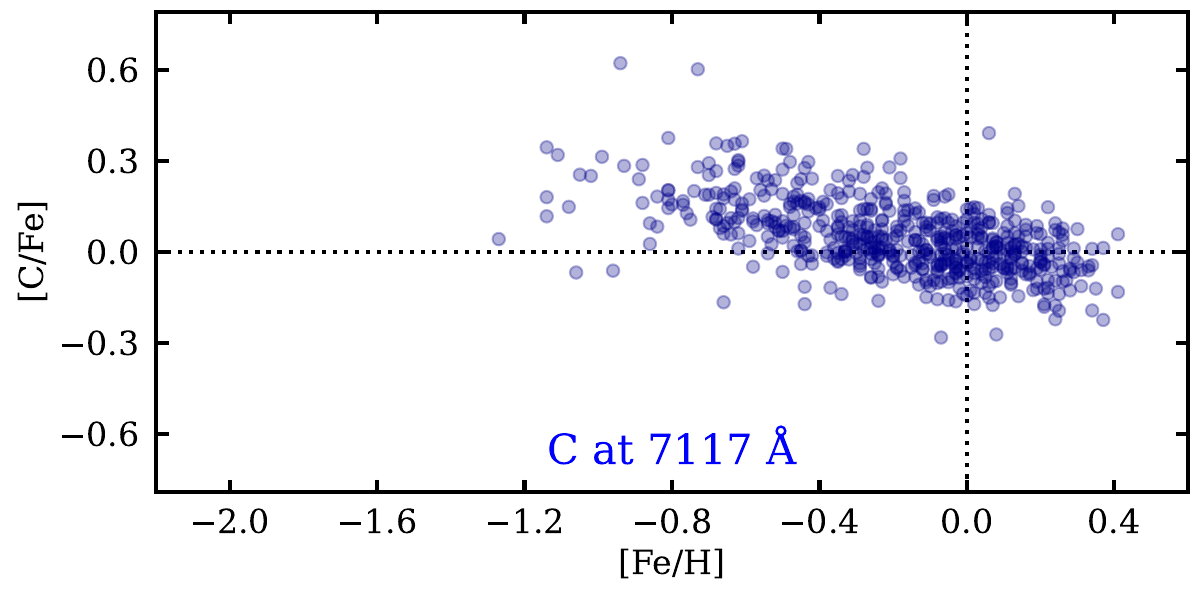}}
\resizebox{0.9\hsize}{!}{
\includegraphics{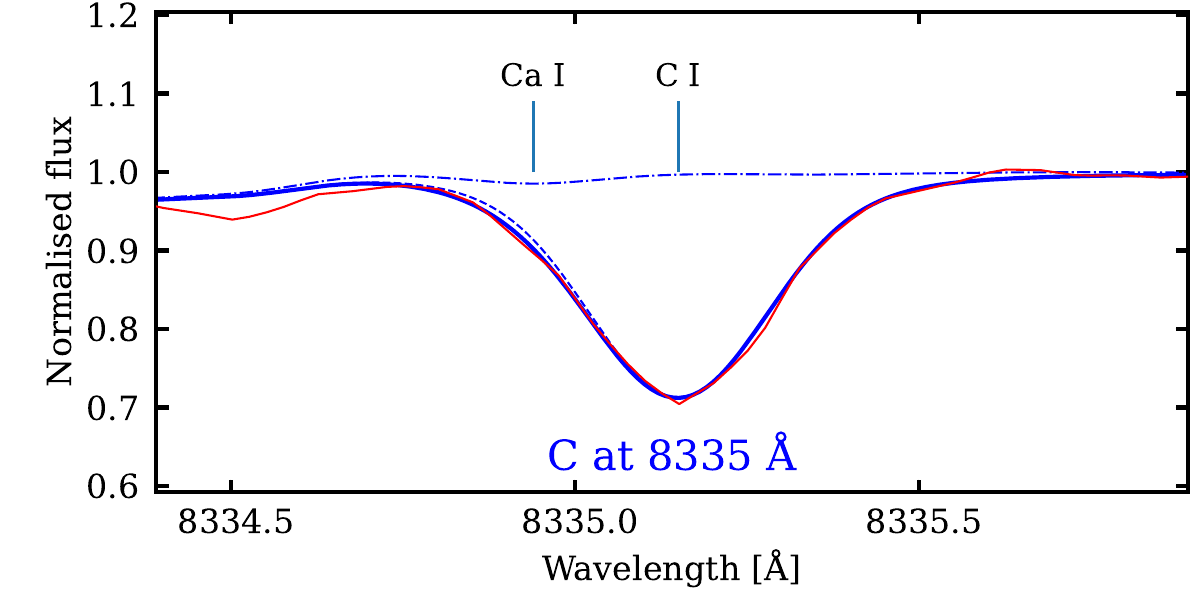}
\hspace{15mm}
\includegraphics{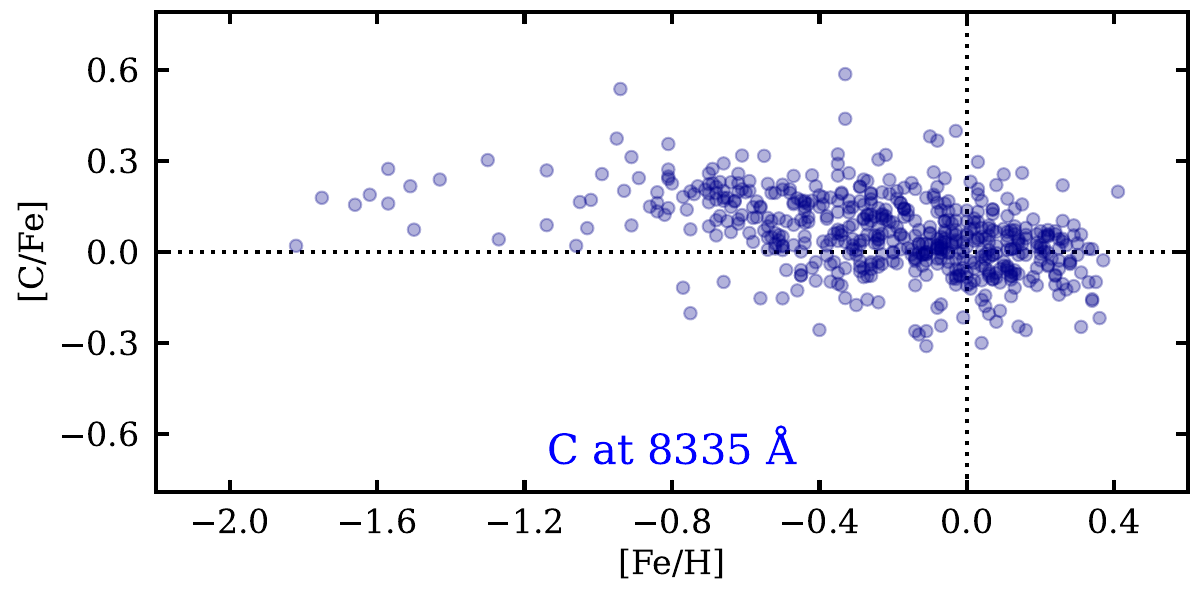}}
\caption{{\it continued}
}
\end{figure*}
\begin{figure*}
\centering
\resizebox{0.9\hsize}{!}{
\includegraphics[trim={0 9.5mm 0 0}]{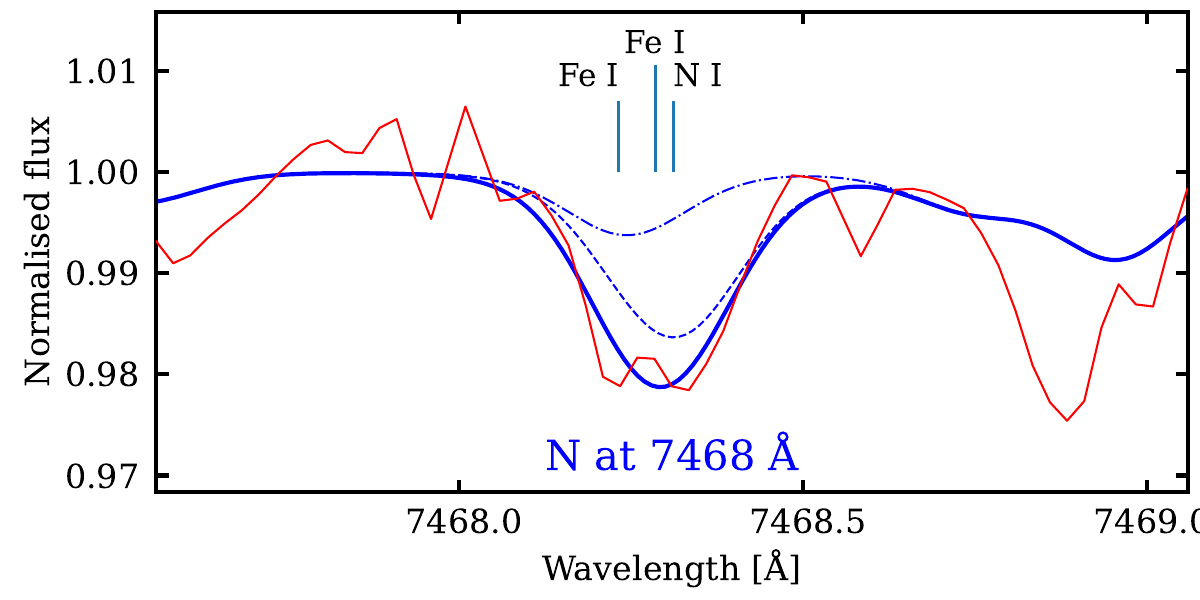}
\hspace{15mm}
\includegraphics[trim={0 9.5mm 0 0}]{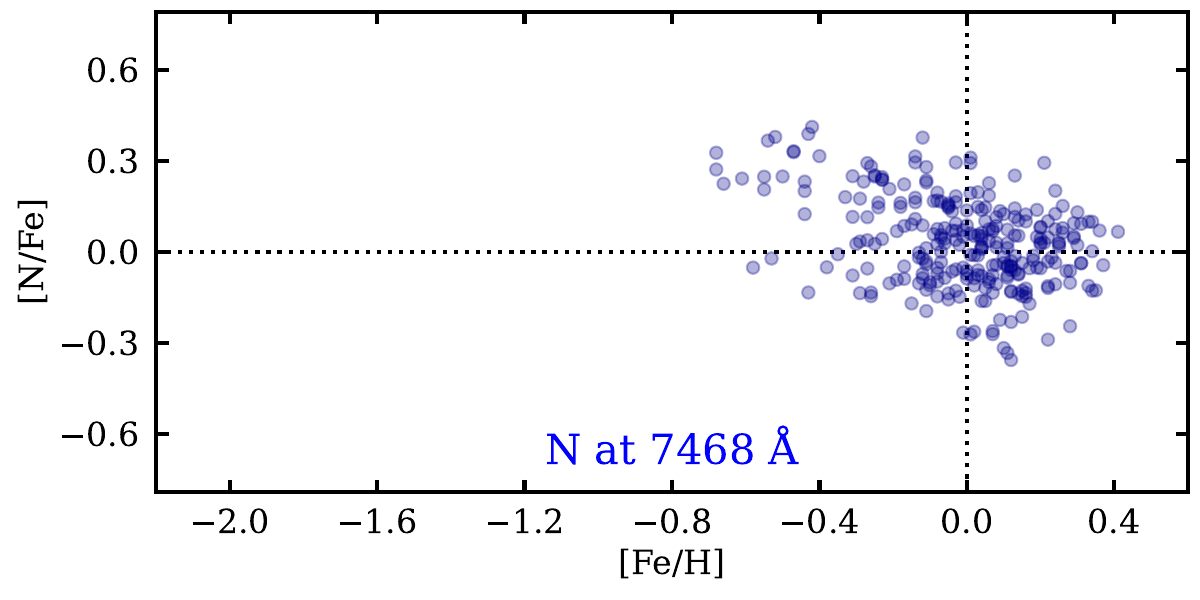}}
\resizebox{0.9\hsize}{!}{
\includegraphics{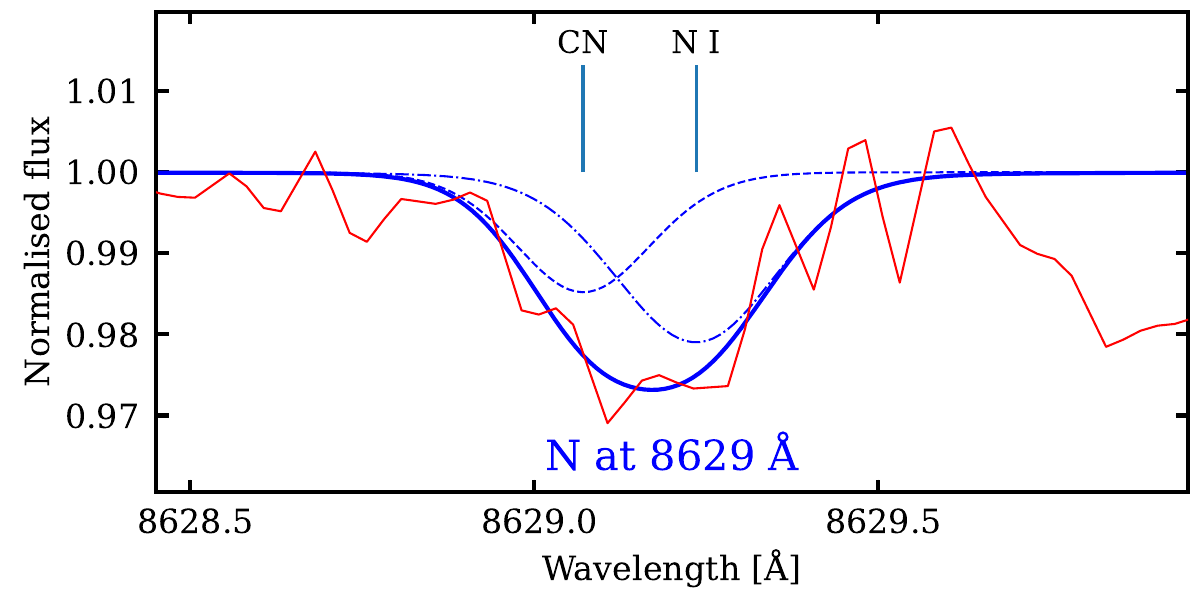}
\hspace{15mm}
\includegraphics{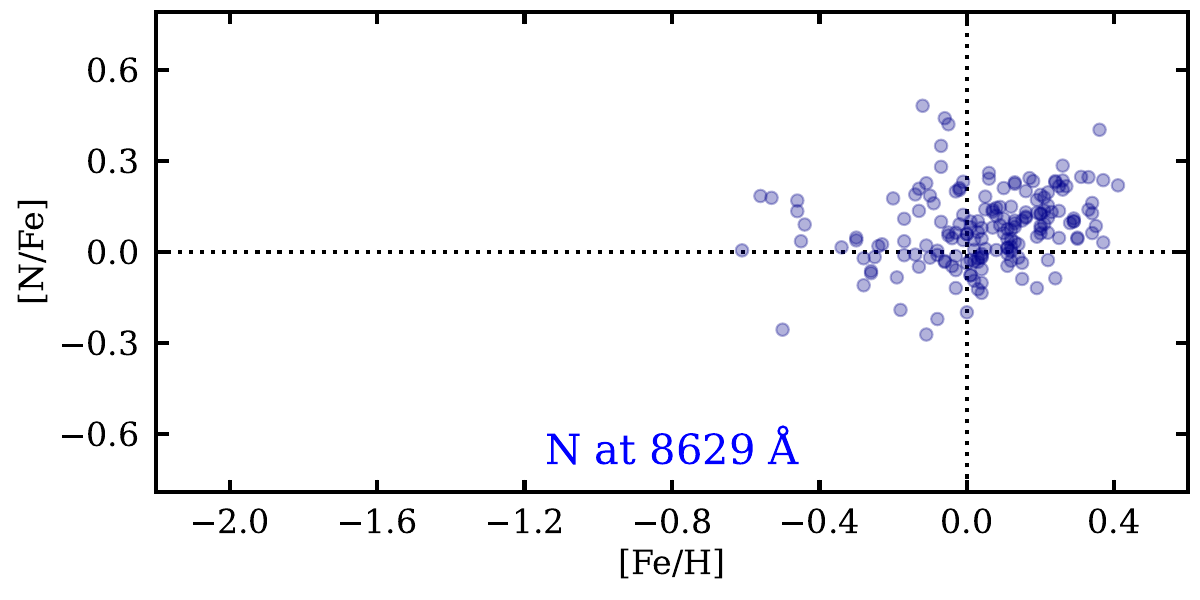}}
\caption{Same as Fig.~\ref{fig:clines}, but for nitrogen. The panels show abundances derived from the individual \ion{N}{i} and CN+N features used in the analysis.
\label{fig:nlines}
}
\end{figure*}
\begin{figure*}
\centering
\resizebox{0.9\hsize}{!}{
\includegraphics[trim={0 9.5mm 0 0}]{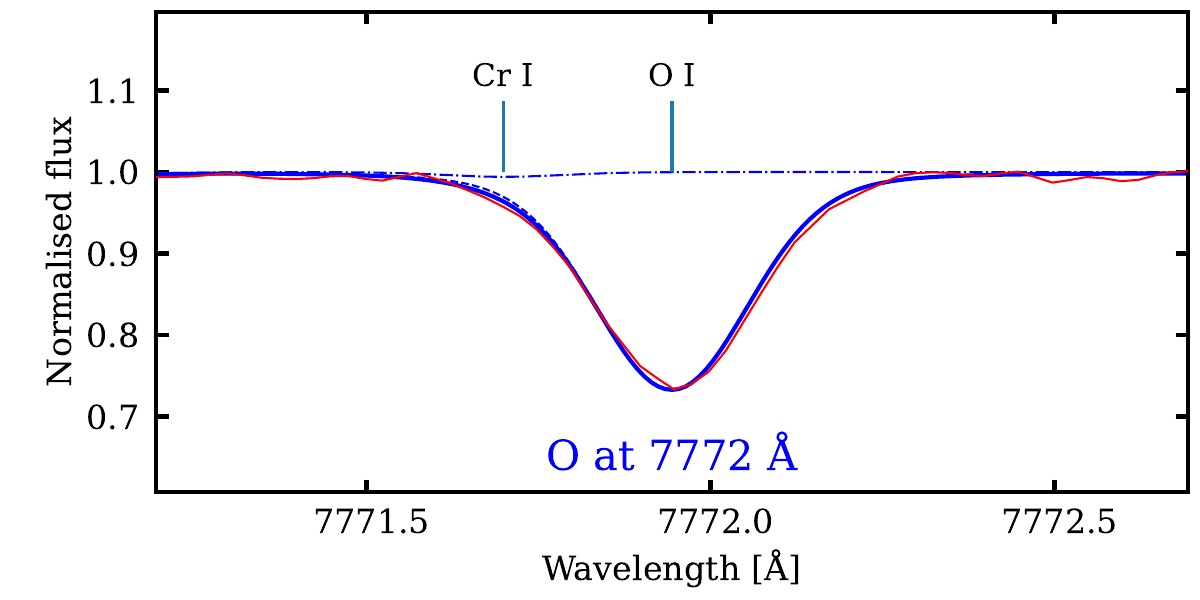}
\hspace{15mm}
\includegraphics[trim={0 9.5mm 0 0}]{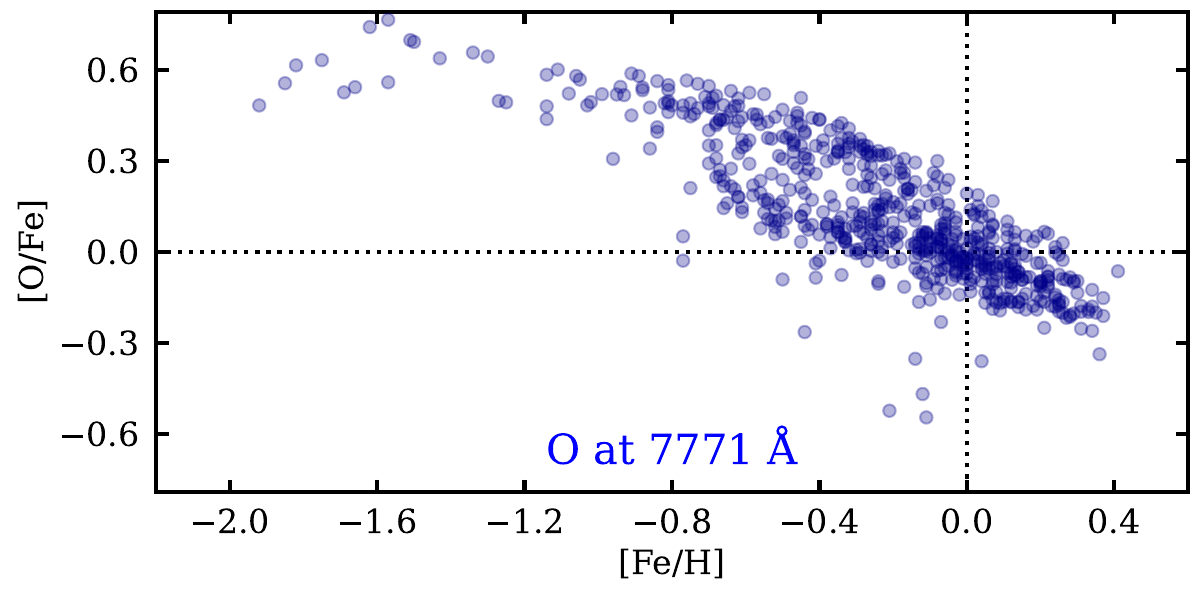}}
\resizebox{0.9\hsize}{!}{
\includegraphics[trim={0 9.5mm 0 0}]{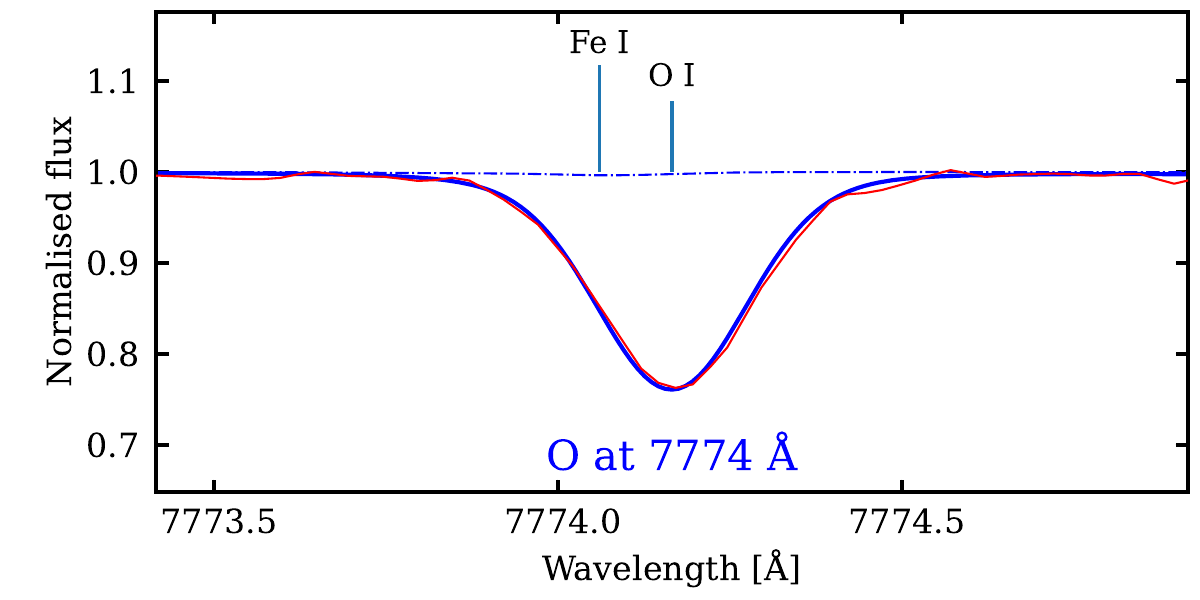}
\hspace{15mm}
\includegraphics[trim={0 9.5mm 0 0}]{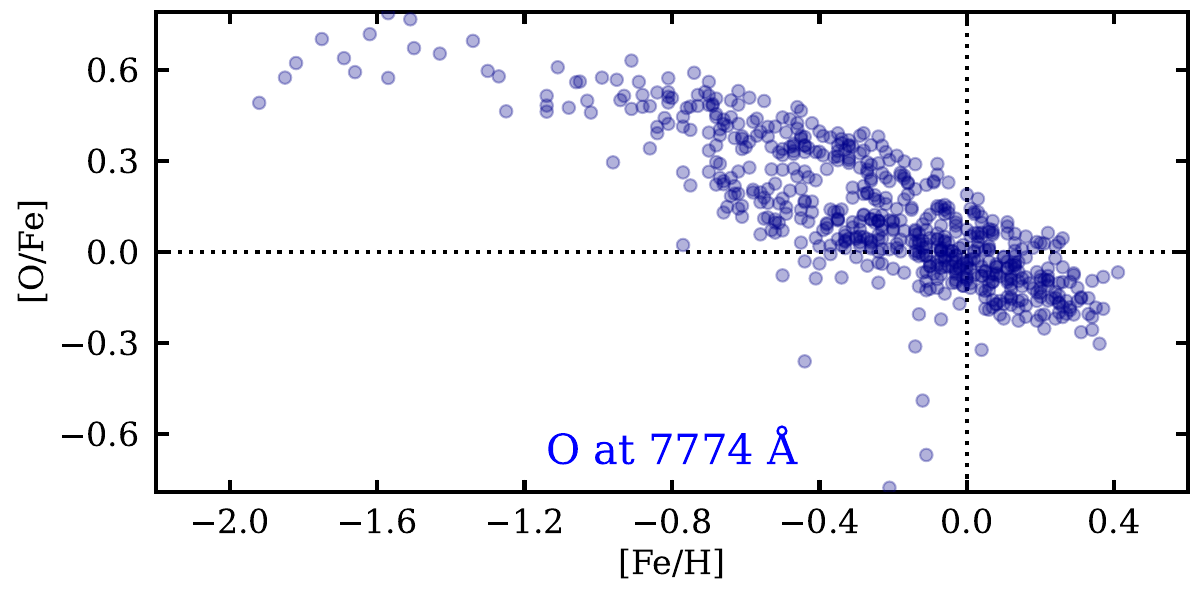}}
\resizebox{0.9\hsize}{!}{
\includegraphics[trim={0 0 0 0}]{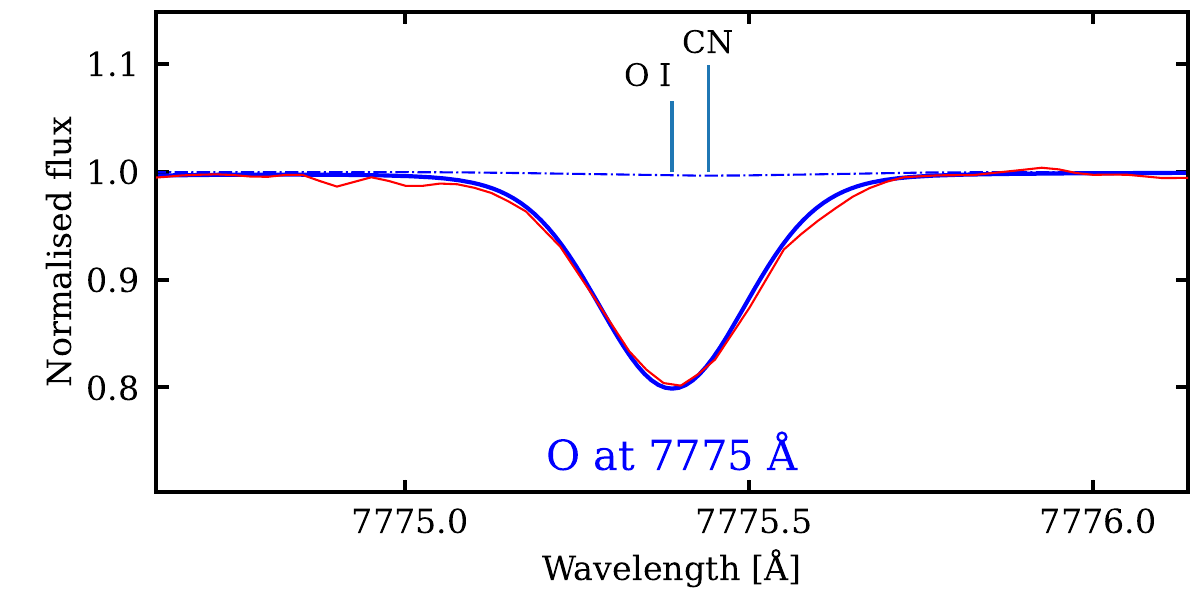}
\hspace{15mm}
\includegraphics[trim={0 0 0 0}]{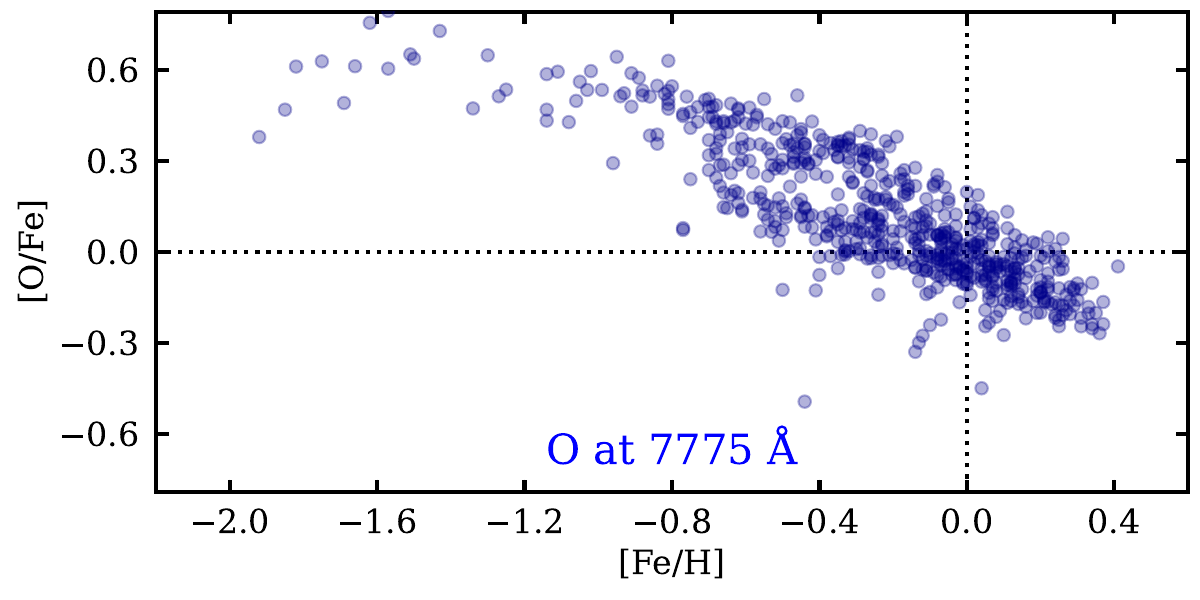}}
\caption{Same as Fig.~\ref{fig:clines}, but for oxygen. The panels show abundances derived from the three components of the \ion{O}{i} triplet at 7771–7775\,\AA.
\label{fig:olines}
}
\end{figure*}
\begin{figure*}
\centering
\resizebox{0.9\hsize}{!}{
\includegraphics[trim={0 9.5mm 0 0}]{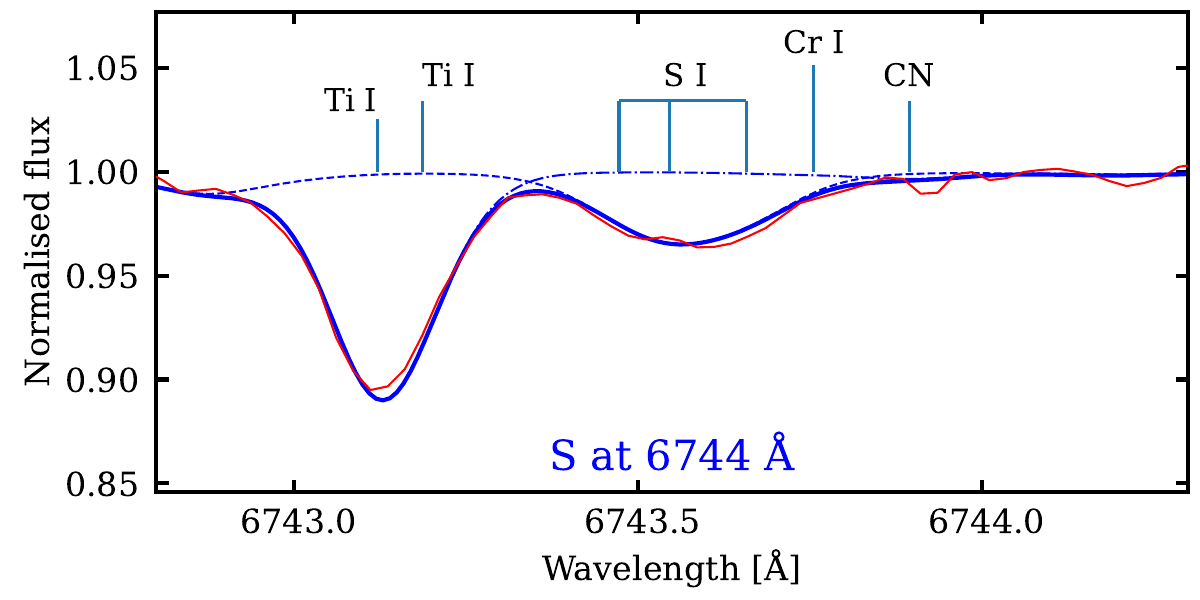}
\hspace{15mm}
\includegraphics[trim={0 9.5mm 0 0}]{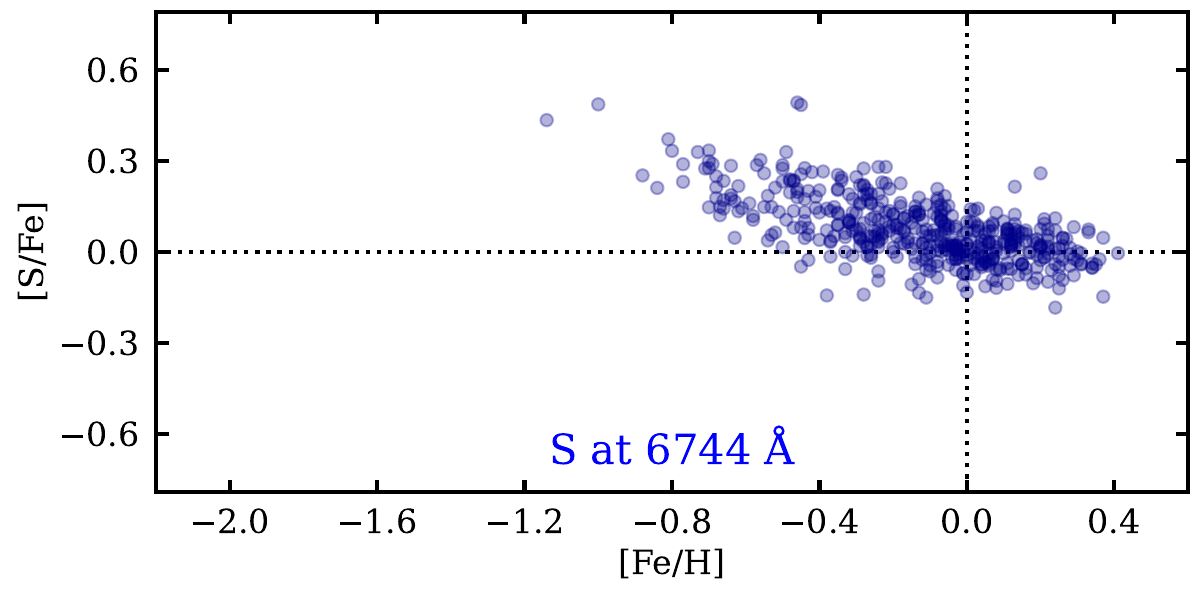}}
\resizebox{0.9\hsize}{!}{
\includegraphics[trim={0 9.5mm 0 0}]{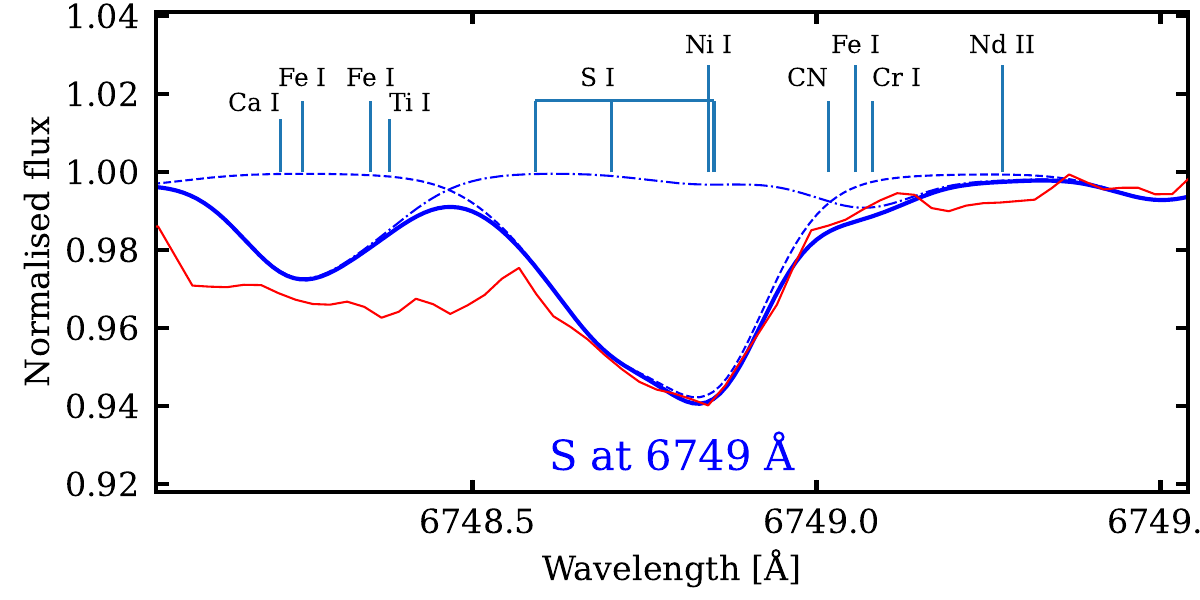}
\hspace{15mm}
\includegraphics[trim={0 9.5mm 0 0}]{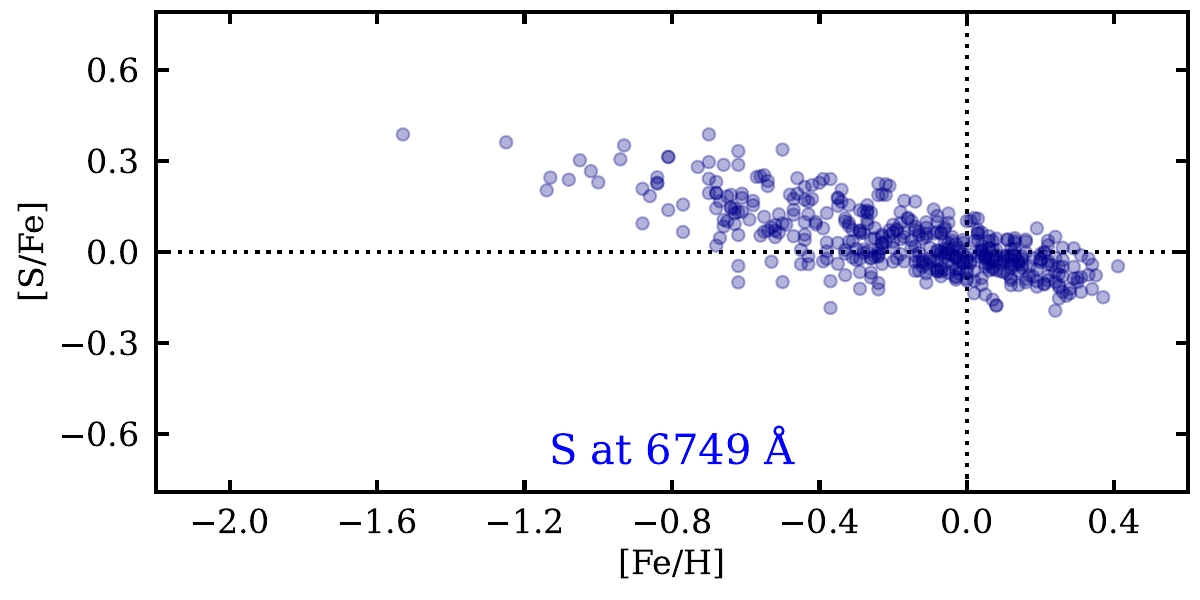}}
\resizebox{0.9\hsize}{!}{
\includegraphics[trim={0 0 0 0}]{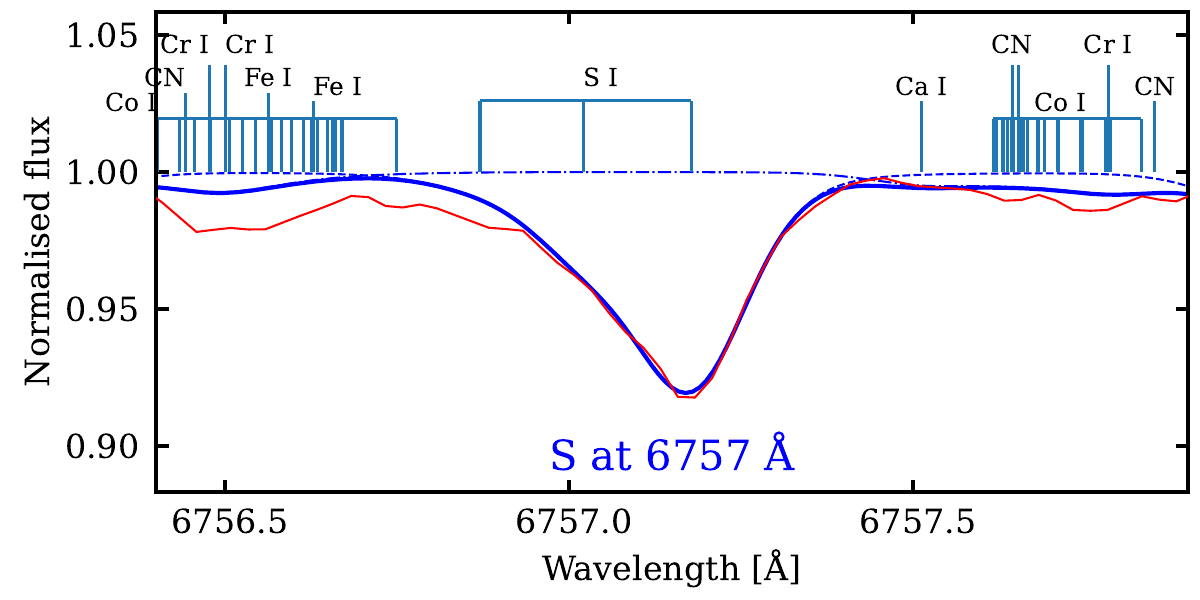}
\hspace{15mm}
\includegraphics[trim={0 0 0 0}]{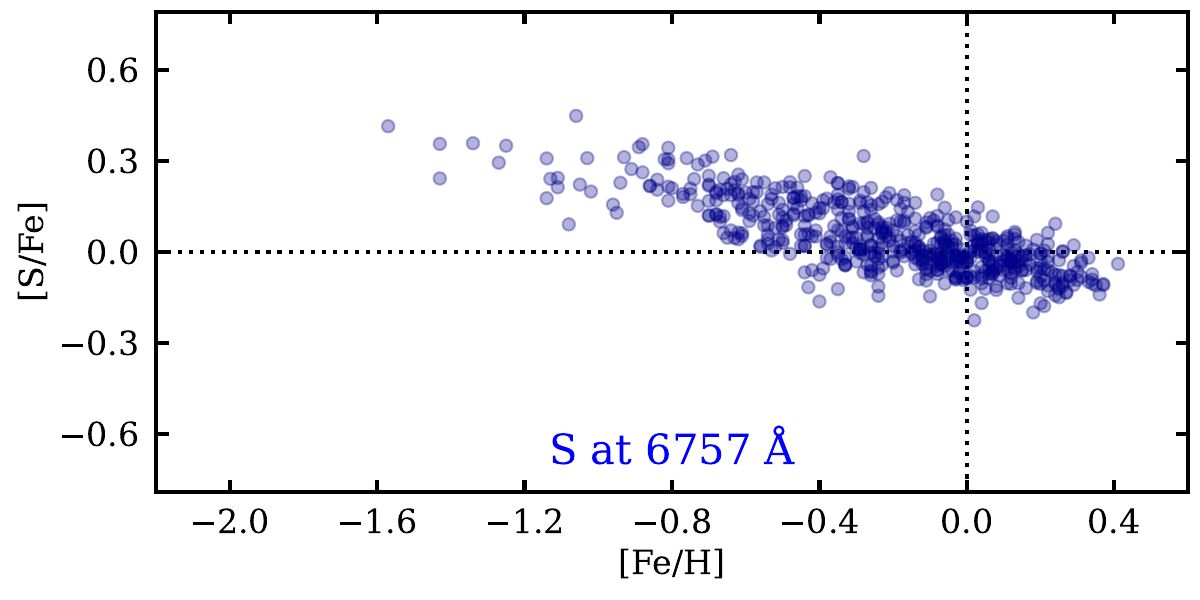}}
\caption{Same as Fig.~\ref{fig:clines}, but for sulphur. The panels show abundances derived from the individual \ion{S}{i} lines analysed in this work.
\label{fig:slines}
}
\end{figure*}
\begin{figure*}
\centering
\resizebox{0.9\hsize}{!}{
\includegraphics{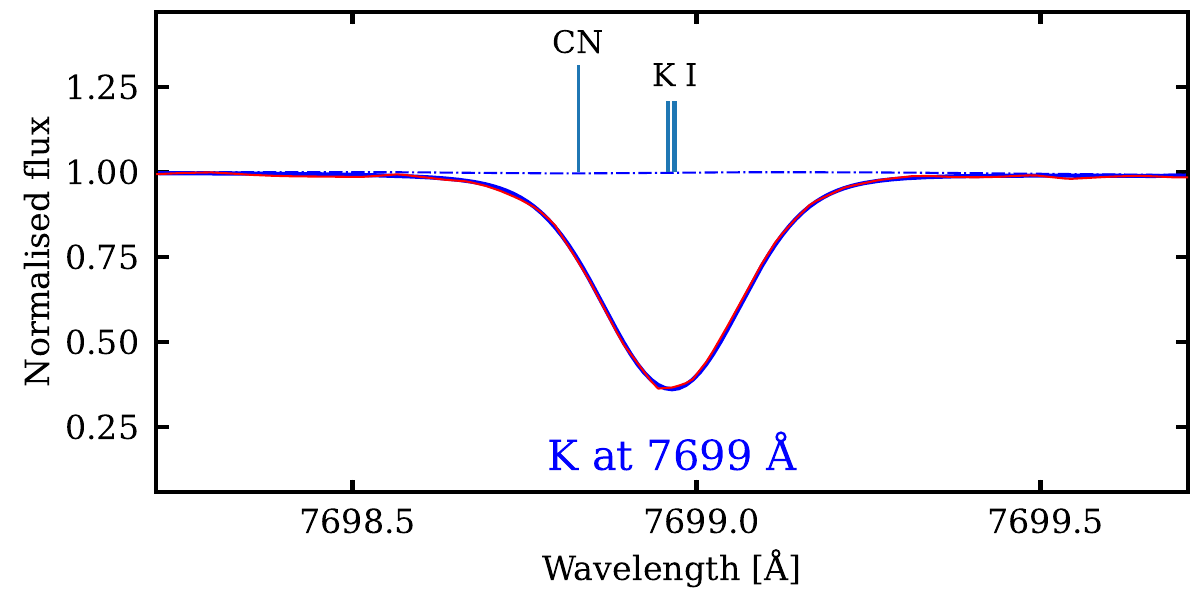}
\hspace{15mm}
\includegraphics{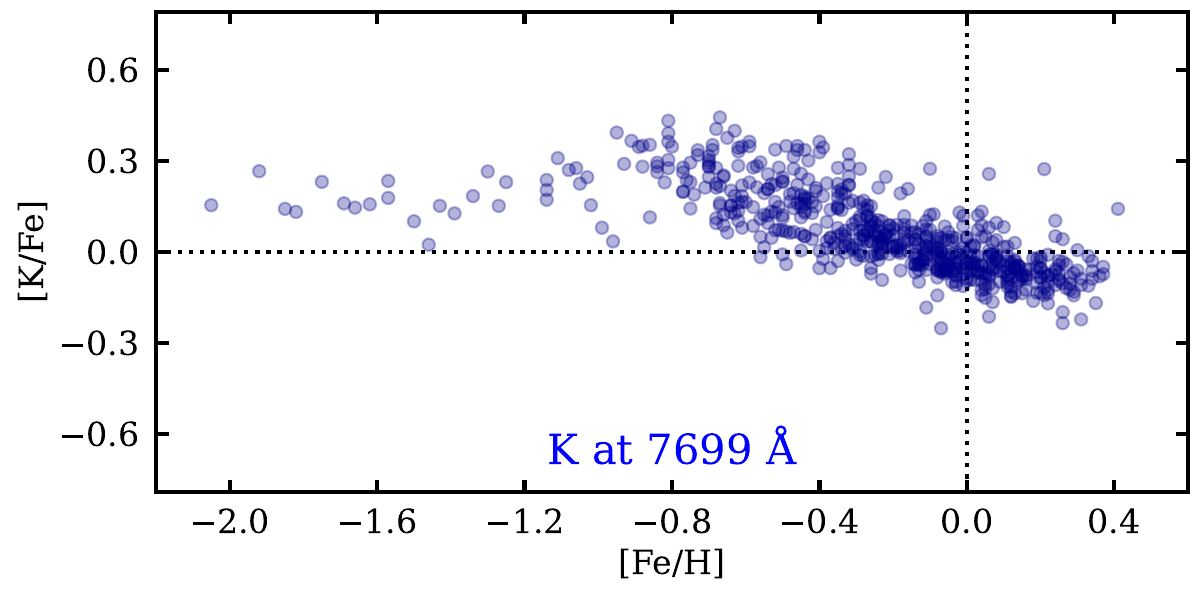}}
\caption{Same as Fig.~\ref{fig:clines}, but for potassium. The panels show abundances derived from the \ion{K}{i} line at 7699\,\AA.
\label{fig:klines}
}
\end{figure*}
\begin{figure*}
\centering
\resizebox{0.9\hsize}{!}{
\includegraphics[trim={0 9.5mm 0 0}]{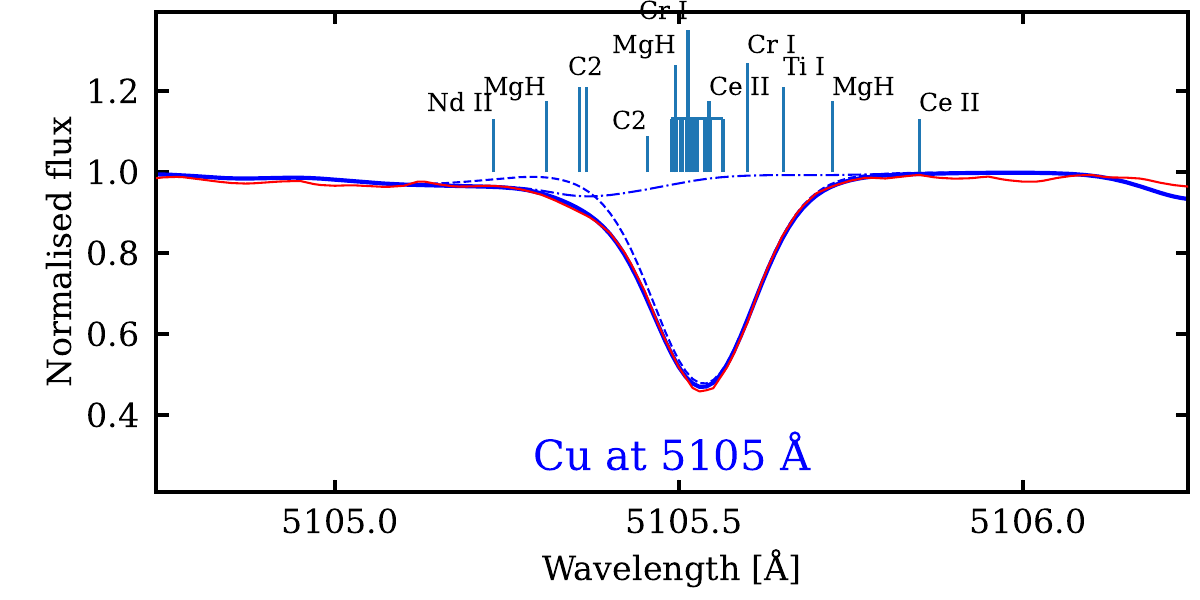}
\hspace{15mm}
\includegraphics[trim={0 9.5mm 0 0}]{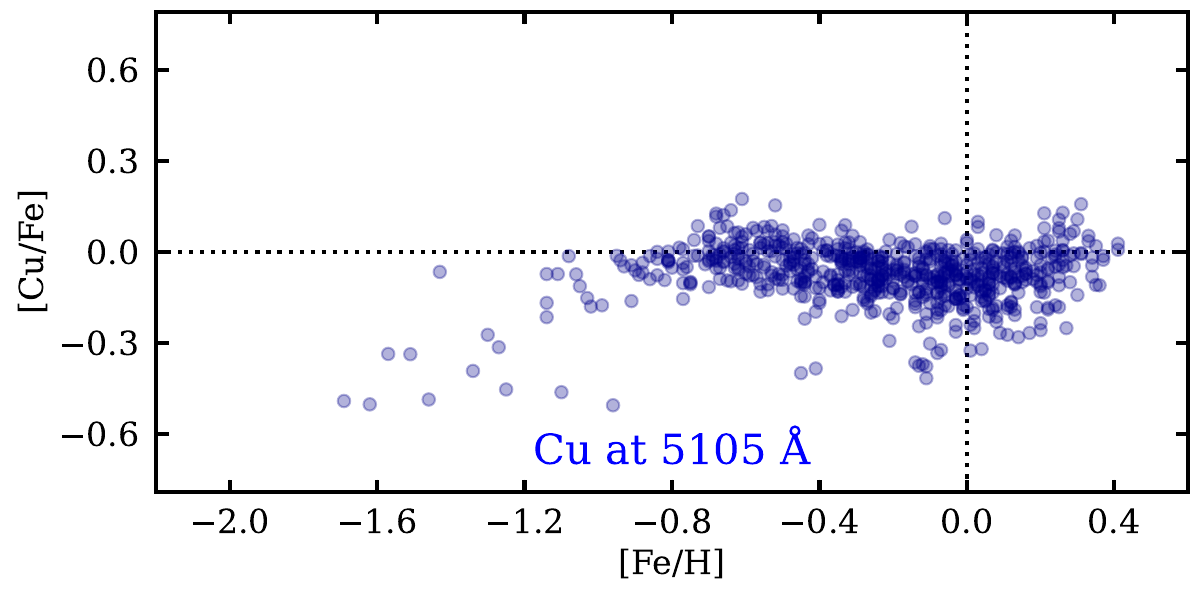}}
\resizebox{0.9\hsize}{!}{
\includegraphics[trim={0 9.5mm 0 0}]{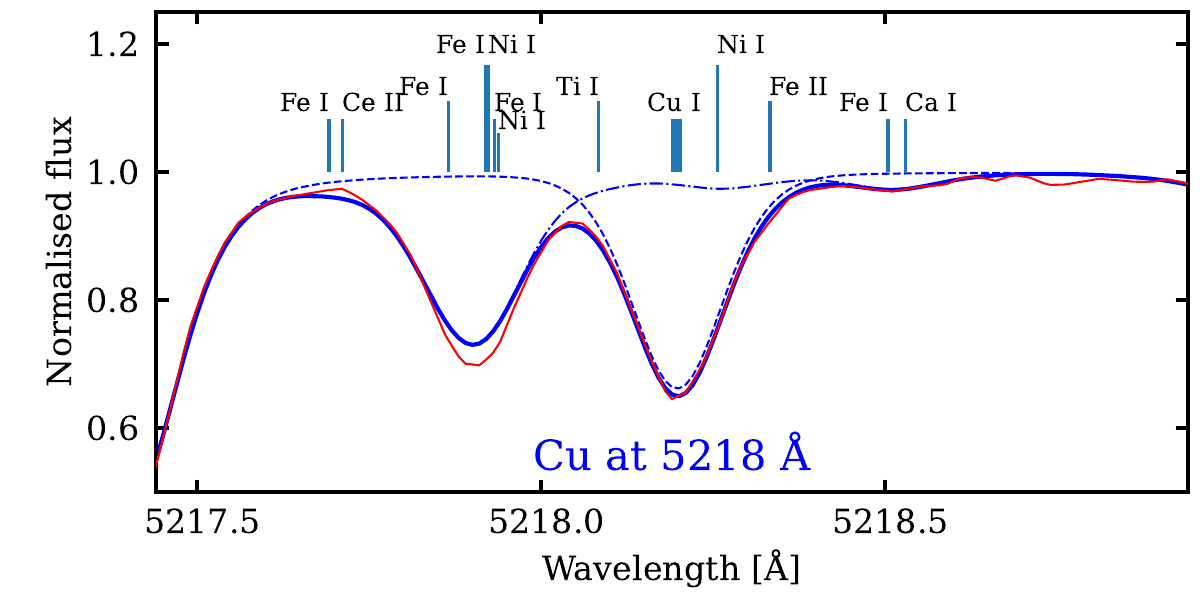}
\hspace{15mm}
\includegraphics[trim={0 9.5mm 0 0}]{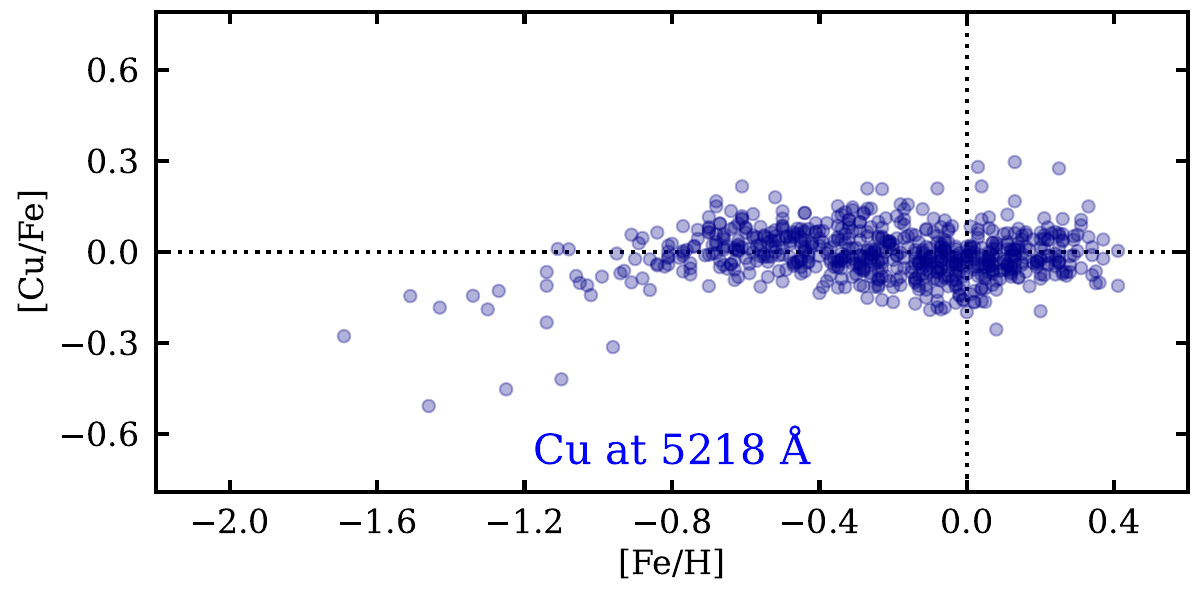}}
\resizebox{0.9\hsize}{!}{
\includegraphics{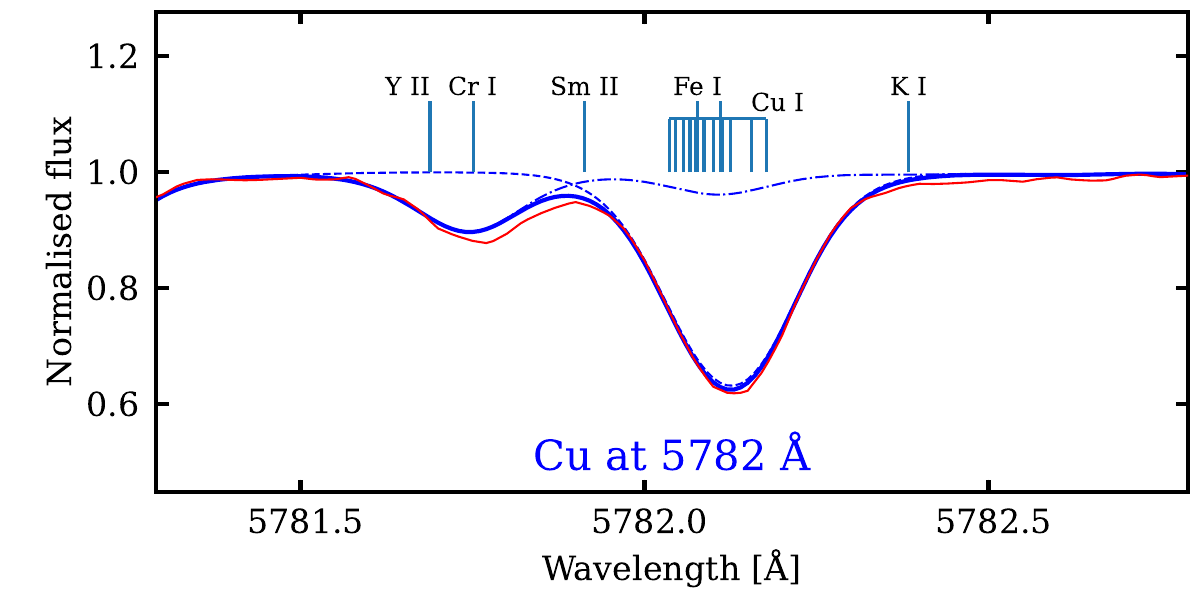}
\hspace{15mm}
\includegraphics{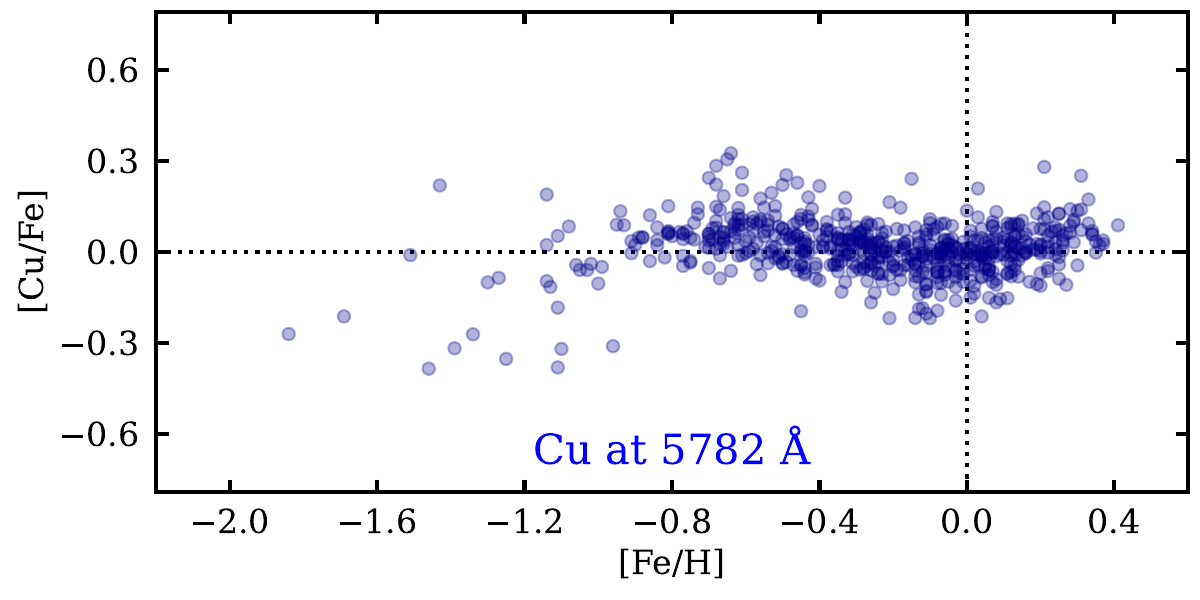}}
\caption{Same as Fig.~\ref{fig:clines}, but for copper. The panels show abundances derived from the individual \ion{Cu}{i} lines at 5105, 5218, and 5782\,\AA.
\label{fig:culines}
}
\end{figure*}

\end{appendix}

\end{document}